\documentclass[aps,prx,reprint,
superscriptaddress,floatfix,longbibliography,footinbib]{revtex4-1}
\usepackage{mathtools}
\usepackage{amsfonts}
\usepackage{amsmath}
\usepackage{physics}
\usepackage{float}
\usepackage{bbold}
\usepackage{bm}
\usepackage{xcolor}

\newcommand{\detune}{\Delta_C}
\newcommand{\wz}{\omega_z}

\newcommand{\Nm}{{N_m}}
\newcommand{\Nc}{N}
\newcommand{\Na}{M}

\newcommand{\loss}{\kappa}
\newcommand{\eigmax}{{\lambda_{\mathrm{max}}}}
\newcommand{\evec}{\mathbf{v}}
\newcommand{\ex}[1]{{\langle #1 \rangle}}
\newcommand{\pos}{\mathbf{r}}

\newcommand{\be}{\begin{equation}}
\newcommand{\ee}{\end{equation}}
\newcommand{\bea}{\begin{eqnarray}}
\newcommand{\eea}{\end{eqnarray}}

\begin{document}

\title{Entanglement and replica symmetry breaking in a driven-dissipative\\quantum spin glass}

\author{Brendan P.~Marsh}
\affiliation{Department of Applied Physics, Stanford University, Stanford CA 94305, USA}
\affiliation{E.~L.~Ginzton Laboratory, Stanford University, Stanford, CA 94305, USA}
\author{Ronen M.~Kroeze}
\affiliation{Department of Physics, Stanford University, Stanford CA 94305, USA}
\affiliation{E.~L.~Ginzton Laboratory, Stanford University, Stanford, CA 94305, USA}
\author{Surya Ganguli}
\affiliation{Department of Applied Physics, Stanford University, Stanford CA 94305, USA}
\author{\\Sarang Gopalakrishnan}
\affiliation{Department of Electrical and Computer Engineering, Princeton University, Princeton NJ 08544, USA}
\author{Jonathan Keeling} 
\affiliation{SUPA, School of Physics and Astronomy, University of St. Andrews, St. Andrews KY16 9SS, United Kingdom}
\author{Benjamin L. Lev}
\affiliation{Department of Applied Physics, Stanford University, Stanford CA 94305, USA}
\affiliation{E.~L.~Ginzton Laboratory, Stanford University, Stanford, CA 94305, USA}
\affiliation{Department of Physics, Stanford University, Stanford CA 94305, USA}

\date{\today}

\begin{abstract}

We describe simulations of the quantum dynamics of a confocal cavity QED system that realizes an intrinsically driven-dissipative spin glass. A close connection between open quantum dynamics and replica symmetry breaking is established, in which individual quantum trajectories are the replicas. We observe that entanglement plays an important role in the emergence of replica symmetry breaking in a fully connected, frustrated spin network of up to fifteen spin--1/2 particles.  Quantum trajectories of entangled spins reach steady-state spin configurations of lower energy than that of semiclassical trajectories.  Cavity emission allows monitoring of the continuous stochastic evolution of spin configurations, while backaction from this projects entangled states into states of broken Ising and replica symmetry. The emergence of spin glass order manifests itself through the simultaneous absence of magnetization and the presence of nontrivial spin overlap density distributions among replicas. Moreover, these overlaps reveal incipient ultrametric order, in line with the Parisi RSB solution ansatz for the Sherrington-Kirkpatrick model.  A nonthermal Parisi order parameter distribution, however, highlights the driven-dissipative nature of this quantum optical spin glass. This practicable system could serve as a testbed for exploring how quantum effects enrich the physics of spin glasses.

\end{abstract}

\maketitle

\section{Introduction}

In a spin glass, quenched disorder and the resulting frustration of spin-spin interactions generate a rugged free energy landscape with many minima.  This means that in some cases, below a critical temperature, the single paramagnetic thermodynamic state fractures into a multitude of distinct possible thermodynamic states~\cite{Nishimori2001spo}.  The number of such states is exponential in the system size. A consequence of this is that exact copies---replicas---of such a system may cool into distinct thermodynamic states. This is replica symmetry breaking (RSB), which Parisi invoked~\cite{Parisi1980top,Parisi1983opf} to solve the Sherrington-Kirkpatrick (SK) model~\cite{Sherrington1975smo}.  The SK model describes a network of spins with all-to-all couplings with random signs. The Parisi solution showed how RSB arises by studying the distribution of spin overlaps between different replicas, as captured by the Parisi order parameter and the ultrametric, clustered tree-like structure of the distances between replicas.  Since these features depend on details of the different thermodynamic states, they cannot be identified purely by looking at averaged properties~\cite{Stein2013sga}.
 
Replica symmetry breaking is one example of the idea of ergodicity breaking.  In an ergodic system, the dynamics of the system explores all allowed states, such that time averages are equivalent to configuration-space averages; this equivalence between time- and configuration-averages fails in states with RSB.
This has important consequences when considering the relation between individual quantum trajectories of the system and the trajectory-averaged density matrix.  
Theoretically, studying individual trajectories corresponds to stochastic unraveling of the density matrix equation of motion~\cite{Breuer2007tto,Daley2014qta,Verstraelen2018gqt}. Physically, unraveling the dynamics into trajectories corresponds to treating the system-environment interaction as a generalized measurement of the system by the environment. Each measurement projects the system into a specific state, conditional on the measurement outcome. The sequence of measurement outcomes (and associated states) is called a quantum trajectory.

We show that quantum trajectories can act as replicas to directly probe RSB. To see this link, we first discuss a simpler case, that of symmetry breaking in a standard second-order phase transition. When a perfectly isolated quantum system undergoes spontaneous symmetry breaking, it enters a macroscopic superposition, or `cat state,' of the symmetry-broken states. This cat state is extremely fragile: Any interaction between the system and the environment allows the environment to learn which state the system chose. Backaction from measuring the environment stochastically collapses the system into one of the symmetry-broken states. Thus, each run of the experiment yields a symmetry-broken state, although the ensemble of states, averaged over experimental runs, remains symmetric. 

The above simple picture also extends to the case where, rather than a small number of symmetry-broken states, one has many complex ergodicity-breaking patterns, as in a spin glass.  In the case of a cavity QED system, each thermodynamic state emits a characteristic pattern of photons into the environment. On each run of the experiment, the backaction from observing this field collapses the system into a distinct thermodynamic state. This corresponds to the notion of a `weak' symmetry that is broken in individual experimental runs but not in the ensemble-averaged density matrix~\cite{lieu2020symmetry}. Thus, because there is a one-to-one correspondence between thermodynamic states and emission patterns, the overlap distribution is accessible through the correlations between the photon measurement records on distinct runs of the experiment.

We investigate the emergence of RSB in the open quantum system dynamics of confocal cavity QED, an experimentally practicable setting~\cite{Vaidya2018tpa,Guo2019spa,Guo2021aol,Kroeze2023rsb}. In this system, atoms represent individual spins, while the cavity provides an all-to-all but sign-changing random interaction, dependent on the position of the atoms.  This position dependence means it is possible to achieve random but repeatable interactions by controlling the placement of the atoms. By monitoring the spatiotemporal correlations of the light leaking out of the cavity, one can reconstruct the dynamics along individual trajectories.  Because monitoring provides access to these correlations, the cavity QED setting gives us a powerful way to study RSB. In the RSB phase, the dynamics along each trajectory reaches a specific nonergodic state, so the spin configuration (and hence that of the emitted light) is stable over time on that trajectory. We quantify the distribution of overlaps among the patterns from different trajectories and the resulting Parisi order parameter. Together these show the distinctive features predicted by the Parisi ansatz for the SK spin glass.

We consider a realization where spins correspond to single atoms, giving spin-1/2 (and thus quantum) degrees of freedom, allowing entanglement to play a role in the emergent spin organization. We show that the quantum dynamics are distinct from the semiclassical limit, in which a semiclassical energy barrier severely inhibits passage to a low-energy manifold of states. By transitioning via entangled states, the trajectory avoids semiclassical energy barriers that would otherwise bar access to the low-energy spin manifold where RSB occurs. 

\begin{figure}[t!]
    \centering
    \includegraphics[width=\columnwidth]{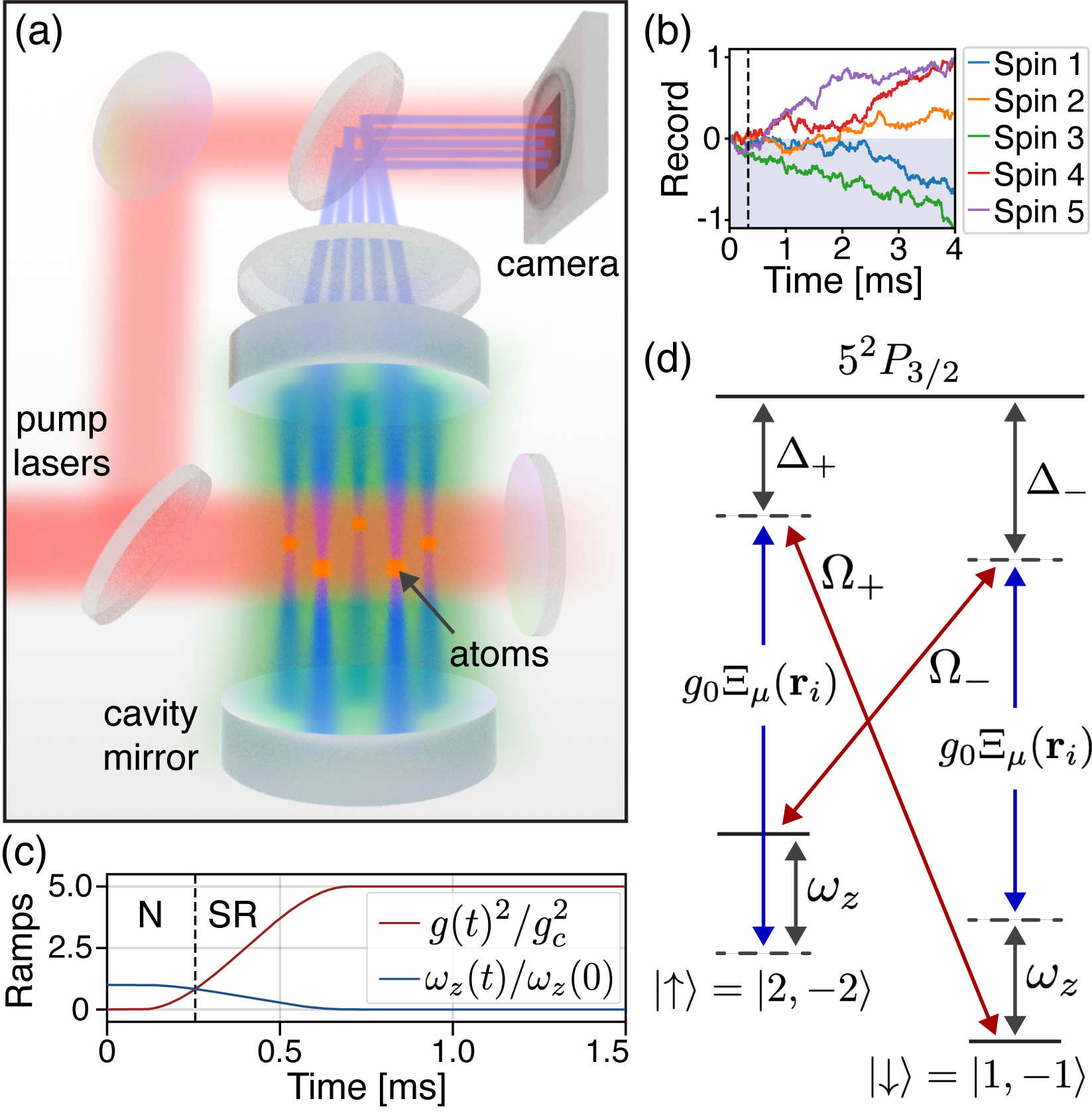}
    \caption{(a) Sketch of the confocal cavity QED system. Transverse pump lasers (red) illuminate a network of atomic spin ensembles (orange) and scatter light into the cavity. The atomic spin ensembles at each node create either a collective spin or an effective spin--$1/2$ via Rydberg-blockade.  Either way, the atomic spin ensembles are held in place by optical dipole `tweezer' lasers (not shown). The confocal cavity field is composed of a local field (blue) at each spin ensemble and a nonlocal field (green) that mediates interactions between spin ensembles. The spin states are read out by imaging the cavity emission on a camera via spatial heterodyne detection~\cite{Guo2019spa}. 
    (b) Example simulated detection traces of integrated cavity emission for five spins; for clarity, the $y$-axis is normalized to the maximum signal magnitude after 4~ms. Each spin organizes into one of two orientations above the semiclassical superradiant threshold, indicated by a dashed line. (c) Plot of the ramp schedules versus time. `Normal' (N) and 'superradiant' (SR) regimes are to either side of the semiclassical threshold (dashed vertical line).  (d)  Pumping scheme for a $^{87}$Rb atom.  Balanced Raman transitions realize a pseudospin--$1/2$ degree of freedom. See text for details.}
    \label{fig1}
\end{figure}

Previously, we considered this same setup, but in the limit far above threshold where semiclassical approaches are valid, and considered the process of memory recall in such a device~\cite{Marsh2021eam}.  Here, we address a very different question, focusing on the (necessarily) quantum dynamics near threshold, and the resulting distribution of low-energy states found. A key point of this paper is that the final states at the end of the pumping sequence are classical, yet the ability to recover them relies on quantum dynamics.  

The theoretical possibility of a spin glass phase in a multimode cavity QED system was suggested in Refs.~\cite{Gopalakrishnan2011fag,Strack2011dqs,Gopalakrishnan2012emo}. We note that the driven-dissipative nature of the confocal cavity QED quantum spin glass discussed differs from previous theoretical investigations of transverse SK models in the closed-system context~\cite{Ray1989smi,Buttner1990rbf,Rotondo2015rsb,Leschke2021eor,Schindler2022vtg}. Experimental observations of RSB in physical settings have been reported in the spectra of semiclassical systems such as random lasers~\cite{ghofraniha2015eeo,gomes2016ool} and nonlinear wave propagation~\cite{pierangeli2017oor}.  Recent   {experimental} results indicate that RSB in a confocal cavity QED system has been realized using a network with many XY spins per node~\cite{Kroeze2023rsb}.   {Recent theoretical work has noted that there can be phase transitions in the entanglement and correlations along individual quantum trajectories~\cite{Skinner2019mpt,Li2018qze}, even when such transitions are absent in the trajectory-averaged density matrix~\footnote{This possibility arises because entanglement and connected correlation functions are nonlinear functions of the state along an individual trajectory.}.} 

The paper is organized as follows.  The next section describes the physical system we aim to simulate, its model Hamiltonian, and the Lindbladian dynamics to be unraveled by the trajectory simulations.  Section~\ref{sec:qtraj} presents the quantum trajectory simulation method we employ, followed by results from individual trajectories in Sec.~\ref{sec:qtrajexample}.  Also discussed in Sec.~\ref{sec:qtrajexample} are the evolution of entanglement entropy per spin and the difference in the lowest reachable configuration energy between entangled and semiclassical trajectories.   {The connection between quantum trajectories and replicas is presented in Sec.~\ref{sec:olap}.} Evidence for RSB using the spin overlap order parameter is shown in Sec.~\ref{RSBsec}.  Section~\ref{sectemp} discusses the nonequilibrium nature of the system's overlap distribution via comparisons to equilibrium distributions. Section~\ref{phases} provides evidence for spin glass, ferromagnetic, and paramagnetic phases. The emergence of an ultrametric structure between replicas is presented in Sec.~\ref{sec:ultra}. A summary and discussion of broader implications are in Sec.~\ref{DiscussSec}. Eight appendices provide information on:~\ref{app:J}, the form of the effective spin connectivity matrices in confocal cavity QED;~\ref{app:crit}, the derivation of the semiclassical critical coupling strength;~\ref{app:atomonly}, the derivation of the atom-only theory;~\ref{app:record}, the stochastic unraveling of the master equation;~\ref{app:semiclassical}, the semiclassical limit of the spin dynamics;  {~\ref{ParisiSupp}, the Parisi distribution in terms of quantum trajectories;}~\ref{app:bootstrap}, error bar estimation via bootstrap analysis; and~\ref{app:allOlaps}, the full set of overlap distributions used in forming the Parisi overlap order parameter.


\section{The confocal cavity QED\\spin system}\label{sec:ccQED}

We now provide a description of a practicable system upon which we base this numerical study.  All parameters have been experimentally realized or are plausible using existing technology~\cite{Kroeze2023hcu}. Photon-mediated spin interactions in the confocal cavity QED system were previously discussed in~\cite{Marsh2021eam} and experimentally explored in~\cite{Guo2019spa,Guo2019eab,Kroeze2023rsb}. The system is depicted in Fig.~\ref{fig1}. Two transversely orientated lasers of pump strength $\Omega_\pm$ scatter photons off a network of $\Nc$ spin ensembles, each with $\Na$ spins per ensemble, where $\Na$ can vary between one (realizing the spin--$1/2$ quantum limit) and $10^5$ (describing current experiments
~\cite{Vaidya2018tpa,Guo2019spa,Guo2019eab,Kroeze2023hcu,Kroeze2023rsb}); see Sec.~\ref{DiscussSec} and Ref.~\cite{Marsh2023rydberg} for description of a practicable spin--$1/2$ scheme. These experiments employ $^{87}$Rb in a 1-cm-long confocal cavity.  To realize the network, each spin may be trapped at a position $\pos_i$ in the midplane of the cavity using an array of optical tweezers~\cite{Kaufman2021qsw} (or optical dipole traps of larger waist~\cite{Guo2019spa,Kroeze2023rsb}). 

Similar to recent experimental work in which spins couple to a transversely pumped cavity~\cite{Kroeze2018sso,Kroeze2019dsc,Zhang2018dsv}, the (pseudo)spin states considered here correspond to $^{87}$Rb hyperfine states $\ket{\downarrow}=\ket{F=1,m_F=-1}$ and $\ket{\uparrow}=\ket{F=2,m_F=-2}$. The two pumps scatter light into the cavity via Raman transitions~\cite{Dimer2007pro}. We will consider an atomic detuning of $\Delta_\pm\approx-2\pi{\times}100$~GHz from the $5^2P_{3/2}$ atomic excited state and a controllable two-photon detuning $\omega_z\approx 2\pi{\times} 10$~kHz. The maximum single-atom, single-mode light-matter coupling strength can reach a magnitude of $g_0=2\pi\times1.5$~MHz; see Sec.~\ref{DiscussSec} for more details.  We consider a confocal cavity of even symmetry under reflection in the cavity center axis~\footnote{Confocality occurs when the cavity length $L$ equals its mirrors' radius of curvature $R$~\cite{Siegman1986l}.}. This restricts the possible cavity modes to the set of Hermite-Gauss TEM$_{lm}$ modes with indices $l+m=0\mod 2$~\cite{Guo2019eab}. An even confocal cavity retains modes of sine and cosine longitudinal character, and trapping the atoms in one of these two longitudinal quadratures with optical tweezers further restricts the set of participating modes to $l+m=0\mod 4$.  This results in the effective Ising coupling we consider here; see Ref.~\cite{Marsh2021eam} and App.~\ref{app:atomonly}. We denote the remaining mode functions by $\Xi_\mu(\pos)$, indexed by $\mu$ for brevity. A total of $\Nm$ modes participate in the near-degenerate family of confocal-cavity modes to which the atoms couple~\footnote{The effective $\Nm$ can be greater than 1,000, yielding a single atom, synthetic mode cooperativity of more than 110 in practicable systems.  See Ref.~\cite{Kroeze2023hcu} for the definition of synthetic mode cooperativity and more details.}. The modes are detuned from the mean pump frequency by $\Delta_\mu=\omega_p-\omega_\mu\approx -2\pi{\times} 80$~MHz. 

The Hamiltonian in the rotating frame of the pump, after adiabatic elimination of the atomic excited state, is a multimode Hepp-Lieb-Dicke model~\cite{Gopalakrishnan2009eca,Kirton2018itt,Mivehvar2021cqw}:
\begin{align}\label{eqn:Ham}
    H_0 &= - \sum_\mu \Delta_\mu a_\mu^\dag a_\mu+\wz\sum_{i=1}^{\Nc}S_i^z  
     \\&\quad + g\sum_{\mu}\sum_{i=1}^{\Nc} \Xi_\mu(\mathbf{r}_i)S_i^x(a^\dag_\mu + a_\mu) \nonumber.
\end{align}
The cavity modes are described by the bosonic operators $a_\mu$, while the spin ensembles are described by the collective spin operators $S_i^{x/y/z}$ to facilitate generalization to the $\Na>1$ case. In the spin--$1/2$ limit, $S_i^{x/y/z}=\sigma_i^{x/y/z}/2$. The effective coupling strength $g=\sqrt{3}g_0\Omega_\pm/12\Delta_\pm$ is the same for each pump laser, which can be achieved by controlling their pump intensities
~\cite{Kroeze2018sso,Kroeze2019dsc,Marsh2021eam}. Dissipation of the field of each cavity mode is incorporated using the Lindblad master equation $\dot \rho = -i[H_0,\rho] + \loss \sum_{\mu}^{\Nm} \mathcal{D}[a_\mu]$, where $\mathcal{D}[a]=2a\rho a^\dag -\{a^\dag a,\rho\}$. We consider a uniform cavity loss rate $\kappa= 2\pi {\times} 260$~kHz, similar to recent experiments~\cite{Kroeze2023hcu}.

The cavity-mediated interaction $J_{ij}$ between spin ensembles $i$ and $j$ may be derived using a polaron transformation of the Hamiltonian~\cite{Guo2019eab,Marsh2021eam}. In the ideal confocal limit, it takes the form
\begin{align}\label{eqn:Jij}
    J_{ij} &= \frac{\detune^2+\loss^2}{\detune}\sum_{\mu} \frac{\Xi_\mu(\mathbf{r}_i)\Xi_\mu(\mathbf{r}_j)\Delta_\mu}{\Delta_\mu^2+\loss^2} \\ &= \delta(\mathbf{r}_i-\mathbf{r}_j) + \delta(\mathbf{r}_i+\mathbf{r}_j) + \frac{1}{\pi}\cos\Big(2\frac{\mathbf{r}_i\cdot \mathbf{r}_j }{w_0^2}\Big), \nonumber
\end{align}
where $\detune$ is the detuning from the fundamental mode. The first two terms are local and mirror-image interactions. The local interactions are shown in blue in Fig.~\ref{fig1}(a); for clarity, the mirror images are not shown. They arise from the constructive interference of cavity modes at the positions of the spins and their image across the cavity axis (due to the modes' even parity). The finite spatial extent of the spin ensemble and cavity imperfections regularize the delta functions to form short, but finite-range interactions with tunable length scale ${\gtrsim} 2$~$\mu$m in realistic cavities~\cite{Vaidya2018tpa,Kroeze2023hcu}. The interaction range is much smaller than the waist of the fundamental mode $w_0=35$~$\mu$m. We provide in App.~\ref{app:J} formulas for the $J$ matrix that incorporates finite $\Nm$ and size effects. 

The third, nonlocal term generates all-to-all, sign-changing interactions~\cite{Vaidya2018tpa,Guo2019spa}.  The nonlocal field is depicted in green in Fig.~\ref{fig1}(a). By choosing positions $\pos_i$ either close to or far from the cavity center, the nonlocal interaction can yield $J$ matrices that interpolate between ferromagnetic (all $J_{ij}>0$) and spin glass regimes~\cite{Marsh2021eam}. The glass regime results from $J$s with randomly signed off-diagonal elements.  That is, these have approximately independent and identically distributed off-diagonal elements, roughly equal parts positive and negative.  When atoms are distributed over a sufficiently large area, the confocal cavity-induced $J$ connectivity matrices exhibit eigenvalue statistics that approximately those of a SK spin glass, the Gaussian orthogonal ensemble~\cite{Marsh2021eam}. Glassiness might also be achievable in a confocal cavity without position disorder~\cite{Erba2021sgp}.

The transversely pumped system realizes a nonequilibrium Hepp-Lieb-Dicke phase transition~\cite{Mivehvar2021cqw}.  At $t=0$, the system is in the `normal' state with all cavity modes in the vacuum state and the spins pointed along $S_i^z$ in $\ket{\downarrow}$.  As discussed  below, we consider a protocol where the coupling strength is ramped up as a function of time. At threshold, the system transitions into to a `superradiant' phase characterized by macroscopically populated cavity modes and spins that spontaneously break the global $\mathbb{Z}_2$ symmetry to align along $\pm S_i^x$.  A sharp increase in cavity emission heralds the superradiant transition.  Concomitantly, the spins order and the phase of the light emitted from each spin ensemble is locked to the spin orientation.  Thus, the spins can be imaged in real time by spatially measuring the emitted phase of the cavity light.  Holographic (spatial heterodyne) imaging has already been demonstrated~\cite{Guo2019spa}.

In App.~\ref{app:crit}, we provide a derivation of a general expression for the critical coupling strength $g_c$ at which the superradiant threshold is reached in the semiclassical limit. This is given by
\begin{equation}\label{eqn:gcrit}
    \Na g_c^2 = \frac{\wz(\detune^2+\loss^2)}{\eigmax|\detune|}.
\end{equation}
Note that it depends on the $J$ matrix through its largest eigenvalue $\eigmax$: The form of the $J$ matrix determines both the threshold and the character of the ordered phase---e.g., ferromagnetic versus spin glass~\cite{Marsh2021eam}.  

The experimental protocol we intend to model involves ramping the transverse pump strength through the threshold of the superradiant phase transition.   We consider pump intensities $\Omega^2_\pm(t)$ that lead to an effective Ising coupling strength $g^2(t)=5g^2_c f(t)$, where $f(t)$ is a sigmoidal function that smoothly interpolates from $0$ to $1$ over a 600-$\mu$s timescale:
\begin{equation}
    f(t) = \begin{cases}
    0 & t<100\,\mu\mathrm{s} \\
    3(\frac{t-100\,\mu\mathrm{s}}{600\,\mu\mathrm{s}})^2 - 2(\frac{t-100\,\mu\mathrm{s}}{600\,\mu\mathrm{s}})^3 & 100\,\mu\mathrm{s}\leq t \leq 700\,\mu\mathrm{s} \\
    1 & t > 700\,\mu\mathrm{s}.
    \end{cases}
\end{equation}
The ramp is the lowest-order polynomial that smoothly interpolates between two points with vanishing first derivative;   {the results that follow are insensitive to the precise functional form.} The final pump intensity is $5{\times}$ the critical value in the semiclassical limit. Given the experimentally relevant parameters employed, the timescales chosen are sufficient to allow the spins to reach an organized steady state before the onset of spontaneous emission, which occurs on approximately the 10-ms timescale. In addition, we choose to simultaneously ramp down the transverse field as $\wz(t)=2\pi\times 10[1-f(t)]$~kHz.  This could be accomplished by changing the two-photon detuning of the pumps~\cite{Kroeze2018sso}.  Ramping $\omega_z$ to zero turns off spin flips between different $S^x$ states because the Hamiltonian becomes diagonal in the $S^x$ basis. The ramps for $g(t)^2$ and $\omega_z(t)$ are plotted in Fig.~\ref{fig1}(c).

\section{Quantum trajectory simulations}\label{sec:qtraj}

\begin{figure*}[t!]
    \centering
    \includegraphics[width=\textwidth]{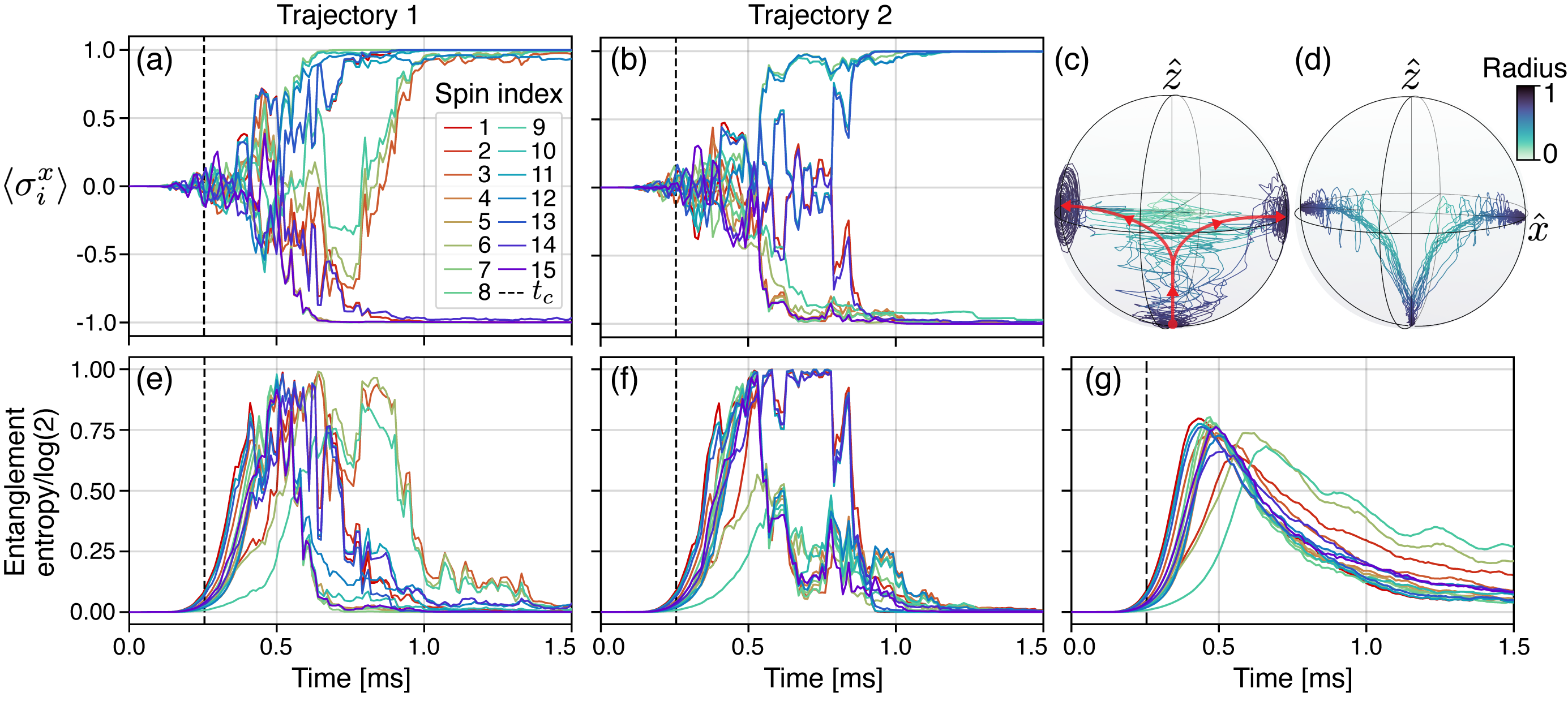}
    \caption{ (a--b) Example of two independent quantum trajectory simulations for the same system with a glassy $J$ connectivity.  
    The top panels show the dynamics of the $\Nc=15$ spin--$1/2$ system pumped through threshold. The spins begin to organize in the $\sigma_i^x$ quadrature when the pump strength approaches the semiclassical normal-to-superradiant threshold at the time $t_c$ indicated by a vertical dashed line. Note, the quantities plotted differ from the simulated measurement records in Fig.~\ref{fig1}b. (c) The same spin quantum trajectories from panel (a) are shown on (and within) the Bloch sphere. Note that here, the 15 traces are colored by radius rather than spin index. The red arrows show the general flow of the spin trajectories. (d) The averaged paths of the 15 spins, taken over 200 quantum trajectories. For each of the 15 spins, only trajectories with the same steady state are averaged together. (e-f) The bottom panels show entanglement entropy versus time of each spin for the trajectory simulations above; e.g., panel (e) pairs with panel (a).  (g) The entanglement entropy per spin averaged over all 200 trajectories.}
    \label{fig:trajectories}
\end{figure*}

Exact numerical simulations of (open) quantum many-body dynamics can be computationally expensive, especially in the confocal system due to the large number of modes in play.  To explore the spin dynamics throughout the superradiant transition, we simplify the full dynamics to an atom-only Lindblad master equation whose derivation is a multimode generalization of the method of J\"ager \textit{et al.}~\cite{Jager2022lme}; see App.~\ref{app:atomonly} for derivation. The atom-only Hamiltonian has the form 
\begin{equation}\label{eqn:HA}
    H = \wz \sum_{i=1}^\Nc S_i^z - \frac{g^2}{4\detune}\sum_{i,j=1}^\Nc J_{ij}S_i^x \big(\alpha_+ S_j^x +i\alpha_-S_j^y +\mathrm{H.c.} \big),
\end{equation}
where $J_{ij}$ is the same matrix as in Eq.~\eqref{eqn:Jij}. The other coefficients are 
\begin{equation}\label{eqn:alphas}
    \alpha_\pm = \frac{\detune}{-\detune+\wz-i\loss} \pm \frac{\detune}{-\detune-\wz-i\loss},
\end{equation}
where we restrict the treatment to the case of a completely degenerate cavity with uniform detuning $\Delta_\mu=\detune$ for all modes.  This is not an unreasonable approximation in the far-detuned regime $|\detune|\gg\kappa,\wz$~\cite{Kroeze2023hcu}.  In this limit, $|\alpha_+|\approx 2$ while $|\alpha_-|\approx 2|\wz/\detune|\ll 1$. The Hamiltonian thus resembles a transverse-field Ising model with an additional term $S_i^xS_j^y$ that is sufficiently small to play little role in the present simulations. The full atom-only master equation has the Lindblad form $\dot \rho = -i[H,\rho] + \sum_{k}^{\Nc} \mathcal{D}[C_k]$. The diagonalized atom-only collapse operators are non-Hermitian, in general, and given by
\begin{equation}\label{eqn:collapse}
    C_k=\frac{g\sqrt{\lambda_k\loss}}{2\detune}\sum_{i=1}^\Nc \evec^k_i\big(\alpha_+S_i^x +i\alpha_- S_i^y\big).
\end{equation}
Here, $\evec^k_i$ is the $i$'th element of the $k$'th eigenvector of the $J$ matrix; all eigenvalues $\lambda_k\geq0$. Each of the $\Nc$ collapse operators represents an orthogonal superposition of spin operators.    Appendix~\ref{app:atomonly} presents its derivation.

As noted above, the experimental protocol we consider involves ramping $\wz$ to zero at late times.  In this limit, $\alpha_-$ goes to zero, so the final Hamiltonian has the simple Ising form $H\propto-\sum_{ij}J_{ij}S_i^xS_j^x$. Likewise, the collapse operators contain only $S_i^x$ operators. As such, any $S_i^x$ eigenstate may become a steady state above threshold (though some are energetically preferred; see Sec.~\ref{sec:qtrajexample}). 

Quantum trajectory simulations of the atom-only master equation provide a continuous record of the state of each spin.  These trajectories arise from simulating a sequence of balanced homodyne measurements of the field emitted from each spin ensemble; see App.~\ref{app:record} for details.  This mimics experimentally practicable heterodyne measurements: Emitted cavity light is interfered on a camera with local oscillator (LO) light derived from the pumps to provide a phase reference~\cite{Kroeze2018sso}. This procedure enables holography of the spin states; see Fig.~\ref{fig1}(a) for illustration. Figure~\ref{fig1}(b) shows what such data would look like as the homodyne signal for each spin is integrated in time. Each signal is dominated by noise below the superradiance threshold. Above threshold, the homodyne signals become phase-locked to the spins and undergo a bifurcation. The sign of each homodyne signal thus serves as a measurement of the corresponding superradiant spin state. 

A single quantum trajectory evolves under the non-Hermitian Hamiltonian $H-i\sum_k C_k^\dag C_k/2$ and is interrupted by quantum jumps with displaced collapse operators $(C_k\pm i\beta)/\sqrt{2}$; see App.~\ref{app:record} for details. These operators represent the two quadratures of the balanced homodyne scheme.  The real number $\beta$ is proportional to $\sqrt{\kappa}$ multiplied by the coherent state amplitude of the LO and the spatial overlap of the LO and emitted cavity light.  Each collapse operator can induce quantum jumps.  To simulate the detection of cavity emission, these are stochastically generated to independently occur at rates $||(C_k\pm i\beta)\ket{\psi}||^2/2$. The trajectories of each spin are derived from these simulated detections, as shown in App.~\ref{app:record}.  While the quantum state diffusion method is simpler to define, the quantum jump method with high LO strength (large $\beta$) provides similar results with greater numerical stability at late times.   {While $\beta$ influences the timescale over which the global $\mathbb{Z}_2$ symmetry is broken, we find it does not affect the ensemble of steady-state spin configurations that are found.}

\section{Quantum spin dynamics, entanglement, and energy barriers }\label{sec:qtrajexample}

We now explore the dynamics of the spin trajectories and elucidate the role of quantum entanglement therein. Figure~\ref{fig:trajectories}(a,b) plots two independent quantum trajectories for a network of $\Nc=15$, $\Na =1$ (i.e., spin--$1/2$) particles that share the same $J$ matrix. A glassy $J$ matrix is selected by assigning the spins to random positions in the cavity midplane according to a 2D Gaussian distribution with a $2w_0$-wide standard deviation~\cite{Marsh2021eam}. 

We observe that a spin-aligned $\sigma_i^x$ steady-state configuration emerges within a few milliseconds of crossing the semiclassical transition threshold.  Beforehand, a combination of unitary quantum dynamics and stochastic projections from the continuous measurement drive the spins away from their initial $\ex{\sigma_i^z}=-S$ configuration. Measurement acts to break the spin's $\mathbb{Z}_2$ symmetry along $\sigma_i^x$.  The  rate at which this happens is proportional to $\kappa g^2/\Delta_C^2$ and is time-dependent through $g$. The timescale is approximately 5~ms at threshold and decreases to approximately 200~$\mu$s at the end of the ramp schedule. However, the organization timescale also depends on the structure of eigenvectors $\evec^k$ of the $J$ matrix. 

We also observe that collective spin-flip events between different low-energy states occur beyond this timescale. A diverse range of collective spin behavior occurs. For example, Fig.~\ref{fig:trajectories}(a) exhibits a group of three spins approaching a $\ex{\sigma_i^x}=-1$ steady state before collectively flipping toward $\ex{\sigma_i^x}=1$ at around 750~$\mu$s. By contrast, Fig.~\ref{fig:trajectories}(b) shows another behavior in which a group of four spins undergo an extended period of unbroken $\mathbb{Z}_2$ symmetry before rapid organization into a steady-state configuration.

The spin trajectory in Fig.~\ref{fig:trajectories}(a) may be visualized using the Bloch sphere representation in Fig.~\ref{fig:trajectories}(c). As the many-body quantum state remains pure within a single trajectory, paths through the interior of the Bloch sphere indicate entanglement between spins. We see that the quantum spins first take a non-classical trajectory of unbroken global $\mathbb{Z}_2$ symmetry through the interior of the Bloch sphere.  After initially moving upward towards the center of the Bloch sphere, the continuous measurement breaks spin-flip symmetry.  The spins then emerge from the interior of the Bloch sphere to reach a steady-state spin configuration. Figure~\ref{fig:trajectories}(d) shows the average of the paths the spins take.

Entanglement is present during both the initial organization near threshold and during subsequent spin-flip events. We consider the entanglement entropy for each spin, given by $-\tr[\rho_i\log(\rho_i)]$, where $\rho_i$ is the reduced density matrix for spin $i$. In general, the entanglement entropy can be nonzero for either entangled states or mixed states. Thus, the entropy would not be a good measure of entanglement when applied to the density matrix of the system. However, the entropy does provide a good measure of entanglement when applied at the level of individual quantum trajectories because each trajectory remains globally pure at all times. We choose the entropy over other measures of entanglement, such as the negativity~\cite{Vidal2002cmo}, as it is more computationally tractable while faithfully capturing entanglement. 

The entanglement entropy per spin is shown in Figs.~\ref{fig:trajectories}(e,f) for the trajectories in panels (a) and (b), respectively. An initial increase near threshold can result from unbroken global $\mathbb{Z}_2$ symmetry and transitions to other low-energy states.   At first, the superposition of low-energy states largely preserves the global $\mathbb{Z}_2$ symmetry.  Measurement then begins to lead to superposition projection around $0.5$~ms, resulting in decreasing entanglement and the breaking of $\mathbb{Z}_2$ symmetry. Subsequent spikes in the entanglement accompany the spin-flip events. Last, the entanglement decays to zero as the spins reach a steady-state configuration that corresponds to a classical state. Figure~\ref{fig:trajectories}(g) plots the entanglement entropy for each spin averaged over 200 trajectories.  The initial peak slowly decays to zero, reflecting the occurrence of later spin-flip events.  

\begin{figure}[t!]
    \centering
   \includegraphics[width=\columnwidth]{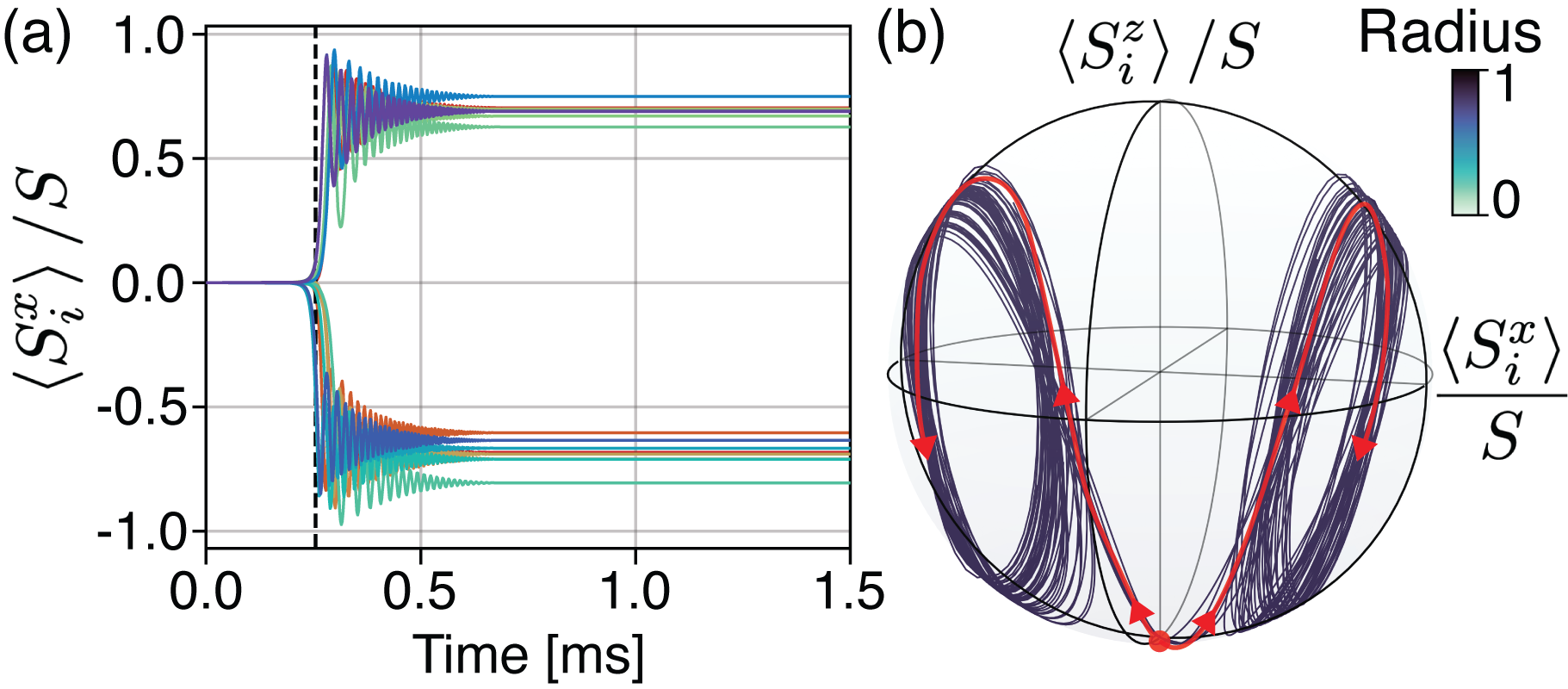}
    \caption{Example semiclassical trajectory simulation for the same frustrated $J$ matrix and the same pump ramp schedule as in Fig.~\ref{fig:trajectories}(a).  (a) A sharper transition occurs with  (b) dynamics restricted to the surface of the Bloch sphere. Panel (a) shows that the semiclassical transition, indicated by a vertical dashed line, is reached at time expected from Eq.~\eqref{eqn:gcrit}. The red arrows in (b) show the general flow of spin trajectories.}
    \label{fig:semictraj}
\end{figure}

Figure~\ref{fig:semictraj} contrasts these quantum spin trajectories with semiclassical trajectories for the same $J$ matrix used in Fig.~\ref{fig:trajectories}. The spin--$1/2$ degrees of freedom are now replaced with semiclassical, collective spins, each comprised of $\Na=10^5$ spin--$1/2$ atoms. The semiclassical equations of motion are derived in   {App.~\ref{app:semiclassical}.} We see that, in contrast to the quantum dynamics, the semiclassical trajectories exhibit a rapid organization at the semiclassical transition threshold but are confined to the surface of the Bloch sphere, indicating the lack of entanglement. Unlike the quantum limit, large oscillations are observed around the $x$-axis on the Bloch sphere.  

\begin{figure}[t!]
    \centering
    \includegraphics[width=\columnwidth]{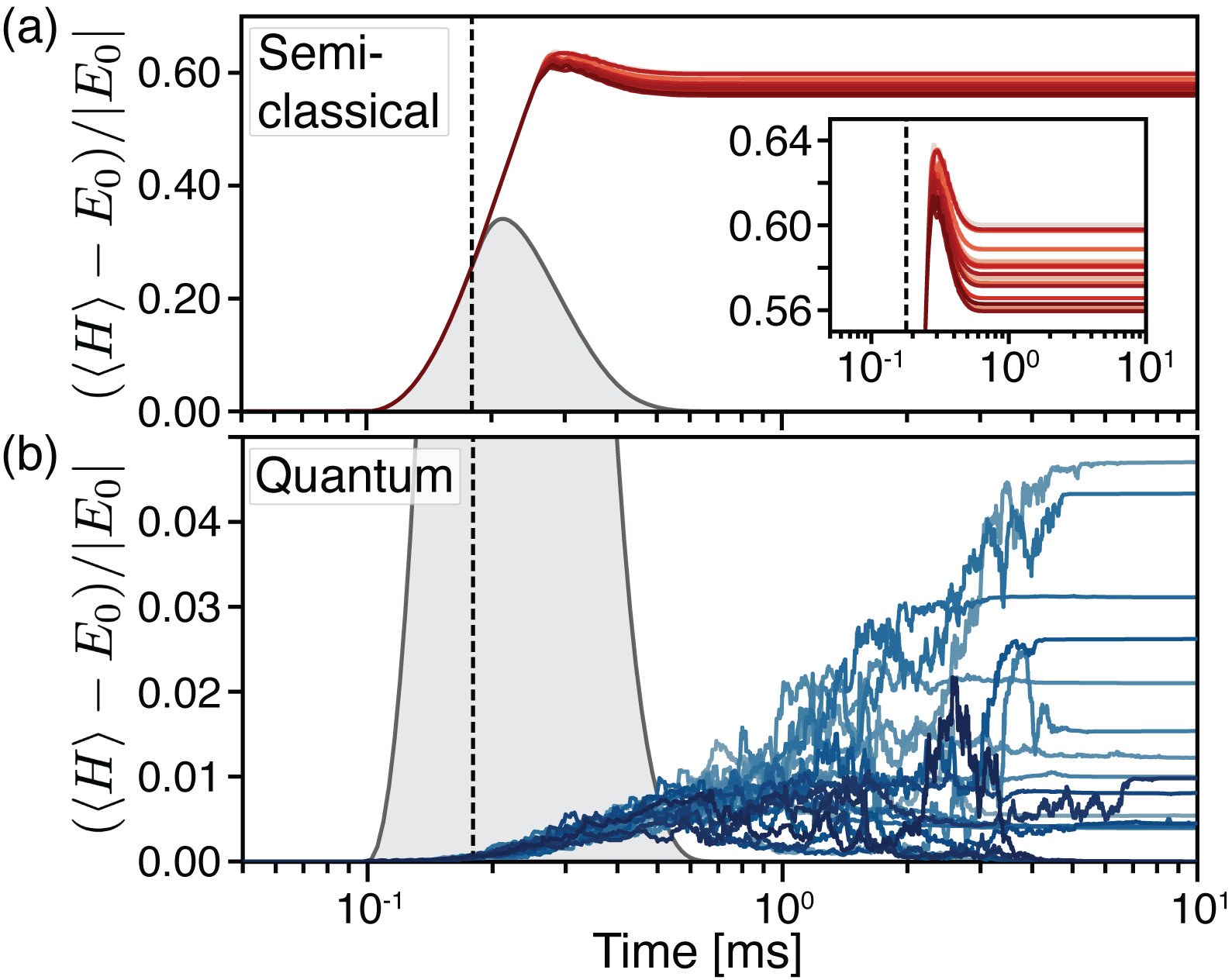}
\caption{(a) Plot of the energy of 20 independent semiclassical trajectories for the same $J$ matrix as in Fig.~\ref{fig:trajectories}. The energy is normalized to the instantaneous quantum ground-state energy $E_0$.  The $\Nc=15$ network has $\Na=10^5$ spins per node. The shaded region is inaccessible to any unentangled state, and thus the semiclassical trajectories must climb over the barrier before reaching steady state. (b) By contrast, quantum trajectories can pass through the semiclassical energy barrier via entanglement between spins.  This provides access to lower-energy steady states.  Plotted are the normalized energies of 20 independent quantum trajectories for the same $J$ matrix as above. Note the change in $y$-axis scale. The vertical dashed line marks the semiclassical threshold. The red (blue) colors in panel a (b) are chosen to be different shades to only help distinguish one trace from another.  }
    \label{fig:barriers}
\end{figure}

To investigate the role entanglement might play in the evolution toward low-energy, steady-state spin configurations, we plot in Fig.~\ref{fig:barriers} the energy of 20 quantum and semiclassical trajectories for the same $J$ matrix and ramp schedule considered in Fig.~\ref{fig:trajectories}. The shaded region is inaccessible to any unentangled spin state constrained to the surface of the Bloch sphere.  We identified the boundaries of this semiclassically forbidden region through Monte-Carlo sampling of semiclassical spin states (i.e., those states constrained to the surface of the Bloch sphere) followed by gradient descent to the lowest possible energy state.  We find that entanglement enables the quantum spins to follow trajectories (through the Bloch sphere interior) that bypass this semiclassical energy barrier.  This allows the quantum spin network to reach lower-energy steady-state configurations. Slower ramps could allow the semiclassical trajectories to follow a more adiabatic path back downward to similarly low-energy states.  However, we find that such ramps must be at least an order-of-magnitude slower than those considered here.   {The addition of noise to the initial state can also yield a ${\sim} 25\%$ decrease in the semiclassical steady-state energies, but this remains an order-of-magnitude higher compared to quantum trajectories.}

The steady-state energy of the quantum trajectories seems to be primarily controlled by the ramp rate. Evolution through the superradiance transition has the form of a many-body Landau-Zener problem with many-body gaps controlling the adiabatic timescale of the transition. The many-body gap near the transition is on the order of $\wz$ for a ferromagnetic $J$. Thus, $\wz^{-1}$ sets the timescale for adiabatic evolution through the transition to either of the two ferromagnetic ground states. By contrast, spin glasses are characterized by nearly degenerate spin configurations that become exponentially numerous with $\Nc$. This results in much smaller gaps near the transition and nonadiabatic evolution is more likely to occur, as we see in Fig.~\ref{fig:barriers}(b). The chosen ramp rate is slow enough to prevent nonadiabatic transitions to highly excited states, but not enough to prevent transitions to the nearly degenerate local minima states. Unitary evolution through the transition then produces an entangled superposition of low-energy states, as seen in Fig.~\ref{fig:trajectories}, before projection into a single spin state occurs. The final energy of the trajectories is thus controlled by the nonadiabatic transitions experienced during the ramp as well as the measurement projection before $\wz$ is ramped to zero.

  {
\section{The overlap order parameter}\label{sec:olap}
We now establish the link between quantum trajectories and the spin-glass order parameter. Order in glassy systems can be identified through correlations between the many symmetry-broken thermodynamic states. This is captured by the replica overlap~\cite{Parisi1983opf}, defined classically as $q_{\alpha\beta}=\sum_i^N \langle{s}_i^\alpha\rangle \langle{s}_i^\beta\rangle/N$ where $\bm{s}^{\alpha,\beta}$ are replica spin states and brackets denote a time average. The overlap distribution is given by
\begin{equation}\label{PJ_classical}
    P_J(q) = \frac{1}{n_R^2}\sum_{\alpha,\beta}^{n_R} \delta\left( q_{\alpha\beta}-q\right),
\end{equation}
where $n_R$ is the number of replicas.
Each $P_J(q)$ can have structure that varies depending on the disorder realization $J$, even in the thermodynamic limit; this is the lack of self-averaging inherent in spin glass~\cite{Stein2013sga}. Disorder-dependent fluctuations are averaged out in the Parisi distribution $P(q)\equiv [P_J(q)]_J$, where $[\cdot]_J$ denotes an average over disorder realizations. We discuss the central features of $P(q)$ in Sec.~\ref{RSBsec}. The overlap distribution for glassy \textit{quantum} systems is defined similarly after performing a Suzuki-Trotter mapping to an equivalent classical system~\cite{Suzuki1991gto,Trotter1959otp,Lai1990mcs}; see App.~\ref{ParisiSupp} for details.  }

\begin{figure*}[t!]
    \centering
    \includegraphics[width=\textwidth]{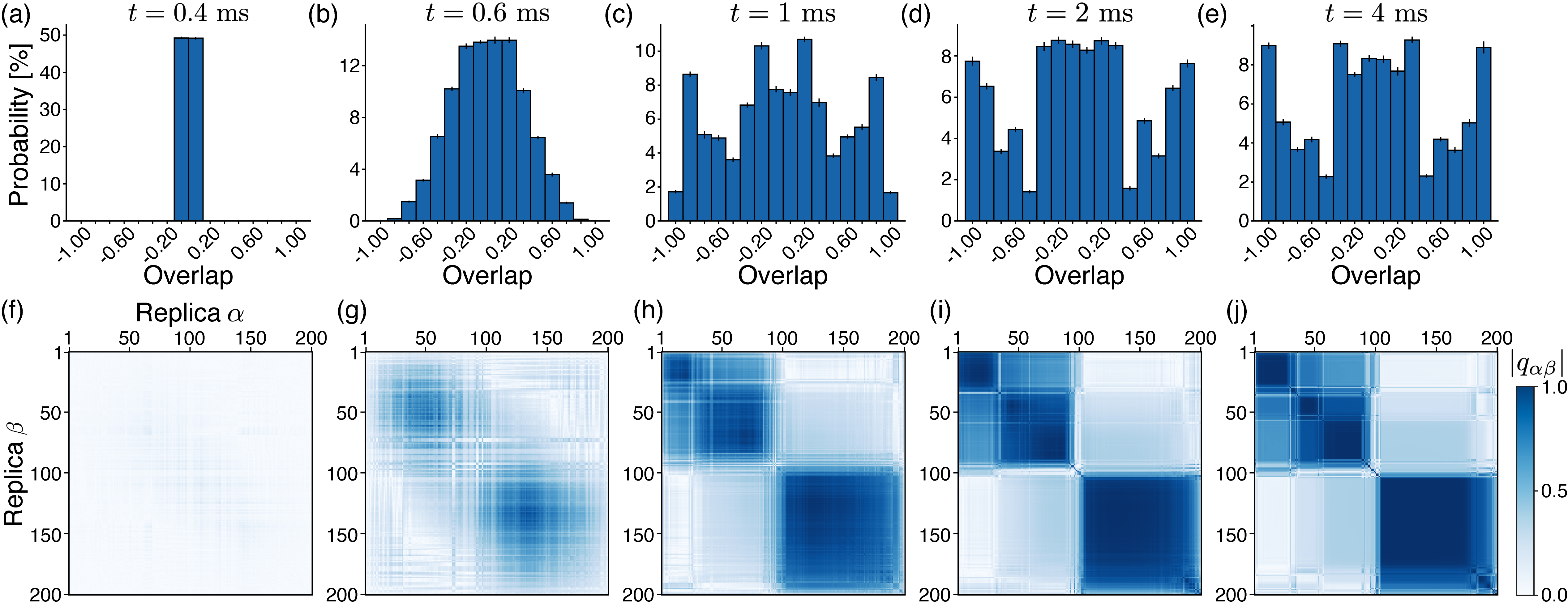}
    \caption{(a-e) Time evolution of   {$q_{\alpha\beta}$} for 200 quantum trajectories corresponding to the $J$ matrix in Figs.~\ref{fig:trajectories}--\ref{fig:barriers}. Each panel shows the probability for each spin overlap value to occur at each given time during the ramp sequence. Error bars in this figure and in subsequent figures are estimated from bootstrap analysis; see App.~\ref{app:bootstrap} for details. (f-j) Time evolution of the full overlap matrix $q_{\alpha\beta}$.  The time that corresponds to each panel is the same as the panel above. The histograms in (a-e) can be recovered by binning the off-diagonal values in the overlap matrix.  Each overlap matrix is ordered via an independent hierarchical clustering of spin states at each time.}
    \label{fig:singleOlap}
\end{figure*}

  {
To connect the overlap distribution to quantum trajectories, we first cast the overlap into a particular form that applies directly at the quantum level. In App.~\ref{ParisiSupp} we show that the overlap distribution is closely related to the operator  
\begin{equation}\label{olapOp}
    \mathcal{O} = \frac{1}{N}\sum_{i=1}^N \sigma_i^x \otimes \sigma_i^x,
\end{equation}
where the $\sigma_i^x$ are Pauli operators for each site.
We refer to the above as the overlap operator given its close correspondence to the classical overlap $q_{\alpha\beta}$. It acts on the doubled Hilbert space of $\rho_J\otimes\rho_J$, where $\rho_J$ is the density matrix for a given $J$ realization.
The statistical moments of the Parisi distribution are shown to be given by $q^{(k)}=[\langle\mathcal{O}^k\rangle]_J$. This close relation allows for a simple expression of the overlap distribution in terms of $\mathcal{O}$. To do so, we use the eigenstate representation $\mathcal{O}=\sum_q q \mathcal{P}_q$, where the sum over $q$ includes all $\Nc+1$ overlap values linearly spaced in $[-1,1]$. The operators $\mathcal{P}_q$ are projections onto the space of spin states with overlap $q$. The overlap is then given by
\begin{equation}\label{PjDensity}
    P_J(q) = \sum_{q'}\delta(q-q')\Tr \Big[ (\rho_J\otimes\rho_J)   \mathcal{P}_{q'} \Big].
\end{equation}
The connection to quantum trajectories is now established using the pure state representation $\rho_J=\sum_{\alpha=1}^{n_T}\dyad{\psi_J^\alpha}/n_T$, where $n_T$ is the total number of pure states $\ket{\psi_J^\alpha}$. Each quantum trajectory is one of these pure states. Inserting this form into Eq.~\eqref{PjDensity} yields
\begin{equation}\label{PJ_Traj}
    P_J(q) = \frac{1}{n_T^2}\sum_{\alpha,\beta}^{n_T}\sum_{q'} \delta(q-q')  \bra{\psi_J^\alpha}\otimes\langle\psi_J^\beta|  \mathcal{P}_{q'} \ket{\psi_J^\alpha}\otimes|\psi_J^\beta\rangle.
\end{equation}
To summarize, trajectories of the same disorder realization can find different symmetry-broken states of the glassy landscape due to stochastic evolution induced by the environment. Each pair of symmetry-broken states $\ket{\psi_J^\alpha}$ and $|\psi_J^\beta\rangle$ then contribute to $P_J(q)$ through their projection onto the subspace with overlap $q$.
} 

  {The classical expression for $P_J(q)$ is recovered when the trajectories $\ket{\psi_J^\alpha}$ are in spin eigenstates. Each state then corresponds to a classical spin vector with elements $\bm{s}^\alpha_i=\bra{\psi_J^\alpha}\sigma_i^x\ket{\psi_J^\alpha}$. Equation~\eqref{PJ_Traj} then reduces to Eq.~\eqref{PJ_classical} with the replica overlap given by
\begin{equation}\label{olap}
    q_{\alpha\beta}=\frac{1}{N}\sum_{i=1}^N\bra{\psi_J^\alpha}\sigma_i^x\ket{\psi_J^\alpha}\langle\psi_J^\beta|\sigma_i^x|\psi_J^\beta\rangle.
\end{equation}
We refer to $q_{\alpha\beta}$ above as the mean-field overlap, as it corresponds to the overlap operator $\mathcal{O}$ in the mean-field limit---i.e., when trajectories are spin eigenstates and thus the wavefunction factorizes between sites. The fundamental difference between the classical and quantum overlap is how entanglement between spins can allow for superpositions of spin states. This allows a pairs of quantum states $\ket{\psi_J^\alpha}$ and $|\psi_J^\beta\rangle$ to contribute to multiple values of the overlap distribution at once, which does not occur in the classical limit.}

\section{Replica symmetry breaking}\label{RSBsec}

We now explore the emergence of RSB as the system is pumped through the transverse Ising transition. To do so, we first analyze the correlations between independent quantum trajectories of a system with the same quenched disorder for all trajectories. That is, a frustrated spin system with the same $J$ matrix for all trajectory simulations.   {Because the trajectories ultimately reach classical steady-state spin configurations, despite entangled quantum dynamics at intermediate times, we study the mean-field overlap $q_{\alpha\beta}$ in Eq.~\eqref{olap}. This $q_{\alpha\beta}$ directly yields to the overlap distribution predicted by replica theory once steady-state spin configurations are reached, which occurs after ${\sim}$2~ms; it provides a mean-field estimate at earlier times. This $q_{\alpha\beta}$ depends on only first-order expectation values, and thus has the significant advantage of being directly observable from the trajectory measurement record.
}

  {Once a steady-state configuration is found,} the overlap takes on one of $\Nc+1$ possible values $\in[-1,1]$.  The overlap distribution is always symmetric about $0$ due to the global $\mathbb{Z}_2$ symmetry, in the absence of a longitudinal field.  An ordered phase will exhibit an overlap distribution containing `goalpost' peaks 
  {at $q=\pm q_{\mathrm{EA}}$, where $q_{\mathrm{EA}}=q_{\alpha\alpha}$ is the Edwards-Anderson order parameter, also known as the self-overlap.}
(Paramagnets do not have such peaks, but ferromagnets and spin glasses do.)  Peaks may also arise associated with overlaps $q_{\alpha\beta}$ between replicas that settle into   {different spin states.}
These additional non-vanishing peaks between the goalposts indicate RSB and arise from the smaller overlap between distinct, low-energy states~\cite{Stein2013sga}. 

We first consider the overlap distribution for the same $J$ matrix considered in Figs.~\ref{fig:trajectories}--\ref{fig:barriers}. Figure~\ref{fig:singleOlap}(a-e) shows the time evolution of   {$q_{\alpha\beta}$} as it approaches steady state, around 4~ms. To construct the overlap distribution of a fixed $J$ matrix, we consider 200 quantum trajectories with identical initial conditions and the same ramp schedule as shown in Fig.~\ref{fig1}(c). We then compute the overlap   {$q_{\alpha\beta}$}
between every pair of the 200 replicas and bin the results as a function of time. We exclude the self-overlaps $q_{\alpha\alpha}$ because, while they have vanishing weight in the limit of an infinite number of replicas, they provide an asymmetric bias to   {only the positive $q=q_{\mathrm{EA}}$} peak of the overlap for finite sample sizes.  At $t=0$, the system is in the normal (paramagnetic) state and the overlap between any two replicas is zero because $\ex{\sigma_{i\alpha}^x}=0$ for all spins in the initial $\sigma_i^z$ state. A nonzero overlap emerges as the spins transition to the superradiant regime and align along $\pm\sigma_i^x$. The final overlap distribution shows goalpost peaks at   {$\pm q_{\mathrm{EA}}$. We find $q_{\mathrm{EA}}\approx 1$ in the parameter regime of our simulation, but note that $q_{\mathrm{EA}}$ is commonly overestimated due to finite-size effects~\cite{Young1983ddo}.} Interior peaks indicate RSB.  These interior peaks arise from correlations between distinct spin configurations that are local minima of the Ising energy $E=-\sum_{ij}J_{ij}s_is_j$, where all $s_i=\pm 1$. By local minimum, we refer to spin states for which flipping any single spin raises the total energy.   {We note that the steady-state overlap distribution is independent of the exact measurement scheme; for any LO strength $\beta>0$ the same ensemble of spin states are found, leading to the same overlap distribution.  }

Figure~\ref{fig:singleOlap}(f-j) shows the full overlap matrix $q_{\alpha\beta}$ versus time before we bin into the above histograms. In the thermodynamic limit, the Parisi ansatz for the solution to the SK model predicts an ultrametric overlap structure emerging from RSB. Specifically, the ansatz predicts a nested block-diagonal structure where the overlap magnitudes are larger in the diagonal blocks than in the off-diagonal blocks. The diagonal blocks are then expected to further divide into smaller diagonal blocks with larger overlap and off-diagonal blocks with smaller overlap, and so on. The ansatz thus predicts a self-similar overlap matrix in the limit $\Nc\to\infty$, while for finite-size systems the self-similarity truncates at a finite depth. For the $\Nc=15$ case in Fig.~\ref{fig:singleOlap}(j), we find evidence for up to three levels of the RSB block structure. A primary $2{\times} 2$ block structure emerges that approximately separates replicas 1--100 from 101-200. The primary block of replicas 1--100 is further subdivided into $2{\times} 2$ blocks separated near replica 30, distinguishing regions of higher overlap from lower. Evidence for tertiary block structure may be found in the sub-block containing replicas 30--100; a final subdivision may be seen near replica 55. We leave to future work quantitative analyses of self-similarity, the depth of RSB, and how RSB scales with $\Nc$. However, we later quantify the degree of ultrametricity in the overlap distribution in Sec.~\ref{sec:ultra}.

\begin{figure}[t]
    \centering
    \includegraphics[width=\columnwidth]{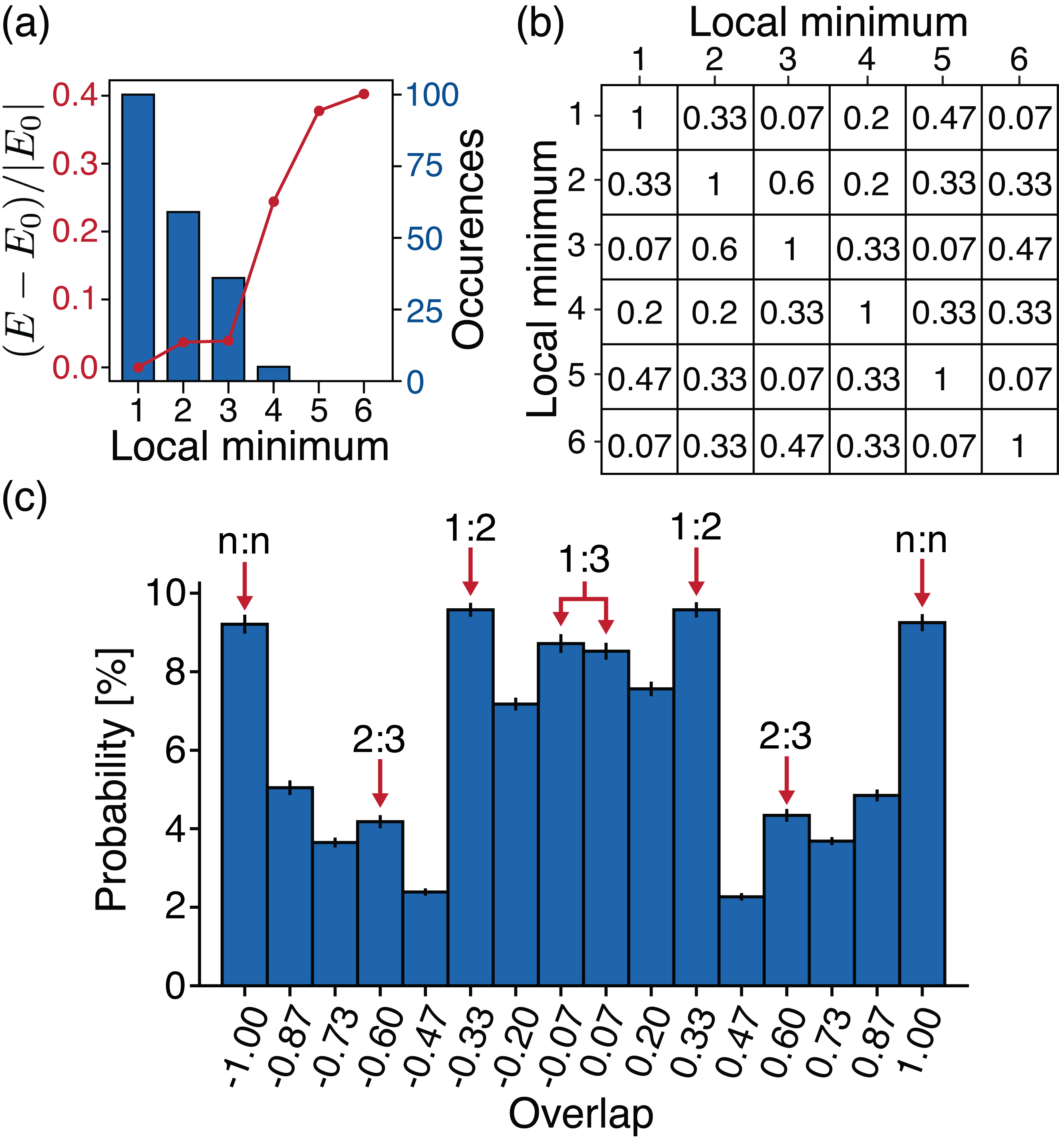}
    \caption{(a) The energy of the six local minima of the Ising energy for the $J$ matrix considered in Figs.~\ref{fig:trajectories}-\ref{fig:singleOlap}. Also plotted are the occurrence probabilities of those local minima as steady states for the 200 quantum trajectories. (b) The spin overlap matrix of these six local minima. 
    (c) The same spin overlap distribution as in Fig.~\ref{fig:singleOlap}(j), but with annotated peaks. The notation $X{:}Y$ indicates that the peak arises from the overlap of the local minima $X$ and $Y$ in the list of panel (a). Finite probability in the unlabeled bins is due to fluctuations of the overlap peaks around the labeled local minima.  }
    \label{fig:olapAnalysis}
\end{figure}

To provide further insight, we delve into the structure of the steady-state overlap distribution produced by the $J$ matrix considered in Figs.~\ref{fig:trajectories}-\ref{fig:singleOlap}. This $J$ matrix induces a rugged Ising energy landscape that contains six local minima not related by the global $\mathbb{Z}_2$ symmetry.  These were found by numerically enumerating all spin states. Of the 200 quantum trajectories in Fig.~\ref{fig:singleOlap}, 66\% reached one of these six local minima in steady state.  An additional 20\% were within one spin flip of a local minimum, while the remaining 14\% were between two-to-four spin flips away.  To show the relation of each minima's occurrence probability to its energy, we plot these together in Fig.~\ref{fig:olapAnalysis}(a), binning each trajectory by its nearest local minimum.   The result shows a clear anticorrelation:  The trajectories with lowest-energy, steady-state spin configurations are observed most frequently.  Though the system is not in thermal equilibrium, this tendency to low-energy states is reminiscent of a low-temperature system; see Sec.~\ref{sectemp} for a discussion of system temperature.

The overlap matrix between local minima is plotted in Fig.~\ref{fig:olapAnalysis}(b).  The numbers in the matrix entries are the absolute value of the overlap values between the indicated minima. The diagonal entries correspond to the self-overlap, which is always unity.  
The overlap matrix allows us to pinpoint the pairs of spin configurations that create each peak in the overlap distribution of Fig.~\ref{fig:singleOlap}(j).  This plot is reproduced in Fig.~\ref{fig:olapAnalysis}(c). Every peak in the distribution can be understood by considering the overlaps between the first 3 local minima in Fig.~\ref{fig:olapAnalysis}(b). Each peak in the distribution is annotated with the pair of minima $x{:}y$ that produce that value of the overlap. The remaining local minima were found too infrequently to produce any distinct peaks in the overlap distribution.

\begin{figure}[t]
    \centering
    \includegraphics[width=\columnwidth]{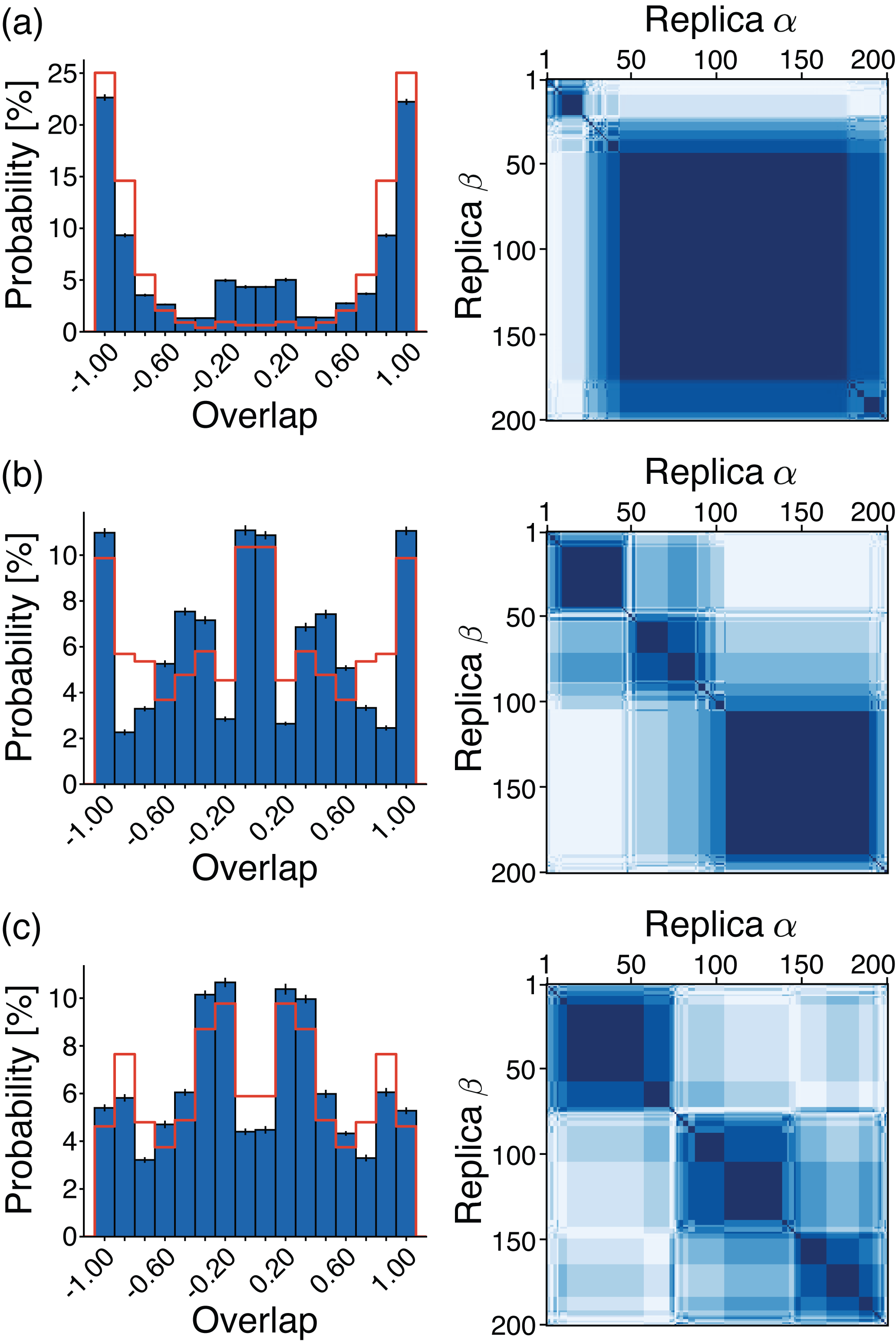}
    \caption{Steady-state overlap distribution and overlap matrix for 200 quantum trajectories of three different confocal $J$ matrices. The best-fit thermal distributions are shown in red. (a) Example of a $J$ matrix that produces a dominant low-energy state; most trajectories find the same steady-state spin configuration. The temperature for the  thermal fit is $T_{\mathrm{fit}}=0.17(1)\, \overline{T}_c$. (b) A $J$ matrix with multiple peaks and levels of structure in the overlap. The thermal fit yields $T_{\mathrm{fit}}=0.23(4)\, \overline{T}_c$. (c) A more complex $J$ matrix. Quantum trajectories find a large set of steady-state configurations. The thermal fit yields $T_{\mathrm{fit}}=0.17(3)\, \overline{T}_c$.}
    \label{fig:manyOlaps}
\end{figure}

Each $J$ matrix produces a different set of local minima, and thus different overlap distributions. This is evident in an ensemble of 100 confocal $J$ matrices produced by assigning spins to different random locations in the cavity midplane with standard deviation $2w_0$, which lies in the spin glass regime~\cite{Marsh2021eam}. The overlap distribution for each $J$ matrix is constructed from 200 quantum trajectories as in Fig.~\ref{fig:singleOlap}. Resulting steady-state overlap distributions for three representative $J$'s are shown in Fig.~\ref{fig:manyOlaps}.  Appendix~\ref{app:allOlaps} has plots of all 100 overlap distributions. 

The overlap distributions all exhibit   {$\pm q_{\mathrm{EA}}$ peaks with $q_{\mathrm{EA}}$ equal to unity for most disorder realizations.} Variation within the interior demonstrates that the correlations between low-energy minima vary between $J$ matrices.  This is the non-self-averaging phenomenon inherent to SK spin glasses~\cite{Stein2013sga}. The three $J$ matrices are chosen to display a representative diversity of structure found in the overlap distributions. The $J$ in Fig.~\ref{fig:manyOlaps}(a) produces an overlap distribution that is dominated by a single low-energy spin configuration.  Other peaks (from different configurations) occur in only ${\sim}10\%$ of the trajectories. Figure~\ref{fig:manyOlaps}(c) shows the other extreme in which many different spin configurations are found with multiple levels of clustering between states. This overlap matrix is indicative of a far glassier system. The character of most overlap matrices falls between these two for our system size of $\Nc=15$, such as the $J$ in Fig.~\ref{fig:manyOlaps}(b). 

In the large-size limit, high peaks should be sparse in the overlap distribution because only a small set of thermodynamic states have significant weight.  The peak positions do not average out into a smooth distribution between the goalposts.  This is indicative of the lack of self-averaging manifest in these order parameter observables of the spin glass state.  The order parameter that does average is the Parisi distribution~\cite{Stein2013sga}, which we discuss in Sec.~\ref{phases} below. 

\section{Effective temperature}\label{sectemp}

The overlap distributions do not appear to be consistent with an effective thermal equilibrium model. This is not surprising in this driven-dissipative quantum optical setting.  Nevertheless, it is instructive to compare these distributions to those expected at equilibrium.  Equilibrium overlap distributions can be constructed by assigning probabilities to spin states according to a Boltzmann factor $\exp(-E/k_B T)$, where $E$ is the Ising energy of the spin state and $T$ is an effective equilibrium temperature that serves as a fit parameter. The overlap between all pairs of states is then binned and weighted by their Boltzmann factors. We perform a least-squares fit to each of the overlap distributions in Fig.~\ref{fig:manyOlaps} to extract $T_{\mathrm{fit}}$.  The corresponding distributions are shown in red. The extracted temperatures are provided in units of $\overline{T}_c$, the largest eigenvalue of the $J$ matrix averaged over all $J$ matrices. This quantity corresponds to the critical temperature for the SK spin glass transition in the thermodynamic limit~\cite{Sherrington1975smo}. In our finite system, $\overline{T}_c$ corresponds to the average crossover temperature.  

While the equilibrium model does capture the location of peaks in the overlap distribution, it is not able to quantitatively match their heights; instead, they seem to often be underestimated (overestimated) near the center (wings) of the distribution. The average $T_{\mathrm{fit}}$ is $0.21(8) \overline{T}_c$. Despite the lack of quantitative correspondence, the fitted temperatures are well-enough below $\overline{T}_c$ to infer the presence of a low-energy ordered phase in this system, even when using realistic parameters.  Indeed, the authors have observed such states in a related experimental system~\cite{Kroeze2023rsb}.

\section{Spin glass, ferromagnetic, and paramagnetic phases}\label{phases}

Two order parameters are needed to distinguish between the different types of phases described by the Ising Hamiltonian.  These are the $J$-averaged spin overlap distribution, also known as the Parisi order parameter, and the usual magnetization. The magnetization order parameter $m=\sum_i\ex{\sigma_i^x}/N$ is used to discriminate ferromagnetic from glass or paramagnetic ordering.  It should   {approach $\pm1$ at low temperature} in a ferromagnetic state but be close to zero in the spin glass and paramagnetic phases.  

The spin glass is distinguished from the paramagnet via the Parisi order parameter,   {the $J$-averaged overlap distribution.} The average should form a smooth distribution for the SK spin glass. The paramagnet has a Parisi order-parameter distribution that is peaked around zero, while the spin glass and ferromagnet are   {peaked around $\pm q_{\mathrm{EA}}$.} Last, while the ferromagnet's Parisi distribution has no support between   {$\pm q_{\mathrm{EA}}$,} the spin glass has a ``net-with-goalposts" structure of smooth interior support.

\begin{figure}[t!]
    \centering
    \includegraphics[width=0.95\columnwidth]{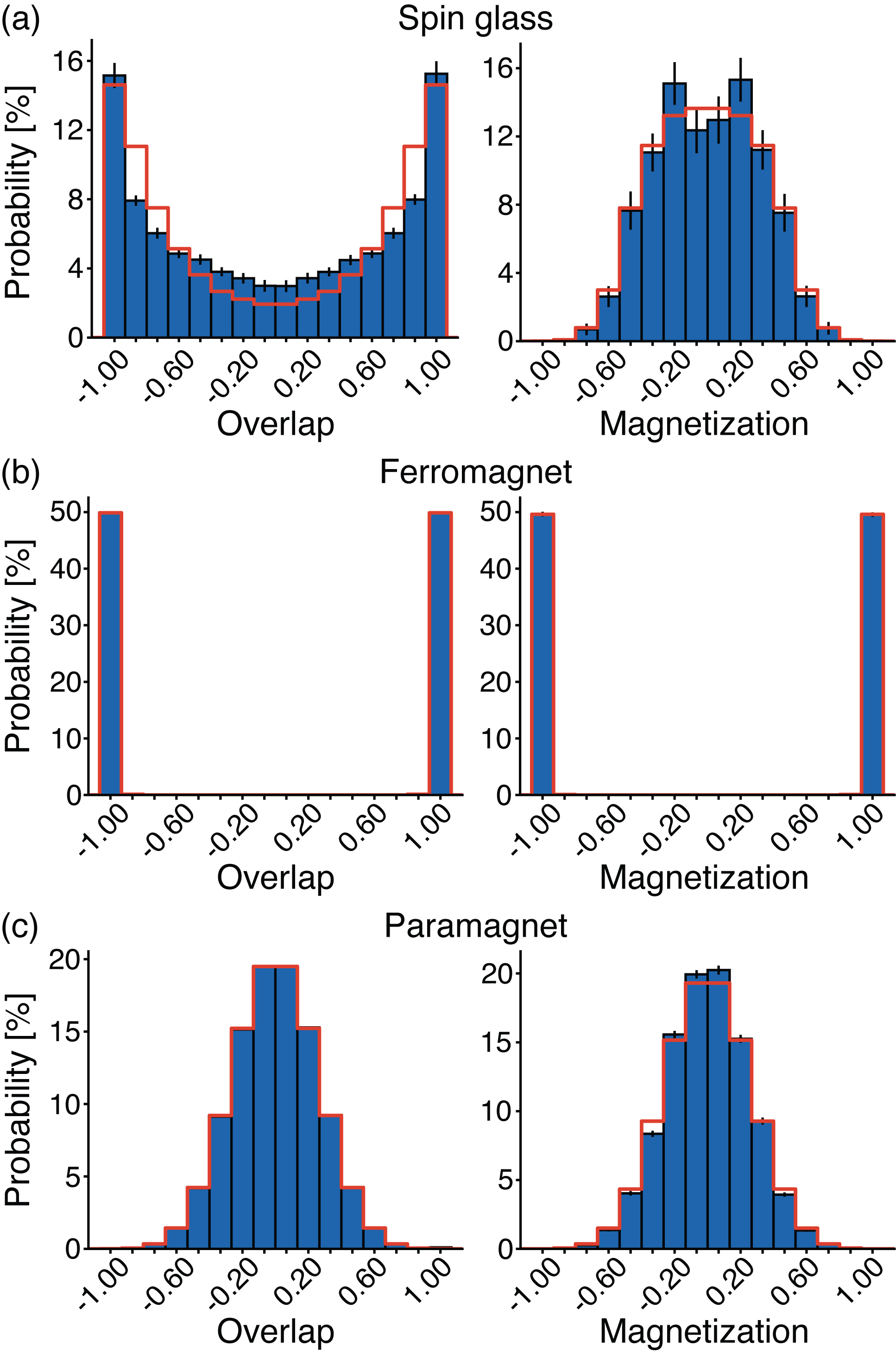}
    \caption{ Aggregate overlap and magnetization distributions in spin glass, ferromagnetic, and paramagnetic phases. The best-fitting thermal approximation is shown in red for all panels. (a) Confocal $J$ matrices in the spin glass regime showing a smooth Parisi-like distribution and a magnetization peaked near zero. The thermal fit yields $T_{\mathrm{fit}}=0.21(2)\, \overline{T}_c$. (b) Ferromagnetic regime showing the absence of interior support in the overlap and strong peaks in the magnetization. The thermal fit yields $T_{\mathrm{fit}}=0.011(2)\, \overline{T}_c$. (c) The same set of spin glass $J$ matrices as in panel (a), but the system is rapidly quenched into the superradiant regime rather than ramped according to $f(t)$. The overlap and magnetization both cluster around zero, indicative of a paramagnetic phase. The thermal fit is poorly constrained in this regime: Shown is the distribution for $T_{\mathrm{fit}}=5\, \overline{T}_c$, which bears similarity to the data. 
    }
    \label{fig:aggOlaps}
\end{figure}

\begin{figure*}[t!]
    \centering
    \includegraphics[width=\textwidth]{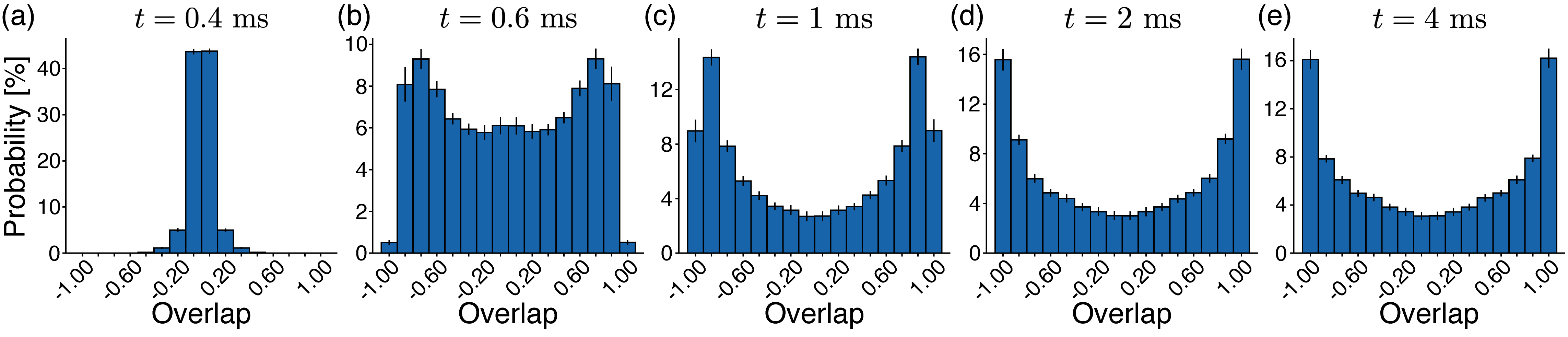}
    \caption{(a-e) Time evolution of the $J$-averaged Parisi distribution. The overlap distribution is averaged over all 100 $J$ matrices in the confocal spin glass regime. The times are chosen to match those in Fig.~\ref{fig:singleOlap}. The steady-state distribution emerges after approximately 2~ms, demonstrating the distinctive goalpost peaks with a continuously filled interior.}
    \label{fig:aggOlapEvo}
\end{figure*}

We average the overlap and magnetization distributions for 100 confocal $J$ matrices to yield the aggregate distributions in Fig.~\ref{fig:aggOlaps}(a). Distributions characteristic of a spin glass arise, consisting of extremal peaks   {at $\pm q_{\mathrm{EA}}$} bridged by a continuous interior for the overlap and a magnetization peaked near zero.  This is consistent with the result expected from the Parisi solution to the SK spin glass~\cite{Stein2013sga}.

A fit to the thermal equilibrium model yields a $T_{\mathrm{fit}}=0.21(2)\, \overline{T}_c$, closely matching the average temperature found from the individual fits discussed above. The magnetization is well approximated by a centered binomial distribution, indicating that the local minima are uncorrelated with a ferromagnetic state.  The standard deviation $0.31$ of this distribution is close to that expected of the SK model at this system size, $1/\sqrt{\Nc}\approx 0.26$. The difference from a Gaussian may indicate a small ferromagnetic remnant in the confocal $J$ matrices that could be eliminated by placing spins further from the cavity midpoint~\cite{Marsh2021eam}. 

By contrast, the \textit{ferromagnetic} confocal $J$ matrices reveal a very different behavior in Fig.~\ref{fig:aggOlaps}(b). The ferromagnetic ensemble is constructed by using Gaussian-distributed spin positions with standard deviation of only $0.5w_0$ in the cavity midplane. This leads to ferromagnetic $J$ matrices with predominantly positive matrix elements and two global ground states corresponding to the two fully aligned spin states~\cite{Marsh2021eam}. Two-hundred quantum trajectories with the same ramp schedule as in Fig.~\ref{fig1}c are used to construct the overlap and magnetization distribution per $J$ matrix. The distributions are then averaged over the 100 matrices in the ensemble to produce the aggregate distributions in Fig.~\ref{fig:aggOlaps}(b). The lack of support in the interior of the overlap and magnetization distributions indicates that only these two $\mathbb{Z}_2$-related spin states are found with high probability. This is consistent with a ferromagnetic phase. A least-squares fit to the thermal model yields a temperature $T_{\mathrm{fit}}=0.011(2)\, \overline{T}_c$, where $\overline{T}_c$ is twice the maximum eigenvalue for $J$ matrices in the ferromagnetic regime~\cite{Sherrington1975smo}. The $T_{\mathrm{fit}}$ is  lower than that found for the spin glass ensemble. This may be due to the larger energy gap to the ground state in the unfrustrated ferromagnetic $J$ matrices. This makes it easier to maintain adiabaticity during the ramp, and fewer Landau-Zener transitions means a lower effective temperature.

A paramagnetic regime can be accessed by quenching the system into the superradiant regime rather than slowly ramping through the transition. In this case, adiabaticity is lost, and transitions into many excited states occur. Figure~\ref{fig:aggOlaps}(c) shows the overlap and magnetization distributions that result from such a quench. The same spin glass $J$ matrices as in Fig.~\ref{fig:aggOlaps}(a) are considered, with 200 quantum trajectories per $J$, and all other parameters remain the same. Both the overlap and magnetization distributions are well approximated by centered binomial distributions of standard deviation $1/\sqrt{\Nc}$. This is indicative of a paramagnetic phase in which states are found at random.   

Last, we present in Fig.~\ref{fig:aggOlapEvo} the dynamical evolution of the Parisi order parameter distribution for the spin glass.  The distribution becomes, after around 2~ms, Fig.~\ref{fig:aggOlaps}(a) in steady state.  We also note that a finite-size scaling analysis of the Binder ratio $1-\langle q_{\alpha\beta}^4 \rangle/(3\langle q_{\alpha\beta}^2 \rangle^2)$ is often used to pinpoint the exact location of the spin glass transition~\cite{Binder1981cpf}. However, the Binder ratio is ill-defined in this quantum system at early times because the overlap distribution begins as a delta function at $t=0$, for which both the second and fourth order moments are zero. This happens because the spins begin aligned along $\sigma^z$ rather than $\sigma^x$, a difficulty not encountered in typical equilibrium states of the classical Ising model. This makes the scaling analysis in this system more complicated, which we leave to future work.

\begin{figure}
    \centering
    \includegraphics[width=\columnwidth]{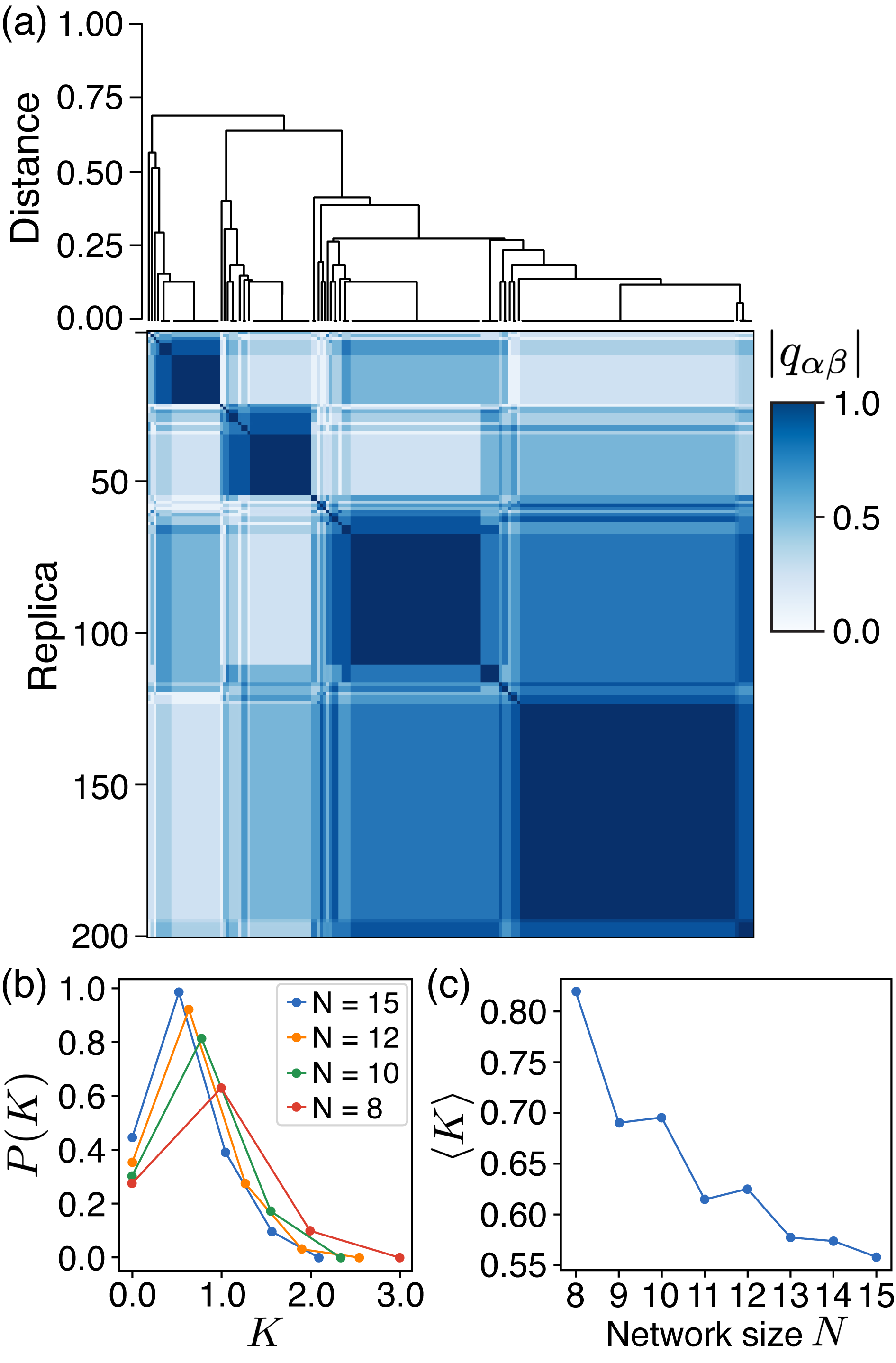}
    \caption{(a) The replica overlap matrix of 200 quantum trajectories for a confocal J matrix in the spin glass regime. The states cluster into one of four primary groups of states, which also appear as four primary blocks on the matrix diagonal. The above dendrogram shows the hierarchical clustering associated with the fracturing of the overlap matrix into four sectors, with various degrees of correlation between sectors. (b) The distribution of the $K$ metric for ultrametricity, averaged over 100 realizations of $J$ matrix for each system size. The distribution becomes increasingly peaked near zero, providing evidence of an ultrametric space  emerging with increased system size. (c) The mean of the $J$-averaged $K$ distribution as a function of system size, further showing the emergence of an ultrametric space.   }
    \label{fig:ultra}
\end{figure}

\section{Ultrametricity}\label{sec:ultra}

A prediction of Parisi's RSB ansatz for the SK spin glass solution is the formation of an ultrametric structure in the space of replicas~\cite{Mezard1984rsb}. An ultrametric space is one satisfying the strong triangle inequality: Given any three points $x$, $y$, and $z$, the distances between those points should satisfy $d(x,z)\leq \max[d(x,y),d(y,z)]$. It can be shown from this inequality that any triplet of points must form an isosceles triangle, either acute or equilateral. In the SK spin glass, replicas cluster into groups corresponding to low-energy local minima. Any triplet of replicas obeys this inequality in the thermodynamic limit, where the distance between replicas is the normalized Hamming distance, or equivalently $d(\alpha,\beta) = 1-|q_{\alpha\beta}|$. 

Numerical studies have verified that ultrametricity slowly emerges in system sizes up to $10^3$~\cite{Bhatt1989uit,Katzgraber2009uac,Katzgraber2012upo}. The approach to ultrametricity is quantified by use of the metric $K= (d_{\mathrm{max}}-d_{\mathrm{med}})/\sigma(d)$, where $d_\mathrm{max}$ is the largest distance in a given triplet of states and $d_\mathrm{med}$ is the second largest (or the median). Their difference should be zero in an ultrametric space due to the isosceles condition. The difference is normalized by $\sigma(d)$, the width of the distribution of distances between all states in the ultrametric space.

The overlap matrices and distributions in Figures~\ref{fig:singleOlap}--\ref{fig:manyOlaps} already exhibit the expected clustering of replicas into groups associated with local minima. Figure~\ref{fig:ultra}(a) demonstrates this even more clearly by plotting the associated dendrogram above the overlap matrix.  The clustering of replicas into four primary groups is visible. 
  
Figure~\ref{fig:ultra}(b) plots the $J$-averaged distribution of $K$ as a function of system size. For each $\Nc$, we generate 100 confocal $J$ matrices in the spin glass regime and perform 200 quantum trajectories per $J$. For each $J$, the $K$ distribution is computed between all triplets of trajectories and the resulting distribution is averaged over $J$ matrices. We begin the analysis at $\Nc=8$ because we find that low-energy local minima are not reliably present in smaller systems, with only a single cluster of states typically found. But even over the restricted range of $\Nc$ available, we find that the $K$ distribution becomes increasingly narrow with increasing $\Nc$. Moreover, Fig.~\ref{fig:ultra}(c) plots the mean of the $K$ distribution with $\Nc$, further showing a  decrease in $K$ with $\Nc$.  These constitute evidence for the emergence of ultrametricity in the system.  Oscillations in $\ex{K}$ may arise from finite-size effects but appear to dampen with increasing $\Nc$. We conclude that there is evidence for an approach to ultrametricity that is consistent with the significant finite-size effects found in SK spin glasses~\cite{Katzgraber2009uac}.   

\section{Discussion}\label{DiscussSec}

In summary, the transient formation of entangled states in a confocal cavity QED system allows RSB, concomitant with Ising symmetry breaking, to arise via the interplay of unitary Hamiltonian evolution and dissipative dynamics.  The resulting Parisi distribution does not appear to be in equilibrium, which is expected given the driven-dissipative nature of the open quantum system.
We note that previous work identified an effective temperature $T_{\mathrm{eff}}=(\detune^2+\loss^2)/4\detune$~\cite{Torre2013kaf,Marsh2021eam} associated with the multimode cavity QED dissipative dynamics. 
Such a temperature was found by considering detailed balance between energy-raising and energy-lowering processes in a system driven far above threshold.
This effective temperature is much larger than $\overline{T}_c$ in the spin--1/2 limit. (Although it is small in the semiclassical limit when $M\gg1$~\cite{Marsh2021eam}.)  Fortunately, however, this poses no obstacle to observing sufficiently low-energy states and RSB because the ramp through threshold reaches a quasi-steady state much faster than the timescale ${\propto}\Delta_c^4/(g^2\omega_z^2 \kappa)$ at which thermalization at $T_{\mathrm{eff}}$ would occur~\cite{Marsh2021eam}.

Last, we note that the light-matter coupling strength required to reach the superradiant threshold for the spin--$1/2$ system is higher by a factor of $\sqrt{\Na}$ compared to the semiclassical limit. We estimate that this interaction strength could be achieved by Rydberg-dressing large atomic ensembles to yield a Rydberg blockade within each~\cite{Marsh2023rydberg}. This allows each spin ensemble to behave as if it were a single spin-1/2 degree of freedom while  retaining the same collectively enhanced coupling strength ${\propto}\sqrt{\Na}g_0$~\cite{Lukin2001dba}.

Coupling to a Rydberg state could be realized through the addition of two pump lasers to the atomic level scheme in Fig.~\ref{fig1}(d); Ref.~\cite{Marsh2023rydberg} provides more details. Briefly, a laser at 780~nm would drive the $\ket{\uparrow}$ state to the atomic excited state $5{^2}P_{3/2}$, while a blue beam at 479~nm would drive a transition from this excited state to the $100^2S_{1/2}$ Rydberg state. (Rydberg dressing inside optical cavities has been achieved~\cite{Ningyuan2016oac}.)  The combined coupling terms produce a dark state that mixes the Rydberg and $\ket{\uparrow}$ states while avoiding the atomic excited state. Conservatively estimating an 8-$\mu$m average inter-atomic separation within a spin ensemble, a Rydberg-Rydberg interaction strength on the order of 200~MHz could be achieved. This should be sufficiently strong to push multiply excited Rydberg states far off resonance, resulting in an effective spin--1/2 degree of freedom for each spin ensemble. The spontaneous emission lifetime of the atoms is estimated to be greater than 10~ms---longer than the time scales shown above for RSB to emerge. This would allow RSB to be observed in a quantum optical context where implementations of, e.g., associative memory~\cite{Gopalakrishnan2012emo,Torggler2017qaw,Rotondo2018oqg,Fiorelli2019qaa,Torggler2019aqn,Fiorelli2020soa,Marsh2021eam} might be realized.  Doing so would provide experimental access to questions regarding how quantum effects might determine memory capacity and fidelity.  

The research data supporting this publication can be
accessed on the Harvard dataverse~\cite{Marsh2023data}.
 
\begin{acknowledgments}
We thank Helmut Katzgraber, David Atri Schuller, Henry Hunt, and Zhendong Zhang for general discussions and Jon Simon for information regarding Rydberg dressing.  We are grateful for funding support from the Army Research Office, NTT Research, and the Q-NEXT DOE National Quantum Information Science Research Center. Surya Ganguli acknowledges funding from NSF CAREER award \#1845166. B.M.~acknowledges funding from the Stanford QFARM Initiative and the NSF Graduate Research Fellowship. A portion of the computing for this project was performed on the Stanford Sherlock cluster.
\end{acknowledgments}

\appendix


\section{Confocal $J$ matrices}\label{app:J}

This section provides a simple approximation to the form of confocal $J$ matrices for realistic cavities. The $J$ matrix is given by the confocal cavity-mediated interaction evaluated at the positions of the atomic ensembles. The interaction is derived from the Green's function $G(\pos,\pos',\alpha)$ for the harmonic oscillator, which is used to compute sums over the cavity's Hermite-Gauss mode functions $\Xi_{lm}(\pos)$, indexed by the integers $l,m>0$. This is given by
\begin{align}
    &G(\pos,\pos',\alpha)\equiv\sum_{l,m} \Xi_{lm}(\pos)\Xi_{lm}(\pos')e^{-(l+m)\alpha} \\
    &= \frac{e^\alpha}{2\pi\sinh(\alpha)}\exp\left[ -\frac{(\pos-\pos')^2}{2w_0^2\tanh(\alpha/2)} -\frac{(\pos+\pos')^2}{2w_0^2\coth(\alpha/2)} \right], \nonumber
\end{align}
where $\alpha$ is any complex number with real part greater than zero and $w_0$ is the waist of the fundamental mode. In a confocal cavity, a given resonance supports only the set of even modes or the set of odd modes.  As such, it is useful to define the Green's function corresponding to either even modes (symmetrized) or odd modes (anti-symmetrized). Choosing the even case, the symmetrized Green's function is defined as $G^+(\pos,\pos',\alpha)=[G(\pos,\pos',\alpha)+G(\pos,-\pos',\alpha)]/2$. The confocal cavity-mediated interaction $\mathcal{D}(\pos,\pos')$ is expressed in terms of Green's functions~\cite{Guo2019eab,Kroeze2023hcu}:
\begin{equation}
    2\mathcal{D}(\pos,\pos') = G^+(\pos,\pos',\alpha)+G^+(\pos,\pos',\alpha+i\pi/2).
\end{equation}
The ideal interaction for a perfectly degenerate cavity with infinite mode support and delta-function-wide atomic ensembles is found by setting $\alpha=0$. The $\alpha=0$ limit corresponds to the $J$ matrix in Eq.~\eqref{eqn:Jij}, where the term on the first line gives rise to delta function local and mirror interactions, while the other terms give rise to the nonlocal interaction. 

Allowing for $\alpha>0$ provides a good approximation for cavities with both mirror aberrations and finite-sized atomic ensembles~\cite{Kroeze2023hcu}. In this work, we use $\alpha=0.02$ to achieve a ratio of approximately ten between the local and nonlocal interactions, which roughly matches observations in recent confocal cavity experiments~\cite{Vaidya2018tpa}. This $\alpha$ yields local and mirror interactions with Gaussian waist much smaller than $w_0$ and a nonlocal interaction that has a large Gaussian envelope of waist much larger $w_0$. While finite $\alpha$ does limit the maximum distance over which the nonlocal interaction can occur, it does not significantly affect the $J$ matrices for this work, which considers atomic positions out to only ${\sim}4w_0$. We note that the precise form of the confocal interaction depends on the details of both the atomic distribution and the nature of the cavity imperfections~\cite{Kroeze2023hcu}. However, these precise details matter little for the present work since they result in only small changes to the already disordered $J$ matrices.

  {The degree of randomness in the $J$ matrices produced by this cavity-mediated interaction was studied in depth in previous work~\cite{Marsh2021eam}. The elements of the $J$ matrix become increasingly uncorrelated as $w$, the standard deviation of the atomic positions in the cavity midplane, becomes large compared with $w_0=35$~$\mu$m, the waist of the fundamental mode of the cavity. The correlation between randomly chosen $J_{ij}$ elements is less than one percent for $w=2w_0$, which is the value of $w$ considered in the main text for generating glassy $J$ matrices. This lack of correlation comes about because of the incommensurate periodic dependence of $J_{ij}$ on the positions $\bm{r}_i, \bm{r}_j$. The confocal $J$ matrices produce eigenvalue spectra that approach a semicircle distribution, as expected for random matrices drawn from the Gaussian orthogonal ensemble (GOE), precisely like those of the SK model. At the system size $N=15$ considered in the main text, the eigenvalue distribution is within 5\% of the GOE semicircle distribution.}

\section{Derivation of semiclassical critical coupling strength}\label{app:crit}

We now derive Eq.~\eqref{eqn:gcrit} for the critical coupling strength of the superradiant phase transition using linear stability analysis. The spin operators $S_i^\alpha$ are first mapped to bosonic operators $b_i$ through the Holstein-Primakoff transformation, 
\begin{equation}
    S_i^z\to-\Na/2+b_i^\dag b_i,\quad S_i^x\to\frac{\sqrt{\Na}}{2}(b_i^\dag+b_i),
\end{equation}
where $\Na$ is the number of atoms per ensemble. This transformation accurately models fluctuations around the normal phase when $\Na$ is large. The original Hamiltonian in Eq.~\eqref{eqn:Ham} is then transformed, up to a constant shift, to
\begin{align}
    H_{0} &= -\sum_{\mu}^\Nm \Delta_\mu a_\mu^\dag a_\mu + \wz\sum_{i=1}^{\Nc} b_i^\dag b_i  \\ &\quad +\sum_{i=1}^{\Nc}\sum_\mu^\Nm g_{i\mu}(a_\mu^\dag + a_\mu)(b_i^\dag + b_i),\nonumber
\end{align}
where $\Nm$ is the total number of cavity modes and $g_{i\mu}=g\sqrt{\Na}\Xi_\mu(\pos_i)/2$ is the effective spin-photon coupling strength. The operator equations of motion, including cavity dissipation, are given by 
\begin{align}
    \dot a_\mu &= (i\Delta_\mu-\loss)a_\mu - i\sum_{i=1}^{\Nc}g_{i\mu}(b_i^\dag +b_i) \\
    \dot b_i &= -i\wz b_i -i\sum_\mu^\Nm g_{i\mu}(a_\mu^\dag + a_\mu) \nonumber. 
\end{align}
The equations of motion are linear and thus directly solvable. To organize the set of operators, we introduce an operator-valued vector 
\begin{equation}
    \mathbf{u} = \big(b_1,b_1^\dag,\cdots,b_{\Nc},b_{\Nc}^\dag,a_1,a_1^\dag,\cdots,a_{\Nm},a_{\Nm}^\dag \big)^T,
\end{equation}
where the first $2\Nc$ elements of $\mathbf{u}$ are the atomic operators followed by $2\Nm$ cavity operators. Using this notation, the equations of motion can be written in the concise form $\dot{\mathbf{u}} = A \mathbf{u}$ for a linear operator $A$. 

The critical coupling strength $g_c$ can be found from the retarded Green's function, which describes the response of the system to an external drive. It takes the matrix form $G_{ij}^R(t)=-i\langle [v_i(t),v_j^\dag(0)]\rangle\theta(t)$, where $\theta(t)$ is the Heaviside step function. The Fourier transform is then defined by $G_{ij}^R(\omega)=\int dt\,e^{i\omega t}G_{ij}^R(t)$. The Green's function is related to the linear operator $A$ by $[G^{R}(\omega) ]^{-1}=S^{-1}(\omega-iA)$ in the case of linear Heisenberg equations~\cite{Kirton2018itt}, where $S_{ij}=\langle [\mathbf{v} _i(0),\mathbf{v}_j^\dag(0)] \rangle$ are the equal time commutation relations. In this case, $S=\mathrm{diag}(+1,-1,+1,-1,\cdots)$ follows from canonical bosonic commutation relations. The full inverse Green's function, while large, has a simple $2{\times} 2$ block form. The first few rows and columns of each block are shown below:

\begin{widetext}
\begin{multline}
    \big[G^{R}(\omega) \big]^{-1} = \\
    -\begin{pmatrix}
    \wz-\omega & 0 & 0 & 0 & \cdots & g_{11} & g_{11} & g_{12} & g_{12} & \cdots \\
    0 & \wz+\omega & 0 & 0 & \cdots & g_{11} & g_{11} & g_{12} & g_{12} & \cdots  \\
    0 & 0 & \wz-\omega & 0 & \cdots & g_{21} & g_{21} & g_{22} & g_{22} & \cdots  \\
    0 & 0 & 0 & \wz+\omega & \cdots & g_{21} & g_{21} & g_{22} & g_{22} & \cdots  \\
    \vdots & \vdots & \vdots & \vdots & \ddots & \vdots & \vdots & \vdots & \vdots & \ddots \\
    g_{11} & g_{11} & g_{21} & g_{21} & \cdots & -\Delta_1-\omega-i\loss & 0 & 0 & 0 & \cdots\\
    g_{11} & g_{11} & g_{21} & g_{21} & \cdots & 0 & -\Delta_1+\omega+i\loss & 0 & 0 & \cdots\\
    g_{12} & g_{12} & g_{22} & g_{22} & \cdots & 0 & 0 & -\Delta_2-\omega-i\loss & 0 & \cdots\\
    g_{12} & g_{12} & g_{22} & g_{22} & \cdots & 0 & 0 & 0 & -\Delta_2+\omega+i\loss  & \cdots\\
    \vdots & \vdots & \vdots & \vdots & \ddots & \vdots & \vdots & \vdots & \vdots & \ddots
    \end{pmatrix}.
\end{multline}
\end{widetext}
We analyze the Green's function by first assigning blocks of the matrix:
\begin{equation}
    \big[G^{R}(\omega) \big]^{-1} = -\begin{pmatrix}
    D_{\mathrm{spin}} & C \\
    C^T & D_{\mathrm{cav}}
    \end{pmatrix}.
\end{equation}
The diagonal matrix $D_{\mathrm{spin}}$ is $2\Nc{\times}2\Nc$ with alternating elements $\wz-\omega$, then $\wz+\omega$. The matrix $D_{\mathrm{cav}}$ is also diagonal of size $2\Nm{\times} 2\Nm$, with elements $-\Delta_\mu-\omega-i\loss$, followed by $-\Delta_\mu+\omega+i\loss$.  In this expression, $\mu$ increases from 1 to $\Nm$. The matrix $C$ describes coupling between the cavity modes and spin modes and is of size $2\Nc{\times} 2\Nm$. It is most easily expressed in terms of $2{\times} 2$ blocks given by
\begin{equation}
    C_{i\mu} =  g_{i\mu}\begin{pmatrix}
    1 & 1\\
    1 & 1
    \end{pmatrix},
\end{equation}
where $i\in[1,\Nc]$ and $\mu\in[1,\Nm]$.

We now determine when an instability in the normal phase occurs by considering the poles of the inverse Green's function. The poles
dictate the characteristic response frequencies of the system, and thus the determination of when $\omega=0$ becomes a pole probes the global stability of the phase. This point can be found by considering when $\det([G^R(\omega=0)]^{-1})$ crosses zero. This procedure relates to a normal mode analysis of the linear equations of motion, in which the system is stable only when all eigenvalues of $A$ are greater than zero. The determinant can be written using the Schur complement~\cite{Zhang2005sca} of $D_{\mathrm{cav}}$:
\begin{equation}
    \det \big[G^{R}(\omega) \big]^{-1} = \det\left( D_{\mathrm{spin}} -CD_{\mathrm{cav}}^{-1}C^T\right)\det\left( D_{\mathrm{cav}}\right).
\end{equation}
Because $D_{\mathrm{cav}}$ is diagonal, we find that $\det\left( D_{\mathrm{cav}}\right) = \prod_\mu (\Delta_\mu^2+\loss^2)$ is always positive. Thus, the instability condition simplifies to 
    $\det\left( D_{\mathrm{spin}} -CD_{\mathrm{cav}}^{-1}C^T\right)=0.$
The matrix inside the determinant has a simple tensor product structure. At $\omega=0$ it is given by
\begin{equation}\label{eqn:spinMatrix}
    \frac{D_{\mathrm{spin}}-CD_{\mathrm{cav}}^{-1}C^T}{\wz} = 
    I -\frac{\Na g^2 |\detune|}{2\wz(\detune^2+\loss^2)} J\otimes\begin{pmatrix}
    1 & 1 \\
    1 & 1
    \end{pmatrix}, 
\end{equation}
where $\otimes$ denotes the tensor product, $I$ is the identity operator, and $J$ is the cavity-mediated interaction connectivity matrix introduced in Eq.~\eqref{eqn:Jij}. 

We must now determine when one of the eigenvalues of the above matrix crosses zero. The identity matrix simply shifts all eigenvalues by one. The eigenvectors of the total matrix now have the form $\mathbf{v}^k\otimes \mathbf{w}$, where $\mathbf{v}^k$ is an eigenvector of $J$ with eigenvalue $\lambda_k\geq 0$ and $\mathbf{w}$ is an eigenvector of the second matrix of ones. The second matrix has a zero eigenvalue and a nonzero eigenvalue $2$ with eigenvector $(1,1)/\sqrt{2}$. We thus find that half of the eigenvalues of the matrix in Eq.~\eqref{eqn:spinMatrix} are degenerate with value one, and the other half are given by
\begin{equation}
    1-\frac{\Na g^2|\detune|\lambda_k}{\wz(\detune^2+\loss^2)}.
\end{equation}

The smallest value of $g$ for which one of the above eigenvalues crossed zero occurs sets the critical coupling strength $g_c$. The critical coupling thus depends on the largest eigenvalue $\eigmax$ of the $J$ matrix. Inserting $\eigmax$, setting the expression to zero, and solving for $g$ yields the critical coupling strength of Eq.~\eqref{eqn:gcrit}.

\section{Derivation of the atom-only theory}\label{app:atomonly}

We apply the method of J\"ager \textit{et al.}~\cite{Jager2022lme} to produce an atom-only theory for spins in a confocal cavity. The method accurately reproduces the low-energy spectrum of the single-mode driven-dissipative Dicke model, both below and above threshold. We extend the method to the multimode, multiple spin ensemble case described by the Hamiltonian in Eq.~\eqref{eqn:Ham}. 

We must first find ``effective fields" corresponding to operators in the spin Hilbert space that best approximate the effect of the cavity modes. There is one effective field $\tilde S_\mu$ for each cavity mode that is eliminated. The field may be time-dependent. The effective fields are chosen to satisfy the differential equations
\begin{equation}\label{eqn:MorProb}
    \frac{d}{dt}\tilde S_\mu = -i[H_0,\tilde S_\mu] + (i\Delta_\mu-\loss) \tilde S_\mu -ig\sum_{i}\Xi_\mu(\pos_i)S_i^x .
\end{equation}
Solving for the effective fields begins with an ansatz of the form
\begin{equation}\label{eqn:effFields}
    \tilde S_\mu = \sum_i\left[x_{\mu i}^+(t)S_i^++x_{\mu i}^-(t)S_i^-\right].
\end{equation}
Inserting the ansatz into Eq.~\eqref{eqn:MorProb} yields the following differential equations for the coefficients:
\begin{equation}
    \frac{d}{dt}x_{mi}^\pm(t)=[i\Delta_\mu\mp i\omega_z(t)-\loss]x_{\mu i}^\pm(t)-i\frac{g(t)}{2}\Xi_\mu(\pos_i).
\end{equation}
This equation can be solved explicitly in the limit that the ramp function $f(t)$ changes slowly over the cavity loss timescale $2\pi/\loss \approx 10\,\mu$s. This limit is well satisfied given that $f(t)$ ramps over a period of $600\,\mu$s. In this limit, the differential equations are given by the solutions
\begin{equation}
    x_{\mu i}^\pm (t) = -\frac{g(t)\Xi_\mu(\pos_i)}{2[-\Delta_\mu\pm\wz(t)-i\loss]}.
\end{equation}
We now restrict ourselves to the degenerate case $\Delta_\mu=\detune$ for all $\mu$. The effective fields can then be written as
\begin{equation}\label{eqn:effFields2}
    \tilde S_\mu=-\frac{g(t)}{2\detune} \sum_{i=1}^\Nc \Xi_\mu(\pos_i)\big(\alpha_+ S_i^x + i\alpha_- S_i^y\big),
\end{equation}
with $\alpha_\pm$ given by Eq.~\eqref{eqn:alphas}.

The effective fields can then be used to write the atom-only master equation. The Hamiltonian part is given by  
\begin{equation}
    H = \wz\sum_{i=1}^\Nc S_i^z + \frac{g(t)}{2}\sum_{i=1}^\Nc S_i^x \sum_\mu \Xi_\mu(\pos_i)(\tilde S_\mu + \tilde S^\dag_\mu).
\end{equation}
Inserting the effective fields from Eq.~\eqref{eqn:effFields2} and recognizing the $J$ matrix then leads to the atom-only Hamiltonian presented in Eq.~\eqref{eqn:HA}. The full form of the master equation is now
\begin{equation}
    \dot\rho = -i[H,\rho] + \loss\sum_\mu \mathcal{D}[\tilde S_\mu].
\end{equation}
While correct, this expression is complicated to use because it involves an infinite sum of dissipation terms.  This can be avoided by noting that there are only $N$ linearly independent jump operators that can be created out of the $N$ operators $\alpha_+ S_i^x + i \alpha_- S_i^y$.  As such, one can rewrite the dissipation term by expanding the effective fields. This yields
\begin{align}
    \dot\rho &= -i[H,\rho] \\&\quad+ \frac{\loss g(t)^2}{4\detune^2}\sum_{ij=1}^\Nc J_{ij} \mathcal{D}[\alpha_+S_i^x +i \alpha_-S_i^y,\alpha_+S_j^x +i \alpha_-S_j^y], \nonumber 
\end{align}
where $\mathcal{D}[X,Y]=2X\rho Y^\dag -\{Y^\dag X,\rho\}$ is the non-diagonal Lindblad superoperator and the Lindblad--Kossakowski matrix $J$ is exactly the cavity-mediated interaction matrix defined in Eq.~\eqref{eqn:Jij}. Diagonalization of $J$ brings the Lindbladian into diagonal form $\dot\rho = -i[H,\rho] +\sum_k \mathcal{D}[C_k]$, with collapse operators
\begin{equation}
    C_k=\frac{g\sqrt{\lambda_k\loss}}{2\detune}\sum_{i=1}^\Nc \evec^k_i\big(\alpha_+S_i^x +i\alpha_- S_i^y\big).
\end{equation}

The square of the coefficients multiplying the collapse operators give the associated decoherence rates. A total decoherence rate per spin can be estimated by approximating the elements $\mathbf{v}^k_i$ of the normalized eigenvectors as uncorrelated random variables. Their variance should be $1/N$ to enforce unit normalization. We also approximate $|\alpha_+|=2$ and $\alpha_-=0$, which is valid well above threshold. The summed decoherence rate per spin can then be approximated as $(\kappa g^2/\detune^2)\sum_{k=1}^N \lambda_k$.

\section{Stochastic unraveling and reconstructing the spin measurement record}
\label{app:record}

The general formalism for homodyne unraveling has been well described by various authors~\cite{Breuer2007tto,Daley2014qta,Verstraelen2018gqt}. We provide a brief discussion of the specific measurement approach, and thus the associated unraveling scheme that we use in this work, which is based on invariance properties of the Lindbladian. 

Spatial heterodyne detection is derived from the interference pattern of the LO and cavity light on a charged-coupled device camera~\cite{Kroeze2018sso,Guo2019spa}. While the measurement is a \textit{spatial} heterodyne detection, meaning that the LO and cavity light have different propagation directions, the LO and cavity light possess the same optical frequency, as in a homodyne detection. We thus model the detection scheme as a balanced homodyne detection of the emitted cavity field. 

To derive the unraveling corresponding to balanced homodyne measurements, we start by considering the atom-only master equation with Lindblad form
\begin{equation}
    \dot \rho = -i[H,\rho] + \sum_{k=1}^\Nc \mathcal{D}[C_k],
\end{equation}
where the collapse operators $C_k$ are given by Eq.~\eqref{eqn:collapse}. We can then manipulate the master equation to cast it in terms of the collapse operators corresponding to balanced homodyne detection. First, we write the Lindblad superoperators as a sum of two equal terms with rescaled collapse operators,
\begin{equation}\label{eqn:twoCollapse}
    \dot \rho = -i[H,\rho] + \sum_{k=1}^\Nc\big( \mathcal{D}\big[C_k/\sqrt{2}\big] + \mathcal{D}\big[C_k/\sqrt{2}\big]\big).
\end{equation}
We then use the shift-invariance property of the master equation, which admits shifts of a collapse operator $C\to C+aI$ if the Hamiltonian is also modified as $H\to H-i(a^*C-aC^\dag)$. We perform shifts $C_k\to C_k\pm i\beta I$ for each $k$ and pair of collapse operators in Eq.~\eqref{eqn:twoCollapse}. Here, $\beta$ represents a real number proportional to the LO amplitude. Performing this shift yields the master equation
\begin{align}
    \dot \rho = -i[H,\rho] + \sum_{k=1}^\Nc\left( \mathcal{D}\left[\frac{C_k+i\beta}{\sqrt{2}}\right] + \mathcal{D}\left[\frac{C_k-i\beta}{\sqrt{2}}\right]\right). 
\end{align}
The collapse operators $(C_k+i\beta)/\sqrt{2}$ and $(C_k-i\beta)/\sqrt{2}$ correspond to measurements of the two field quadratures, respectively. The Hamiltonian contribution from the two above collapse operators cancel out up to a global constant offset. The master equation is now unraveled into quantum trajectories using the standard quantum jump formalism~\cite{Breuer2007tto} but with the above, shifted collapse operators.  The extra field $\beta$ means that the probability of a jump at each time step is higher, but the form of the collapse operator means that in such a jump, there is a smaller change to the state of the system.  As one interpolates between $\beta=0$ and $\beta=\infty$ this approach thus interpolates between quantum jumps (in terms of the original jump operators) and continuous quantum state diffusion.

The value of $\beta$ used in our simulations is $0.1\sqrt{\kappa}$. Recall that $\beta$ corresponds to $\sqrt{\kappa}$ multiplied by the coherent state amplitude of the LO and the overlap between the LO and emitted cavity light. While this sounds small, the relevant quantity for determining how close the system is to the quantum state diffusion limit  is the ratio of the number of detections per unit time to the rate associated with spin dynamics. The quantum state diffusion limit occurs when detections occur much more quickly than system dynamics. 

The detection rate can be approximated for a given value of $\beta$~\cite{Breuer2007tto} by
\begin{equation}
    \sum_{k=1}^\Nc\sqrt{\left(\lambda_k\frac{\kappa\omega_z g^2}{\Delta_c g_c^2}\right) \beta^2}.
\end{equation}
The term in parentheses is the bare detection rate corresponding to the $k$'th collapse operator, where $\lambda_k$ is the $k$'th eigenvector of $J$. The term $\beta^2$ boosts the bare rate by the LO strength. The sum is taken over all collapse operators to approximate the total detection rate for each spin. This yields a detection rate of approximately 2~MHz at full ramp power when using typical $J$ matrices in the spin glass regime;  this translates to a timescale of about 0.5~$\mu$s. On the other hand, the spin dynamics typically occur no faster than ${\sim}10$~$\mu$s. Thus, we conclude that the dynamics are similar to those obtained in the diffusion limit.
 
The measurement records for each spin are constructed from the balanced homodyne signals. Balanced homodyne records $h_k(t)$ measure the difference in the number of jumps between $(C_k+i\beta)/\sqrt{2}$ and $(C_k-i\beta)/\sqrt{2}$ as a function of time. Each record is initialized as $h_k(0)=0$. Every time that a jump occurs in $(C_k+i\beta)/\sqrt{2}$, $h_k(t)$ is increased by one, while it is decreased by one every time a jump occurs in $(C_k-i\beta)/\sqrt{2}$. 
The homodyne records do not yet reflect the spin measurements because the collapse operators $C_k$ are composed of a linear combination of spin operators from different sites. To construct the spin measurements records $s_i(t)$, the relation between homodyne records and spin states must be inverted by taking the linear combination
\begin{equation}
    s_i(t) = \sum_{k=1}^\Nc \mathbf{v}^k_i h_k(t),
\end{equation}
where again $\mathbf{v}^k_i$ is the $i$'th element of the $k$'th eigenvector of the $J$ matrix. The records $s_i(t)$ now change at a rate proportional to $\ex{S_i^x}$, and thus, after sufficient integration time, their signs indicate the steady-state spin configuration. An example of the measurement records for a typical quantum trajectory is shown in Fig.~\ref{fig1}(b).

\section{Semiclassical equations of motion}
\label{app:semiclassical}

We now derive the semiclassical equations of motion that describe a homodyne unraveling of the master equation.  For these semiclassical calculations, we take the quantum state diffusion limit appropriate for large spins~\cite{Breuer2007tto,Verstraelen2018gqt}, corresponding to the limit $\beta\to\infty$ in the previous section. 

The stochastic differential equation describing the expectation value of an observable $A$ is written in {It\^o} form as
\begin{align}
    d\ex{A} &= i\ex{[H,A]}dt \\&\quad +\sum_k \Big[ \big( 2\ex{C_k^\dag A C_k} - \ex{C_k^\dag C_kA} - \ex{AC_k^\dag C_k} \big)dt \nonumber \\&\quad\quad\quad\quad +\sqrt{2}\big( \ex{C_k^\dag A} - \ex{C_k^\dag}\ex{ A}\big)dW_k \nonumber \\&\quad\quad\quad\quad + \sqrt{2}\big( \ex{AC_k} - \ex{A}\ex{C_k}\big)dW_k^* \Big],  \nonumber
\end{align}
where each $dW_k$ is the differential of an independent, Wiener process with $\langle dW_k dW_k^*\rangle=dt$. The Wiener process can be either real, for homodyne detection, or complex for heterodyne detection. We consider a real Wiener process from here on for simplicity and without loss of generality. The semiclassical equations of motion are derived by first evaluating the above exact equation for $S_i^{x/y/z}$ and simplifying the result through commutation relations. Any remaining product of spin operators $A$ and $B$ is then decomposed into a commutator and anticommutator, $AB\to [A,B]/2 + \{A,B\}/2$. This step is necessary to achieve a semiclassical limit that retains stochastic terms from the homodyne detection. We then perform a mean-field decoupling of the anticommutator terms $\ex{\{A,B\}}/2 \to \ex{A}\ex{B}$ to arrive at semiclassical equations of motion:
\begin{widetext}
\begin{align}
    d\ex{S_i^x} &= -\left(\wz +\Re[\alpha_-]\frac{g^2J_{ii}}{4\detune} \right)\ex{S_i^y} dt + \frac{g^2\ex{S_i^z}}{2\detune}\sum_{j=1}^\Nc J_{ij}\Big(\Im[\alpha_-]-\frac{\loss}{\detune}\Re[\alpha_+^*\alpha_-]\Big)\ex{S_j^x}dt \\ &\quad-\frac{g^2\loss J_{ii}}{4\detune^2}\big( \Im[\alpha_+^*\alpha_-]\ex{S_i^y}+|\alpha_-|^2\ex{S_i^x} \big)dt +\frac{g\sqrt{\loss}}{\sqrt{2}\detune}\ex{S_i^z}\sum_{k=1}^\Nc \sqrt{\lambda_k}\mathbf{v}^k_i \Re[\alpha_- dW_k] \nonumber \\
    d\ex{S_i^y} &= \left(\wz +\Re[\alpha_-]\frac{g^2J_{ii}}{4\detune} \right)\ex{S_i^x} dt + \frac{g^2\ex{S_i^z}}{2\detune} \sum_{j=1}^\Nc J_{ij}\Big[2\Re[\alpha_+]\ex{S_j^x} - \Big(\Im[\alpha_-]+\frac{\loss}{\detune}\Re[\alpha_+^*\alpha_-]\Big)\ex{S_j^y}\Big]dt\\ 
    &\quad - \frac{g^2\loss J_{ii}}{4\detune^2} \big( \Im[\alpha_+^*\alpha_-]\ex{S_i^x}+|\alpha_+|^2\ex{S_i^y} \big)dt -\frac{g\sqrt{\loss}}{\sqrt{2}\detune}\ex{S_i^z}\sum_{k=1}^\Nc \sqrt{\lambda_k}\mathbf{v}^k_i \Im[\alpha_+ dW_k]\nonumber\\
    d\ex{S_i^z} &= \frac{g^2}{2\detune}\sum_{j=1}^\Nc J_{ij}\Big[\Im[\alpha_-]\big(\ex{S_i^y}\ex{S_j^y}-\ex{S_i^x}\ex{S_j^x}\big)+\frac{\loss}{\detune}\Re[\alpha_+^*\alpha_-]\big(\ex{S_i^x}\ex{S_j^x}+\ex{S_i^y}\ex{S_j^y}\big)-2\Re[\alpha_+]\ex{S_i^y}\ex{S_j^x} \Big]dt \nonumber\\
    &\quad  - \frac{g^2\loss J_{ii}}{4\detune^2} \big( |\alpha_+|^2+|\alpha_-|^2 \big)\ex{S_i^z}dt +\frac{g\sqrt{\loss}}{\sqrt{2}\detune}\sum_{k=1}^\Nc \sqrt{\lambda_k}\mathbf{v}^k_i\big( \Im[\alpha_+ dW_k]\ex{S_i^y} - \Re[\alpha_- dW_k]\ex{S_i^x}\big).
\end{align}
\end{widetext}
Above, $\alpha_\pm$ are as defined in Eq.~\eqref{eqn:alphas}.

  { 
\section{The Parisi distribution in terms of quantum trajectories}\label{ParisiSupp}

In this section, we first summarize the form of the Parisi distribution in a quantum spin glass, following Refs.~\cite{Goldschmidt1990isg,Charbonneau2023sgt}. We then show how it can be understood in terms of an overlap operator in a doubled Hilbert space. This leads to the trajectory formulation of the overlap distribution presented in Eq.~\eqref{PJ_Traj} of the main text.

\subsection{Parisi order parameter for a quantum spin glass}
For simplicity, we consider here the SK model in a transverse field.  It is a close approximation of the more realistic model derived in Sec.~\ref{sec:ccQED}. The Hamiltonian is given by  
\begin{equation}\label{eq:QIsing}
    H=-h_q\sum_{i=1}^N \sigma_i^z -\sum_{i<j}^N J_{ij}\sigma_i^x\sigma_j^x,
\end{equation}
where $\sigma_i^{x,y,z}$ are Pauli operators acting on one of the $N$ total spins. The $J_{ij}$ couplings here are all-to-all with each element sampled independently from a Gaussian distribution with zero mean and variance $\sigma_J^2/N$. We consider here the equilibrium density matrix $\rho=e^{- H/T}/Z$ with partition function $Z=\Tr[e^{- H/T}]$. Application of the Suzuki-Trotter formula~\cite{Suzuki1991gto,Trotter1959otp} allows for a reformulation of $Z$ in terms of an equivalent classical model in a higher dimension. The classical energy is~\cite{Lai1990mcs}
\begin{equation}\label{eq:QCmap}
    E(\bm{s},J)=-\frac{1}{L}\sum_{\tau=1}^L\sum_{i<j}^N J_{ij} s_{i,\tau}s_{j,\tau} - h_c\sum_{\tau=1}^L\sum_{i=1}^N s_{i,\tau}s_{i,\tau+1},
\end{equation}
where $\bm{s}$ denotes the set of classical spin variables $s_{i,\tau}=\pm 1$. The term $h_c = T\ln[\coth( h_q/LT)]/2$ describes a nearest-neighbor coupling in the Trotter dimension indexed by $\tau$, with periodic boundary conditions. The mapping becomes exact in the limit that the number sites $L$ in the Trotter dimension tends to infinity.

Mapping the quantum partition function to an effective classical one unlocks the tools of replica theory. The free energy is evaluated via the ``replica trick" $\log(Z)=\lim_{n\to 0}(Z^n-1)/n$, reducing the problem to computing the replicated partition function $Z^n$ for $n$ replica spin systems $\bm{s}^\alpha$. The free energy is then averaged over the quenched disorder matrices. This corresponds to the disorder-averaged, replicated partition function $\overline{Z^n}$ given by
\begin{equation}
    \overline{Z^n} = \int \prod_{i<j}^N dJ_{ij}p(J_{ij})\sum_{\{\bm{s}^\alpha\}}\exp\left\{ -\frac{1}{T}\sum_{\alpha=1}^n E(\bm{s}^\alpha,J) \right\},
\end{equation}
where $p(J_{ij})=\exp(-NJ_{ij}^2/2\sigma_J^2)/\sqrt{2\pi}\sigma_J$ and the sum is taken over all replica spin states. The disorder integrals can be computed exactly, which introduces a coupling between replicas. One finds that in the large-$N$ limit the partition function can be expressed in terms of a local effective action with spins at different sites $i$ decoupled, $\overline{Z^n}=\prod_i e^{-S_i}$. This gives a single-site action~\cite{Goldschmidt1990isg}
\begin{align}\label{eq:effAction}
    S_i &= -\frac{\sigma_J^2}{L^2T^2}\sum_{\tau,\tau'}^L \left(\sum_{\alpha<\beta}^n  s_{i,\tau}^\alpha s_{i,\tau'}^\beta Q_{\alpha\beta}+\frac{\chi_{\tau-\tau'}}{2}\sum_{\alpha=1}^n s_{i,\tau}^\alpha s_{i,\tau'}^\alpha \right) \nonumber \\ &\quad -\frac{ h_c}{T}\sum_{\tau=1}^L\sum_{\alpha=1}^n s_{i,\tau}^\alpha s_{i,\tau+1}^\alpha.
\end{align}
The matrix $Q_{\alpha\beta}$ is the overlap order parameter that, after performing the disorder average, describes a coupling between replicas. The order parameter $\chi_{\Delta\tau}$ is a replica-diagonal correlator describing a translation-invariant coupling in the Trotter dimension. Evaluation of $\overline{Z^n}$ by the method of steepest descent gives self-consistent equations for these order parameters,
\begin{equation}\label{eq:selfCon}
    Q_{\alpha\beta} = \frac{1}{N}\sum_{i=1}^N\langle s^\alpha_{i,\tau}s^\beta_{i,\tau'} \rangle_{S},\quad \chi_{\tau-\tau'}=\frac{1}{N}\sum_{i=1}^N\langle s^\alpha_{i,\tau} s^\alpha_{i,\tau'} \rangle_S,
\end{equation}
where $\langle A \rangle_S=\sum_{\{\bm{s}^\alpha\}}Ae^{-S}/(\sum_{\{\bm{s}^\alpha\}}e^{-S})$ denotes an expectation value over the measure defined by $S$. While the overlap matrix may seem an abstract or purely theoretical object, Parisi realized~\cite{Parisi1983opf} that $Q_{\alpha\beta}$ contains clear physical content: It describes the overlaps, or equivalently the distances, between the large number of distinct thermodynamic states in a single disorder realization. Finally, the Parisi order parameter $P(q)$ is the distribution of elements of the $Q_{\alpha\beta}$ matrix,
\begin{equation}\label{eq:ParisiDef}
    P(q)= \lim_{n\to 0} \frac{1}{n(n-1)}\sum_{\alpha\neq \beta}^n \delta(Q_{\alpha\beta}-q).
\end{equation}
The structure of $P(q)$~\cite{Mezard1984rsb,Charbonneau2023sgt} is modified in the quantum case by a transverse field~\cite{Yamamoto1987ape,Lai1990mcs,Young2017sot}.   
  
\subsection{Parisi distribution from the overlap operator}

To establish a connection between $Q_{\alpha\beta}$ as predicted by replica theory and a physical observable of the quantum system, we consider the moments $q^{(k)} \equiv \int dq P(q) q^k $ of the Parisi distribution. Using Eq.~\eqref{eq:ParisiDef} yields
\begin{equation}
    q^{(k)} = \lim_{n\to 0} \frac{1}{n(n-1)}\sum_{\alpha\neq \beta}^n Q_{\alpha\beta}^k.
\end{equation}
We now insert the self-consistency equation~\eqref{eq:selfCon} for $Q_{\alpha\beta}$ and make use of the fact that different sites decouple in the effective action, Eq.~\eqref{eq:effAction}. In the large-$N$ limit, this yields
\begin{equation}
    q^{(k)} =  \lim_{n\to 0} \frac{1}{ n(n-1)N^k}\sum_{\alpha\neq \beta}^n \sum_{i_1\cdots i_k}^N \Bigg\langle \prod_{j=1}^k s_{i_j,\tau}^\alpha s_{i_j,\tau'}^\beta \Bigg\rangle_S. \nonumber
\end{equation}
The expectation value over the effective action $S$, in which the disorder has been integrated out, is now related back to expectation values at the level of individual disorder realizations. Following Ref.~\cite{Binder1986sge}, each distinct replica index appearing in the expectation value under $S$ corresponds to a distinct thermal average under the classical energy~\eqref{eq:QCmap}. The result is then averaged to yield $q^{(k)}=[q^{(k)}_J]_J$, where $[\cdot]_J$ denotes the disorder average over $J$ realizations and $q^{(k)}_J$ is the moment associated with an individual disorder realization, given by
\begin{equation}
    q_J^{(k)} = \frac{1}{N^k}\sum_{i_1\cdots i_k}^N \Bigg\langle  \prod_{j=1}^k s_{i_j,\tau} \Bigg \rangle_E \Bigg\langle  \prod_{l=1}^k s_{i_l,\tau'} \Bigg \rangle_E .
\end{equation}
Here, $\langle \cdot \rangle_E$ denotes a thermal average with respect to the classical energy in Eq.~\eqref{eq:QCmap}. This can be related back to quantum expectation values as $\langle \prod_j s_{i_j,\tau} \rangle_{E}=\Tr[ \rho_J \prod_j\sigma_{i_j}^x]$, where $\rho_J$ is the equilibrium density matrix and the expression holds for any choice of $\tau$~\cite{Lai1990mcs}. The key step is then to apply the simple relation $\Tr[X]^2=\Tr[X\otimes X]$, yielding
\begin{align}
    q_J^{(k)} &= \frac{1}{N^k}\sum_{i_1\cdots i_k}^N \Tr[\Bigg(\rho_J\prod_{j=1}^k\sigma_{i_j}^x\Bigg)\otimes \Bigg(\rho_J \prod_{l=1}^k\sigma_{i_l}^x\Bigg) ]  \nonumber \\ 
    &= \Tr[\big(\rho_J\otimes\rho_J\big)\left(\frac{1}{N}\sum_{i=1}^N \sigma_i^x\otimes\sigma_i^x\right)^k ].  
\end{align}
The term in parentheses to the power $k$ is precisely the overlap operator $\mathcal{O}$ discussed in Eq.~\eqref{olapOp} of the main text, leading to relation $q^{(k)}=[\langle \mathcal{O}^k \rangle]_J$.

The characteristic function $\varphi(t)$ for the overlap distribution of each $J$ is now straightforward to compute given the above simple form for the moments. Direct evaluation yields
\begin{align}
    \varphi(t) &= \sum_{k=0}^\infty \frac{(it)^k q^{(k)}}{k!} \\
    &= \Bigg[ \Tr[ (\rho_J\otimes\rho_J)\sum_{k=0}^\infty \frac{(it\mathcal{O})^k}{k!}] \Bigg]_J\nonumber \\
    &=\Bigg[\Tr\Big[ \big(\rho_J\otimes\rho_J\big) e^{it\mathcal{O}} \Big]\Bigg]_J. \nonumber
\end{align}
We obtain the overlap distribution through a Fourier transform of the characteristic function. Performing the disorder average then yields the Parisi distribution, 
\begin{equation}\label{ParisiExp}
    P(q) = \left[\Tr\Big[ \big(\rho_J\otimes\rho_J\big)\int \frac{dt}{2\pi} e^{it(\mathcal{O}-qI)} \Big] \right]_J,
\end{equation}
where $I$ is the identity matrix in the matrix exponential. The exponential term can be recognized as the integral form of the Dirac delta function. To describe explicitly the action of the delta function, we expand the overlap operator in its eigenbasis as $\mathcal{O} = \sum_{q} q \mathcal{P}_q$, where the sum runs over all $N+1$ allowed values of the spin overlap $q=-1,-1+2/N,\cdots,1$. The operators $\mathcal{P}_q$ are orthogonal projections onto the space of spin state pairs with overlap $q$. They mutually commute, $[\mathcal{P}_q,\mathcal{P}_{q'}]=0$, and have product relations $\mathcal{P}_q\mathcal{P}_{q'}=\mathcal{P}_q \delta_{qq'}$. Furthermore, the projectors span the full space of spin states and thus form a resolution of the identity, $I=\sum_q \mathcal{P}_q$. Inserting these forms into the integral expression of Eq.~\eqref{ParisiExp} and evaluating the operator exponential, we find
\begin{align}
    \int \frac{dt}{2\pi} e^{it(\mathcal{O}-qI)} &= \int \frac{dt}{2\pi} \exp\left[it\sum_{q'}\left(q'-q\right)\mathcal{P}_{q'}\right] \\
    &= \sum_{q'}\int \frac{dt}{2\pi} \exp\left[it\left(q'-q\right)\right]\mathcal{P}_{q'} \nonumber \\
    &= \sum_{q'}\delta(q'-q)\mathcal{P}_{q'}. \nonumber
\end{align}
Inserting this form back into Eq.~\eqref{ParisiExp} produces the form of the overlap distribution presented in the main text,
\begin{equation}
    P(q) = \sum_{q'}\delta(q-q')\Bigg[ \Tr\Big[ (\rho_J\otimes\rho_J)\mathcal{P}_{q'} \Big] \Bigg]_J. 
\end{equation}
}

\begin{figure*}[th!]
    \centering
    \includegraphics[width=\textwidth]{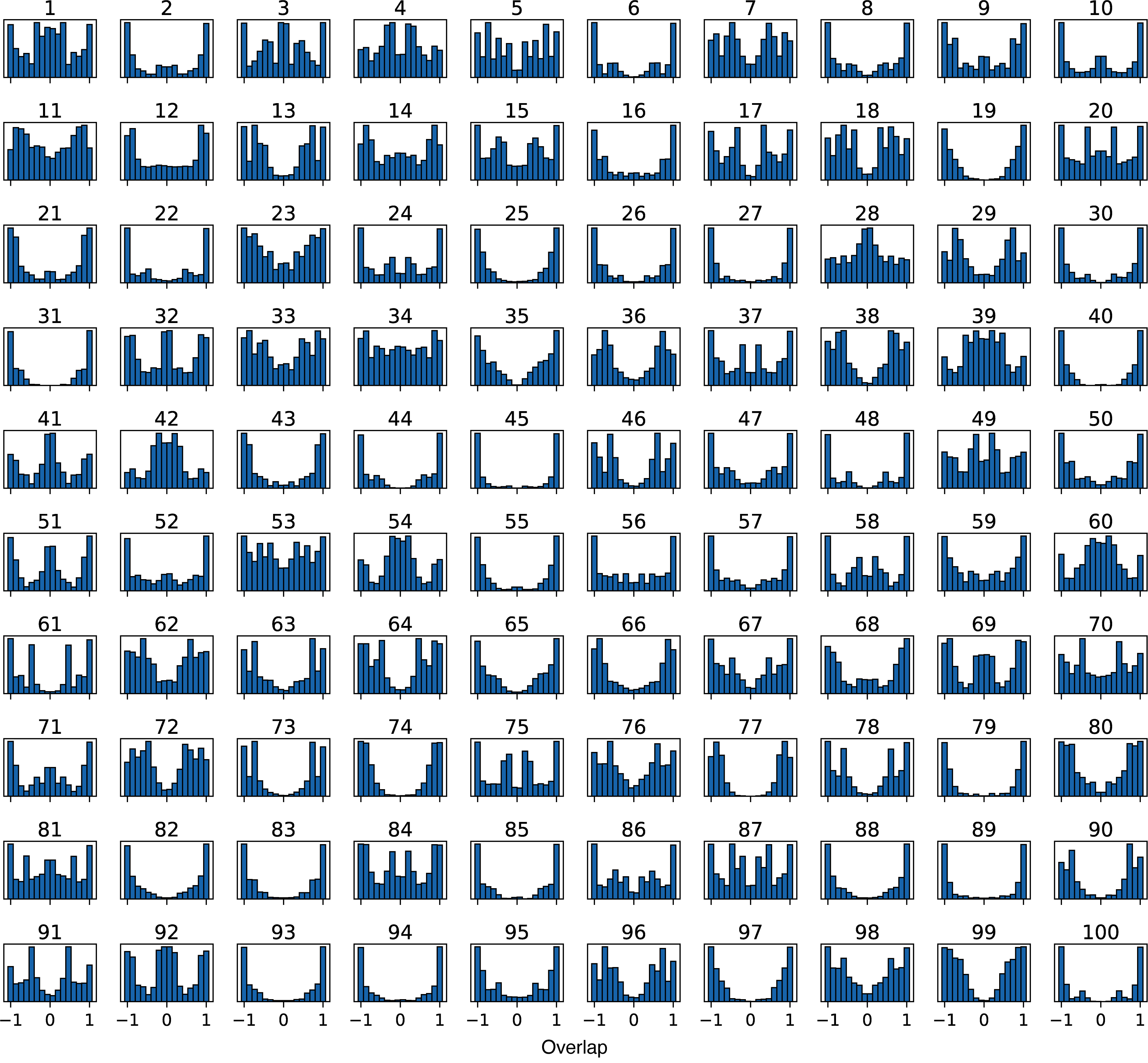}
    \caption{All 100 overlap distributions used in Fig.~\ref{fig:aggOlaps}(a).  The $y$-axis is probability on a linear scale.  The 100 $J$ matrices used are all derived from that realizable in confocal cavities in the spin glass regime, as described in the main text.}
    \label{fig:allOlaps}
\end{figure*}

\section{Bootstrap error analysis}\label{app:bootstrap}

Error bars for overlap and magnetization distributions are estimated via the bootstrap method~\cite{Davison1997bma}. Bootstrap samples are constructed by sampling-with-replacement from the 200 quantum trajectories for each $J$ matrix. For the magnetization distributions, a bootstrap sample of 200 trajectories is generated before computing the magnetization distribution of the sample. The standard deviation across 100 bootstrap samples is computed for each bin in the distribution to provide an error estimate per bin. 

Error estimates for overlap distributions are performed in a different manner to avoid the inclusion of self-overlaps. We choose this method since the self-overlap is not included in any of the overlap distributions in the main figures, which would contribute an asymmetric peak at $q_{\alpha\beta}=1$. Bootstrap samples for the overlap distribution are thus generated by sampling pairs of trajectories, rather than sampling single trajectories like we do for the magnetization distribution. Each pair of trajectories produces a single value of the overlap, which is then binned over all pairs in the bootstrap sample. The self-overlap is avoided by sampling only pairs of two different trajectories. Note that the 200 trajectories we simulate for each $J$ matrix result in $200\times(200-1)/2=19,900$ distinct pairs of trajectories, which each contribute to one value of the overlap distribution. The number of pairs in the bootstrap samples are chosen to match this number. The error estimate for each bin in the overlap distribution is then computed by taking the standard deviation across 100 bootstrap samples.

\vspace{-5mm}
\section{All overlap distributions}\label{app:allOlaps}
\vspace{-5mm}

Figure~\ref{fig:allOlaps} presents all of the overlap distributions for the 100 confocal $J$ matrices in the spin glass regime. Distribution 1 corresponds to the $J$ matrix used for Figs.~\ref{fig:trajectories}-\ref{fig:singleOlap}. Distributions 2-4 are those presented in Fig.~\ref{fig:manyOlaps}. Distribution 5 corresponds to the $J$ matrix considered in Fig.~\ref{fig:ultra}(a). All 100 of the distributions are used to construct the $J$-averaged Parisi distribution in Fig.~\ref{fig:aggOlaps}(a).

A variety of shapes are observed. Many distributions show interior support like peaks or a filling that is smooth. Interior support occurs in the most glassy $J$ matrices where many nearly degenerate minima contribute to the low-energy manifold. Others show less pronounced interior structure. This occurs when the energy landscape is mostly dominated by a single ground state with energy much lower than any other local minimum. Even in these cases, there are often signs of smaller structures in the interior region arising from infrequently found local minima. This distinguishes the system from any ferromagnetic system, where only two goalpost peaks emerge from the single paramagnetic peak as the system cools. 


\begin{thebibliography}{70}%
\makeatletter
\providecommand \@ifxundefined [1]{%
 \@ifx{#1\undefined}
}%
\providecommand \@ifnum [1]{%
 \ifnum #1\expandafter \@firstoftwo
 \else \expandafter \@secondoftwo
 \fi
}%
\providecommand \@ifx [1]{%
 \ifx #1\expandafter \@firstoftwo
 \else \expandafter \@secondoftwo
 \fi
}%
\providecommand \natexlab [1]{#1}%
\providecommand \enquote  [1]{``#1''}%
\providecommand \bibnamefont  [1]{#1}%
\providecommand \bibfnamefont [1]{#1}%
\providecommand \citenamefont [1]{#1}%
\providecommand \href@noop [0]{\@secondoftwo}%
\providecommand \href [0]{\begingroup \@sanitize@url \@href}%
\providecommand \@href[1]{\@@startlink{#1}\@@href}%
\providecommand \@@href[1]{\endgroup#1\@@endlink}%
\providecommand \@sanitize@url [0]{\catcode `\\12\catcode `\$12\catcode `\&12\catcode `\#12\catcode `\^12\catcode `\_12\catcode `\%12\relax}%
\providecommand \@@startlink[1]{}%
\providecommand \@@endlink[0]{}%
\providecommand \url  [0]{\begingroup\@sanitize@url \@url }%
\providecommand \@url [1]{\endgroup\@href {#1}{\urlprefix }}%
\providecommand \urlprefix  [0]{URL }%
\providecommand \Eprint [0]{\href }%
\providecommand \doibase [0]{http://dx.doi.org/}%
\providecommand \selectlanguage [0]{\@gobble}%
\providecommand \bibinfo  [0]{\@secondoftwo}%
\providecommand \bibfield  [0]{\@secondoftwo}%
\providecommand \translation [1]{[#1]}%
\providecommand \BibitemOpen [0]{}%
\providecommand \bibitemStop [0]{}%
\providecommand \bibitemNoStop [0]{.\EOS\space}%
\providecommand \EOS [0]{\spacefactor3000\relax}%
\providecommand \BibitemShut  [1]{\csname bibitem#1\endcsname}%
\let\auto@bib@innerbib\@empty
\bibitem [{\citenamefont {Nishimori}(2001)}]{Nishimori2001spo}%
  \BibitemOpen
  \bibfield  {author} {\bibinfo {author} {\bibfnamefont {H.}~\bibnamefont {Nishimori}},\ }\href {\doibase 10.1093/acprof:oso/9780198509417.001.0001} {\emph {\bibinfo {title} {{Statistical Physics of Spin Glasses and Information Processing}}}}\ (\bibinfo  {publisher} {Oxford University Press},\ \bibinfo {year} {2001})\BibitemShut {NoStop}%
\bibitem [{\citenamefont {Parisi}(1980)}]{Parisi1980top}%
  \BibitemOpen
  \bibfield  {author} {\bibinfo {author} {\bibfnamefont {G.}~\bibnamefont {Parisi}},\ }\enquote {\bibinfo {title} {{The order parameter for spin glasses: a function on the interval 0-1}},}\ \href {\doibase 10.1088/0305-4470/13/3/042} {\bibfield  {journal} {\bibinfo  {journal} {J. Phys. A: Math. Gen.}\ }\textbf {\bibinfo {volume} {13}},\ \bibinfo {pages} {1101} (\bibinfo {year} {1980})}\BibitemShut {NoStop}%
\bibitem [{\citenamefont {Parisi}(1983)}]{Parisi1983opf}%
  \BibitemOpen
  \bibfield  {author} {\bibinfo {author} {\bibfnamefont {G.}~\bibnamefont {Parisi}},\ }\enquote {\bibinfo {title} {{Order Parameter for Spin-Glasses}},}\ \href {\doibase 10.1103/physrevlett.50.1946} {\bibfield  {journal} {\bibinfo  {journal} {Phys. Rev. Lett.}\ }\textbf {\bibinfo {volume} {50}},\ \bibinfo {pages} {1946} (\bibinfo {year} {1983})}\BibitemShut {NoStop}%
\bibitem [{\citenamefont {Sherrington}\ and\ \citenamefont {Kirkpatrick}(1975)}]{Sherrington1975smo}%
  \BibitemOpen
  \bibfield  {author} {\bibinfo {author} {\bibfnamefont {D.}~\bibnamefont {Sherrington}}\ and\ \bibinfo {author} {\bibfnamefont {S.}~\bibnamefont {Kirkpatrick}},\ }\enquote {\bibinfo {title} {{Solvable Model of a Spin-Glass}},}\ \href {\doibase 10.1103/physrevlett.35.1792} {\bibfield  {journal} {\bibinfo  {journal} {Phys. Rev. Lett.}\ }\textbf {\bibinfo {volume} {35}},\ \bibinfo {pages} {1792} (\bibinfo {year} {1975})}\BibitemShut {NoStop}%
\bibitem [{\citenamefont {Stein}\ and\ \citenamefont {Newman}(2013)}]{Stein2013sga}%
  \BibitemOpen
  \bibfield  {author} {\bibinfo {author} {\bibfnamefont {D.~L.}\ \bibnamefont {Stein}}\ and\ \bibinfo {author} {\bibfnamefont {C.~M.}\ \bibnamefont {Newman}},\ }\href {\doibase 10.23943/princeton/9780691147338.001.0001} {\emph {\bibinfo {title} {{Spin Glasses and Complexity}}}},\ Primers in Complex Systems\ (\bibinfo  {publisher} {Princeton University Press},\ \bibinfo {year} {2013})\BibitemShut {NoStop}%
\bibitem [{\citenamefont {Breuer}\ and\ \citenamefont {Petruccione}(2007)}]{Breuer2007tto}%
  \BibitemOpen
  \bibfield  {author} {\bibinfo {author} {\bibfnamefont {H.-P.}\ \bibnamefont {Breuer}}\ and\ \bibinfo {author} {\bibfnamefont {F.}~\bibnamefont {Petruccione}},\ }\href {\doibase 10.1093/acprof:oso/9780199213900.001.0001} {\emph {\bibinfo {title} {{The Theory of Open Quantum Systems}}}}\ (\bibinfo  {publisher} {Oxford University Press},\ \bibinfo {address} {Oxford},\ \bibinfo {year} {2007})\BibitemShut {NoStop}%
\bibitem [{\citenamefont {Daley}(2014)}]{Daley2014qta}%
  \BibitemOpen
  \bibfield  {author} {\bibinfo {author} {\bibfnamefont {A.~J.}\ \bibnamefont {Daley}},\ }\enquote {\bibinfo {title} {{Quantum trajectories and open many-body quantum systems}},}\ \href {\doibase 10.1080/00018732.2014.933502} {\bibfield  {journal} {\bibinfo  {journal} {Adv. Phys.}\ }\textbf {\bibinfo {volume} {63}},\ \bibinfo {pages} {77} (\bibinfo {year} {2014})}\BibitemShut {NoStop}%
\bibitem [{\citenamefont {Verstraelen}\ and\ \citenamefont {Wouters}(2018)}]{Verstraelen2018gqt}%
  \BibitemOpen
  \bibfield  {author} {\bibinfo {author} {\bibfnamefont {W.}~\bibnamefont {Verstraelen}}\ and\ \bibinfo {author} {\bibfnamefont {M.}~\bibnamefont {Wouters}},\ }\enquote {\bibinfo {title} {{Gaussian Quantum Trajectories for the Variational Simulation of Open Quantum-Optical Systems}},}\ \href {\doibase 10.3390/app8091427} {\bibfield  {journal} {\bibinfo  {journal} {Applied Sciences}\ }\textbf {\bibinfo {volume} {8}},\ \bibinfo {pages} {1427} (\bibinfo {year} {2018})}\BibitemShut {NoStop}%
\bibitem [{\citenamefont {Lieu}\ \emph {et~al.}(2020)\citenamefont {Lieu}, \citenamefont {Belyansky}, \citenamefont {Young}, \citenamefont {Lundgren}, \citenamefont {Albert},\ and\ \citenamefont {Gorshkov}}]{lieu2020symmetry}%
  \BibitemOpen
  \bibfield  {author} {\bibinfo {author} {\bibfnamefont {S.}~\bibnamefont {Lieu}}, \bibinfo {author} {\bibfnamefont {R.}~\bibnamefont {Belyansky}}, \bibinfo {author} {\bibfnamefont {J.~T.}\ \bibnamefont {Young}}, \bibinfo {author} {\bibfnamefont {R.}~\bibnamefont {Lundgren}}, \bibinfo {author} {\bibfnamefont {V.~V.}\ \bibnamefont {Albert}}, \ and\ \bibinfo {author} {\bibfnamefont {A.~V.}\ \bibnamefont {Gorshkov}},\ }\enquote {\bibinfo {title} {Symmetry breaking and error correction in open quantum systems},}\ \href@noop {} {\bibfield  {journal} {\bibinfo  {journal} {Physical Review Letters}\ }\textbf {\bibinfo {volume} {125}},\ \bibinfo {pages} {240405} (\bibinfo {year} {2020})}\BibitemShut {NoStop}%
\bibitem [{\citenamefont {Vaidya}\ \emph {et~al.}(2018)\citenamefont {Vaidya}, \citenamefont {Guo}, \citenamefont {Kroeze}, \citenamefont {Ballantine}, \citenamefont {Koll\'{a}r}, \citenamefont {Keeling},\ and\ \citenamefont {Lev}}]{Vaidya2018tpa}%
  \BibitemOpen
  \bibfield  {author} {\bibinfo {author} {\bibfnamefont {V.~D.}\ \bibnamefont {Vaidya}}, \bibinfo {author} {\bibfnamefont {Y.}~\bibnamefont {Guo}}, \bibinfo {author} {\bibfnamefont {R.~M.}\ \bibnamefont {Kroeze}}, \bibinfo {author} {\bibfnamefont {K.~E.}\ \bibnamefont {Ballantine}}, \bibinfo {author} {\bibfnamefont {A.~J.}\ \bibnamefont {Koll\'{a}r}}, \bibinfo {author} {\bibfnamefont {J.}~\bibnamefont {Keeling}}, \ and\ \bibinfo {author} {\bibfnamefont {B.~L.}\ \bibnamefont {Lev}},\ }\enquote {\bibinfo {title} {{Tunable-Range, Photon-Mediated Atomic Interactions in Multimode Cavity QED}},}\ \href {\doibase 10.1103/physrevx.8.011002} {\bibfield  {journal} {\bibinfo  {journal} {Phys. Rev. X}\ }\textbf {\bibinfo {volume} {8}},\ \bibinfo {pages} {011002} (\bibinfo {year} {2018})}\BibitemShut {NoStop}%
\bibitem [{\citenamefont {Guo}\ \emph {et~al.}(2019{\natexlab{a}})\citenamefont {Guo}, \citenamefont {Kroeze}, \citenamefont {Vaidya}, \citenamefont {Keeling},\ and\ \citenamefont {Lev}}]{Guo2019spa}%
  \BibitemOpen
  \bibfield  {author} {\bibinfo {author} {\bibfnamefont {Y.}~\bibnamefont {Guo}}, \bibinfo {author} {\bibfnamefont {R.~M.}\ \bibnamefont {Kroeze}}, \bibinfo {author} {\bibfnamefont {V.~D.}\ \bibnamefont {Vaidya}}, \bibinfo {author} {\bibfnamefont {J.}~\bibnamefont {Keeling}}, \ and\ \bibinfo {author} {\bibfnamefont {B.~L.}\ \bibnamefont {Lev}},\ }\enquote {\bibinfo {title} {{Sign-Changing Photon-Mediated Atom Interactions in Multimode Cavity Quantum Electrodynamics}},}\ \href {\doibase 10.1103/physrevlett.122.193601} {\bibfield  {journal} {\bibinfo  {journal} {Phys. Rev. Lett.}\ }\textbf {\bibinfo {volume} {122}},\ \bibinfo {pages} {193601} (\bibinfo {year} {2019}{\natexlab{a}})}\BibitemShut {NoStop}%
\bibitem [{\citenamefont {Guo}\ \emph {et~al.}(2021)\citenamefont {Guo}, \citenamefont {Kroeze}, \citenamefont {Marsh}, \citenamefont {Gopalakrishnan}, \citenamefont {Keeling},\ and\ \citenamefont {Lev}}]{Guo2021aol}%
  \BibitemOpen
  \bibfield  {author} {\bibinfo {author} {\bibfnamefont {Y.}~\bibnamefont {Guo}}, \bibinfo {author} {\bibfnamefont {R.~M.}\ \bibnamefont {Kroeze}}, \bibinfo {author} {\bibfnamefont {B.~P.}\ \bibnamefont {Marsh}}, \bibinfo {author} {\bibfnamefont {S.}~\bibnamefont {Gopalakrishnan}}, \bibinfo {author} {\bibfnamefont {J.}~\bibnamefont {Keeling}}, \ and\ \bibinfo {author} {\bibfnamefont {B.~L.}\ \bibnamefont {Lev}},\ }\enquote {\bibinfo {title} {{An optical lattice with sound}},}\ \href {\doibase 10.1038/s41586-021-03945-x} {\bibfield  {journal} {\bibinfo  {journal} {Nature}\ }\textbf {\bibinfo {volume} {599}},\ \bibinfo {pages} {211} (\bibinfo {year} {2021})}\BibitemShut {NoStop}%
\bibitem [{\citenamefont {Kroeze}\ \emph {et~al.}(2023{\natexlab{a}})\citenamefont {Kroeze}, \citenamefont {Marsh}, \citenamefont {{Atri Schuller}}, \citenamefont {Hunt}, \citenamefont {Gopalakrishnan}, \citenamefont {Keeling},\ and\ \citenamefont {Lev}}]{Kroeze2023rsb}%
  \BibitemOpen
  \bibfield  {author} {\bibinfo {author} {\bibfnamefont {R.~M.}\ \bibnamefont {Kroeze}}, \bibinfo {author} {\bibfnamefont {B.~P.}\ \bibnamefont {Marsh}}, \bibinfo {author} {\bibfnamefont {D.}~\bibnamefont {{Atri Schuller}}}, \bibinfo {author} {\bibfnamefont {H.~S.}\ \bibnamefont {Hunt}}, \bibinfo {author} {\bibfnamefont {S.}~\bibnamefont {Gopalakrishnan}}, \bibinfo {author} {\bibfnamefont {J.}~\bibnamefont {Keeling}}, \ and\ \bibinfo {author} {\bibfnamefont {B.~L.}\ \bibnamefont {Lev}},\ }\enquote {\bibinfo {title} {{Replica symmetry breaking in a quantum-optical vector spin glass}},}\ \href@noop {} {\  (\bibinfo {year} {2023}{\natexlab{a}})},\ \Eprint {http://arxiv.org/abs/arXiv:2311.04216} {arXiv:2311.04216} \BibitemShut {NoStop}%
\bibitem [{\citenamefont {Marsh}\ \emph {et~al.}(2021)\citenamefont {Marsh}, \citenamefont {Guo}, \citenamefont {Kroeze}, \citenamefont {Gopalakrishnan}, \citenamefont {Ganguli}, \citenamefont {Keeling},\ and\ \citenamefont {Lev}}]{Marsh2021eam}%
  \BibitemOpen
  \bibfield  {author} {\bibinfo {author} {\bibfnamefont {B.~P.}\ \bibnamefont {Marsh}}, \bibinfo {author} {\bibfnamefont {Y.}~\bibnamefont {Guo}}, \bibinfo {author} {\bibfnamefont {R.~M.}\ \bibnamefont {Kroeze}}, \bibinfo {author} {\bibfnamefont {S.}~\bibnamefont {Gopalakrishnan}}, \bibinfo {author} {\bibfnamefont {S.}~\bibnamefont {Ganguli}}, \bibinfo {author} {\bibfnamefont {J.}~\bibnamefont {Keeling}}, \ and\ \bibinfo {author} {\bibfnamefont {B.~L.}\ \bibnamefont {Lev}},\ }\enquote {\bibinfo {title} {{Enhancing Associative Memory Recall and Storage Capacity Using Confocal Cavity QED}},}\ \href {\doibase 10.1103/physrevx.11.021048} {\bibfield  {journal} {\bibinfo  {journal} {Phys. Rev. X}\ }\textbf {\bibinfo {volume} {11}},\ \bibinfo {pages} {021048} (\bibinfo {year} {2021})}\BibitemShut {NoStop}%
\bibitem [{\citenamefont {Gopalakrishnan}\ \emph {et~al.}(2011)\citenamefont {Gopalakrishnan}, \citenamefont {Lev},\ and\ \citenamefont {Goldbart}}]{Gopalakrishnan2011fag}%
  \BibitemOpen
  \bibfield  {author} {\bibinfo {author} {\bibfnamefont {S.}~\bibnamefont {Gopalakrishnan}}, \bibinfo {author} {\bibfnamefont {B.~L.}\ \bibnamefont {Lev}}, \ and\ \bibinfo {author} {\bibfnamefont {P.~M.}\ \bibnamefont {Goldbart}},\ }\enquote {\bibinfo {title} {{Frustration and Glassiness in Spin Models with Cavity-Mediated Interactions}},}\ \href {\doibase 10.1103/physrevlett.107.277201} {\bibfield  {journal} {\bibinfo  {journal} {Phys. Rev. Lett.}\ }\textbf {\bibinfo {volume} {107}},\ \bibinfo {pages} {277201 } (\bibinfo {year} {2011})}\BibitemShut {NoStop}%
\bibitem [{\citenamefont {Strack}\ and\ \citenamefont {Sachdev}(2011)}]{Strack2011dqs}%
  \BibitemOpen
  \bibfield  {author} {\bibinfo {author} {\bibfnamefont {P.}~\bibnamefont {Strack}}\ and\ \bibinfo {author} {\bibfnamefont {S.}~\bibnamefont {Sachdev}},\ }\enquote {\bibinfo {title} {{Dicke Quantum Spin Glass of Atoms and Photons}},}\ \href {\doibase 10.1103/physrevlett.107.277202} {\bibfield  {journal} {\bibinfo  {journal} {Phys. Rev. Lett.}\ }\textbf {\bibinfo {volume} {107}},\ \bibinfo {pages} {277202} (\bibinfo {year} {2011})}\BibitemShut {NoStop}%
\bibitem [{\citenamefont {Gopalakrishnan}\ \emph {et~al.}(2012)\citenamefont {Gopalakrishnan}, \citenamefont {Lev},\ and\ \citenamefont {Goldbart}}]{Gopalakrishnan2012emo}%
  \BibitemOpen
  \bibfield  {author} {\bibinfo {author} {\bibfnamefont {S.}~\bibnamefont {Gopalakrishnan}}, \bibinfo {author} {\bibfnamefont {B.~L.}\ \bibnamefont {Lev}}, \ and\ \bibinfo {author} {\bibfnamefont {P.~M.}\ \bibnamefont {Goldbart}},\ }\enquote {\bibinfo {title} {{Exploring models of associative memory via cavity quantum electrodynamics}},}\ \href {\doibase 10.1080/14786435.2011.637980} {\bibfield  {journal} {\bibinfo  {journal} {Philos. Mag.}\ }\textbf {\bibinfo {volume} {92}},\ \bibinfo {pages} {353} (\bibinfo {year} {2012})}\BibitemShut {NoStop}%
\bibitem [{\citenamefont {Ray}\ \emph {et~al.}(1989)\citenamefont {Ray}, \citenamefont {Chakrabarti},\ and\ \citenamefont {Chakrabarti}}]{Ray1989smi}%
  \BibitemOpen
  \bibfield  {author} {\bibinfo {author} {\bibfnamefont {P.}~\bibnamefont {Ray}}, \bibinfo {author} {\bibfnamefont {B.~K.}\ \bibnamefont {Chakrabarti}}, \ and\ \bibinfo {author} {\bibfnamefont {A.}~\bibnamefont {Chakrabarti}},\ }\enquote {\bibinfo {title} {{Sherrington-Kirkpatrick model in a transverse field: Absence of replica symmetry breaking due to quantum fluctuations}},}\ \href {\doibase 10.1103/physrevb.39.11828} {\bibfield  {journal} {\bibinfo  {journal} {Phys. Rev. B}\ }\textbf {\bibinfo {volume} {39}},\ \bibinfo {pages} {11828} (\bibinfo {year} {1989})}\BibitemShut {NoStop}%
\bibitem [{\citenamefont {B\"uttner}\ and\ \citenamefont {Usadel}(1990)}]{Buttner1990rbf}%
  \BibitemOpen
  \bibfield  {author} {\bibinfo {author} {\bibfnamefont {G.}~\bibnamefont {B\"uttner}}\ and\ \bibinfo {author} {\bibfnamefont {K.~D.}\ \bibnamefont {Usadel}},\ }\enquote {\bibinfo {title} {{Replica-symmetry breaking for the Ising spin glass in a transverse field}},}\ \href {\doibase 10.1103/physrevb.42.6385} {\bibfield  {journal} {\bibinfo  {journal} {Phys. Rev. B}\ }\textbf {\bibinfo {volume} {42}},\ \bibinfo {pages} {6385} (\bibinfo {year} {1990})}\BibitemShut {NoStop}%
\bibitem [{\citenamefont {Rotondo}\ \emph {et~al.}(2015)\citenamefont {Rotondo}, \citenamefont {Tesio},\ and\ \citenamefont {Caracciolo}}]{Rotondo2015rsb}%
  \BibitemOpen
  \bibfield  {author} {\bibinfo {author} {\bibfnamefont {P.}~\bibnamefont {Rotondo}}, \bibinfo {author} {\bibfnamefont {E.}~\bibnamefont {Tesio}}, \ and\ \bibinfo {author} {\bibfnamefont {S.}~\bibnamefont {Caracciolo}},\ }\enquote {\bibinfo {title} {{Replica symmetry breaking in cold atoms and spin glasses}},}\ \href {\doibase 10.1103/physrevb.91.014415} {\bibfield  {journal} {\bibinfo  {journal} {Phys. Rev. B}\ }\textbf {\bibinfo {volume} {91}},\ \bibinfo {pages} {014415} (\bibinfo {year} {2015})}\BibitemShut {NoStop}%
\bibitem [{\citenamefont {Leschke}\ \emph {et~al.}(2021)\citenamefont {Leschke}, \citenamefont {Manai}, \citenamefont {Ruder},\ and\ \citenamefont {Warzel}}]{Leschke2021eor}%
  \BibitemOpen
  \bibfield  {author} {\bibinfo {author} {\bibfnamefont {H.}~\bibnamefont {Leschke}}, \bibinfo {author} {\bibfnamefont {C.}~\bibnamefont {Manai}}, \bibinfo {author} {\bibfnamefont {R.}~\bibnamefont {Ruder}}, \ and\ \bibinfo {author} {\bibfnamefont {S.}~\bibnamefont {Warzel}},\ }\enquote {\bibinfo {title} {{Existence of Replica-Symmetry Breaking in Quantum Glasses}},}\ \href {\doibase 10.1103/physrevlett.127.207204} {\bibfield  {journal} {\bibinfo  {journal} {Phys. Rev. Lett.}\ }\textbf {\bibinfo {volume} {127}},\ \bibinfo {pages} {207204} (\bibinfo {year} {2021})}\BibitemShut {NoStop}%
\bibitem [{\citenamefont {Schindler}\ \emph {et~al.}(2022)\citenamefont {Schindler}, \citenamefont {Guaita}, \citenamefont {Shi}, \citenamefont {Demler},\ and\ \citenamefont {Cirac}}]{Schindler2022vtg}%
  \BibitemOpen
  \bibfield  {author} {\bibinfo {author} {\bibfnamefont {P.~M.}\ \bibnamefont {Schindler}}, \bibinfo {author} {\bibfnamefont {T.}~\bibnamefont {Guaita}}, \bibinfo {author} {\bibfnamefont {T.}~\bibnamefont {Shi}}, \bibinfo {author} {\bibfnamefont {E.}~\bibnamefont {Demler}}, \ and\ \bibinfo {author} {\bibfnamefont {J.~I.}\ \bibnamefont {Cirac}},\ }\enquote {\bibinfo {title} {{Variational Ansatz for the Ground State of the Quantum Sherrington-Kirkpatrick Model}},}\ \href {\doibase 10.1103/physrevlett.129.220401} {\bibfield  {journal} {\bibinfo  {journal} {Phys. Rev. Lett.}\ }\textbf {\bibinfo {volume} {129}},\ \bibinfo {pages} {220401} (\bibinfo {year} {2022})}\BibitemShut {NoStop}%
\bibitem [{\citenamefont {Ghofraniha}\ \emph {et~al.}(2015)\citenamefont {Ghofraniha}, \citenamefont {Viola}, \citenamefont {Di~Maria}, \citenamefont {Barbarella}, \citenamefont {Gigli}, \citenamefont {Leuzzi},\ and\ \citenamefont {Conti}}]{ghofraniha2015eeo}%
  \BibitemOpen
  \bibfield  {author} {\bibinfo {author} {\bibfnamefont {N.}~\bibnamefont {Ghofraniha}}, \bibinfo {author} {\bibfnamefont {I.}~\bibnamefont {Viola}}, \bibinfo {author} {\bibfnamefont {F.}~\bibnamefont {Di~Maria}}, \bibinfo {author} {\bibfnamefont {G.}~\bibnamefont {Barbarella}}, \bibinfo {author} {\bibfnamefont {G.}~\bibnamefont {Gigli}}, \bibinfo {author} {\bibfnamefont {L.}~\bibnamefont {Leuzzi}}, \ and\ \bibinfo {author} {\bibfnamefont {C.}~\bibnamefont {Conti}},\ }\enquote {\bibinfo {title} {{Experimental evidence of replica symmetry breaking in random lasers}},}\ \href {\doibase 10.1038/ncomms7058} {\bibfield  {journal} {\bibinfo  {journal} {Nat. Commun.}\ }\textbf {\bibinfo {volume} {6}},\ \bibinfo {pages} {6058} (\bibinfo {year} {2015})}\BibitemShut {NoStop}%
\bibitem [{\citenamefont {Gomes}\ \emph {et~al.}(2016)\citenamefont {Gomes}, \citenamefont {Raposo}, \citenamefont {Moura}, \citenamefont {Fewo}, \citenamefont {Pincheira}, \citenamefont {Jerez}, \citenamefont {Maia},\ and\ \citenamefont {de~Ara\'ujo}}]{gomes2016ool}%
  \BibitemOpen
  \bibfield  {author} {\bibinfo {author} {\bibfnamefont {A.~S.~L.}\ \bibnamefont {Gomes}}, \bibinfo {author} {\bibfnamefont {E.~P.}\ \bibnamefont {Raposo}}, \bibinfo {author} {\bibfnamefont {A.~L.}\ \bibnamefont {Moura}}, \bibinfo {author} {\bibfnamefont {S.~I.}\ \bibnamefont {Fewo}}, \bibinfo {author} {\bibfnamefont {P.~I.~R.}\ \bibnamefont {Pincheira}}, \bibinfo {author} {\bibfnamefont {V.}~\bibnamefont {Jerez}}, \bibinfo {author} {\bibfnamefont {L.~J.~Q.}\ \bibnamefont {Maia}}, \ and\ \bibinfo {author} {\bibfnamefont {C.~B.}\ \bibnamefont {de~Ara\'ujo}},\ }\enquote {\bibinfo {title} {{Observation of Lévy distribution and replica symmetry breaking in random lasers from a single set of measurements}},}\ \href {\doibase 10.1038/srep27987} {\bibfield  {journal} {\bibinfo  {journal} {Sci. Rep.}\ }\textbf {\bibinfo {volume} {6}},\ \bibinfo {pages} {27987} (\bibinfo {year} {2016})}\BibitemShut {NoStop}%
\bibitem [{\citenamefont {Pierangeli}\ \emph {et~al.}(2017)\citenamefont {Pierangeli}, \citenamefont {Tavani}, \citenamefont {Di~Mei}, \citenamefont {Agranat}, \citenamefont {Conti},\ and\ \citenamefont {DelRe}}]{pierangeli2017oor}%
  \BibitemOpen
  \bibfield  {author} {\bibinfo {author} {\bibfnamefont {D.}~\bibnamefont {Pierangeli}}, \bibinfo {author} {\bibfnamefont {A.}~\bibnamefont {Tavani}}, \bibinfo {author} {\bibfnamefont {F.}~\bibnamefont {Di~Mei}}, \bibinfo {author} {\bibfnamefont {A.~J.}\ \bibnamefont {Agranat}}, \bibinfo {author} {\bibfnamefont {C.}~\bibnamefont {Conti}}, \ and\ \bibinfo {author} {\bibfnamefont {E.}~\bibnamefont {DelRe}},\ }\enquote {\bibinfo {title} {{Observation of replica symmetry breaking in disordered nonlinear wave propagation}},}\ \href {\doibase 10.1038/s41467-017-01612-2} {\bibfield  {journal} {\bibinfo  {journal} {Nat. Commun.}\ }\textbf {\bibinfo {volume} {8}},\ \bibinfo {pages} {1501} (\bibinfo {year} {2017})}\BibitemShut {NoStop}%
\bibitem [{\citenamefont {Skinner}\ \emph {et~al.}(2019)\citenamefont {Skinner}, \citenamefont {Ruhman},\ and\ \citenamefont {Nahum}}]{Skinner2019mpt}%
  \BibitemOpen
  \bibfield  {author} {\bibinfo {author} {\bibfnamefont {B.}~\bibnamefont {Skinner}}, \bibinfo {author} {\bibfnamefont {J.}~\bibnamefont {Ruhman}}, \ and\ \bibinfo {author} {\bibfnamefont {A.}~\bibnamefont {Nahum}},\ }\enquote {\bibinfo {title} {{Measurement-Induced Phase Transitions in the Dynamics of Entanglement}},}\ \href {\doibase 10.1103/physrevx.9.031009} {\bibfield  {journal} {\bibinfo  {journal} {Phys. Rev. X}\ }\textbf {\bibinfo {volume} {9}},\ \bibinfo {pages} {031009} (\bibinfo {year} {2019})}\BibitemShut {NoStop}%
\bibitem [{\citenamefont {Li}\ \emph {et~al.}(2018)\citenamefont {Li}, \citenamefont {Chen},\ and\ \citenamefont {Fisher}}]{Li2018qze}%
  \BibitemOpen
  \bibfield  {author} {\bibinfo {author} {\bibfnamefont {Y.}~\bibnamefont {Li}}, \bibinfo {author} {\bibfnamefont {X.}~\bibnamefont {Chen}}, \ and\ \bibinfo {author} {\bibfnamefont {M.~P.~A.}\ \bibnamefont {Fisher}},\ }\enquote {\bibinfo {title} {{Quantum Zeno effect and the many-body entanglement transition}},}\ \href {\doibase 10.1103/physrevb.98.205136} {\bibfield  {journal} {\bibinfo  {journal} {Phys. Rev. B}\ }\textbf {\bibinfo {volume} {98}},\ \bibinfo {pages} {205136} (\bibinfo {year} {2018})}\BibitemShut {NoStop}%
\bibitem [{Note1()}]{Note1}%
  \BibitemOpen
  \bibinfo {note} {This possibility arises because entanglement and connected correlation functions are nonlinear functions of the state along an individual trajectory.}\BibitemShut {Stop}%
\bibitem [{\citenamefont {Kroeze}\ \emph {et~al.}(2023{\natexlab{b}})\citenamefont {Kroeze}, \citenamefont {Marsh}, \citenamefont {Lin}, \citenamefont {Keeling},\ and\ \citenamefont {Lev}}]{Kroeze2023hcu}%
  \BibitemOpen
  \bibfield  {author} {\bibinfo {author} {\bibfnamefont {R.~M.}\ \bibnamefont {Kroeze}}, \bibinfo {author} {\bibfnamefont {B.~P.}\ \bibnamefont {Marsh}}, \bibinfo {author} {\bibfnamefont {K.-Y.}\ \bibnamefont {Lin}}, \bibinfo {author} {\bibfnamefont {J.}~\bibnamefont {Keeling}}, \ and\ \bibinfo {author} {\bibfnamefont {B.~L.}\ \bibnamefont {Lev}},\ }\enquote {\bibinfo {title} {{High Cooperativity Using a Confocal-Cavity–QED Microscope}},}\ \href {\doibase 10.1103/prxquantum.4.020326} {\bibfield  {journal} {\bibinfo  {journal} {PRX Quantum}\ }\textbf {\bibinfo {volume} {4}},\ \bibinfo {pages} {020326} (\bibinfo {year} {2023}{\natexlab{b}})}\BibitemShut {NoStop}%
\bibitem [{\citenamefont {Guo}\ \emph {et~al.}(2019{\natexlab{b}})\citenamefont {Guo}, \citenamefont {Vaidya}, \citenamefont {Kroeze}, \citenamefont {Lunney}, \citenamefont {Lev},\ and\ \citenamefont {Keeling}}]{Guo2019eab}%
  \BibitemOpen
  \bibfield  {author} {\bibinfo {author} {\bibfnamefont {Y.}~\bibnamefont {Guo}}, \bibinfo {author} {\bibfnamefont {V.~D.}\ \bibnamefont {Vaidya}}, \bibinfo {author} {\bibfnamefont {R.~M.}\ \bibnamefont {Kroeze}}, \bibinfo {author} {\bibfnamefont {R.~A.}\ \bibnamefont {Lunney}}, \bibinfo {author} {\bibfnamefont {B.~L.}\ \bibnamefont {Lev}}, \ and\ \bibinfo {author} {\bibfnamefont {J.}~\bibnamefont {Keeling}},\ }\enquote {\bibinfo {title} {{Emergent and broken symmetries of atomic self-organization arising from Gouy phase shifts in multimode cavity QED}},}\ \href {\doibase 10.1103/physreva.99.053818} {\bibfield  {journal} {\bibinfo  {journal} {Phys. Rev. A}\ }\textbf {\bibinfo {volume} {99}},\ \bibinfo {pages} {053818} (\bibinfo {year} {2019}{\natexlab{b}})}\BibitemShut {NoStop}%
\bibitem [{\citenamefont {Marsh}\ \emph {et~al.}(2023{\natexlab{a}})\citenamefont {Marsh} \emph {et~al.}}]{Marsh2023rydberg}%
  \BibitemOpen
  \bibfield  {author} {\bibinfo {author} {\bibfnamefont {B.~P.}\ \bibnamefont {Marsh}} \emph {et~al.},\ }\href@noop {} {\bibfield  {journal} {\bibinfo  {journal} {in preparation}\ } (\bibinfo {year} {2023}{\natexlab{a}})}\BibitemShut {NoStop}%
\bibitem [{\citenamefont {Kaufman}\ and\ \citenamefont {Ni}(2021)}]{Kaufman2021qsw}%
  \BibitemOpen
  \bibfield  {author} {\bibinfo {author} {\bibfnamefont {A.~M.}\ \bibnamefont {Kaufman}}\ and\ \bibinfo {author} {\bibfnamefont {K.-K.}\ \bibnamefont {Ni}},\ }\enquote {\bibinfo {title} {{Quantum science with optical tweezer arrays of ultracold atoms and molecules}},}\ \href {\doibase 10.1038/s41567-021-01357-2} {\bibfield  {journal} {\bibinfo  {journal} {Nat. Phys.}\ }\textbf {\bibinfo {volume} {17}},\ \bibinfo {pages} {1324} (\bibinfo {year} {2021})}\BibitemShut {NoStop}%
\bibitem [{\citenamefont {Kroeze}\ \emph {et~al.}(2018)\citenamefont {Kroeze}, \citenamefont {Guo}, \citenamefont {Vaidya}, \citenamefont {Keeling},\ and\ \citenamefont {Lev}}]{Kroeze2018sso}%
  \BibitemOpen
  \bibfield  {author} {\bibinfo {author} {\bibfnamefont {R.~M.}\ \bibnamefont {Kroeze}}, \bibinfo {author} {\bibfnamefont {Y.}~\bibnamefont {Guo}}, \bibinfo {author} {\bibfnamefont {V.~D.}\ \bibnamefont {Vaidya}}, \bibinfo {author} {\bibfnamefont {J.}~\bibnamefont {Keeling}}, \ and\ \bibinfo {author} {\bibfnamefont {B.~L.}\ \bibnamefont {Lev}},\ }\enquote {\bibinfo {title} {{Spinor Self-Ordering of a Quantum Gas in a Cavity}},}\ \href {\doibase 10.1103/physrevlett.121.163601} {\bibfield  {journal} {\bibinfo  {journal} {Phys. Rev. Lett.}\ }\textbf {\bibinfo {volume} {121}},\ \bibinfo {pages} {163601} (\bibinfo {year} {2018})}\BibitemShut {NoStop}%
\bibitem [{\citenamefont {Kroeze}\ \emph {et~al.}(2019)\citenamefont {Kroeze}, \citenamefont {Guo},\ and\ \citenamefont {Lev}}]{Kroeze2019dsc}%
  \BibitemOpen
  \bibfield  {author} {\bibinfo {author} {\bibfnamefont {R.~M.}\ \bibnamefont {Kroeze}}, \bibinfo {author} {\bibfnamefont {Y.}~\bibnamefont {Guo}}, \ and\ \bibinfo {author} {\bibfnamefont {B.~L.}\ \bibnamefont {Lev}},\ }\enquote {\bibinfo {title} {{Dynamical Spin-Orbit Coupling of a Quantum Gas}},}\ \href {\doibase 10.1103/physrevlett.123.160404} {\bibfield  {journal} {\bibinfo  {journal} {Phys. Rev. Lett.}\ }\textbf {\bibinfo {volume} {123}},\ \bibinfo {pages} {160404} (\bibinfo {year} {2019})}\BibitemShut {NoStop}%
\bibitem [{\citenamefont {Zhang}\ \emph {et~al.}(2018)\citenamefont {Zhang}, \citenamefont {Lee}, \citenamefont {Kumar}, \citenamefont {Arnold}, \citenamefont {Masson}, \citenamefont {Grimsmo}, \citenamefont {Parkins},\ and\ \citenamefont {Barrett}}]{Zhang2018dsv}%
  \BibitemOpen
  \bibfield  {author} {\bibinfo {author} {\bibfnamefont {Z.}~\bibnamefont {Zhang}}, \bibinfo {author} {\bibfnamefont {C.~H.}\ \bibnamefont {Lee}}, \bibinfo {author} {\bibfnamefont {R.}~\bibnamefont {Kumar}}, \bibinfo {author} {\bibfnamefont {K.~J.}\ \bibnamefont {Arnold}}, \bibinfo {author} {\bibfnamefont {S.~J.}\ \bibnamefont {Masson}}, \bibinfo {author} {\bibfnamefont {A.~L.}\ \bibnamefont {Grimsmo}}, \bibinfo {author} {\bibfnamefont {A.~S.}\ \bibnamefont {Parkins}}, \ and\ \bibinfo {author} {\bibfnamefont {M.~D.}\ \bibnamefont {Barrett}},\ }\enquote {\bibinfo {title} {{Dicke-model simulation via cavity-assisted Raman transitions}},}\ \href {\doibase 10.1103/physreva.97.043858} {\bibfield  {journal} {\bibinfo  {journal} {Phys. Rev. A}\ }\textbf {\bibinfo {volume} {97}},\ \bibinfo {pages} {043858} (\bibinfo {year} {2018})}\BibitemShut {NoStop}%
\bibitem [{\citenamefont {Dimer}\ \emph {et~al.}(2007)\citenamefont {Dimer}, \citenamefont {Estienne}, \citenamefont {Parkins},\ and\ \citenamefont {Carmichael}}]{Dimer2007pro}%
  \BibitemOpen
  \bibfield  {author} {\bibinfo {author} {\bibfnamefont {F.}~\bibnamefont {Dimer}}, \bibinfo {author} {\bibfnamefont {B.}~\bibnamefont {Estienne}}, \bibinfo {author} {\bibfnamefont {A.~S.}\ \bibnamefont {Parkins}}, \ and\ \bibinfo {author} {\bibfnamefont {H.~J.}\ \bibnamefont {Carmichael}},\ }\enquote {\bibinfo {title} {{Proposed realization of the Dicke-model quantum phase transition in an optical cavity QED system}},}\ \href {\doibase 10.1103/physreva.75.013804} {\bibfield  {journal} {\bibinfo  {journal} {Phys. Rev. A}\ }\textbf {\bibinfo {volume} {75}},\ \bibinfo {pages} {013804} (\bibinfo {year} {2007})}\BibitemShut {NoStop}%
\bibitem [{Note2()}]{Note2}%
  \BibitemOpen
  \bibinfo {note} {Confocality occurs when the cavity length $L$ equals its mirrors' radius of curvature $R$~\cite {Siegman1986l}.}\BibitemShut {Stop}%
\bibitem [{Note3()}]{Note3}%
  \BibitemOpen
  \bibinfo {note} {The effective ${N_m}$ can be greater than 1,000, yielding a single atom, synthetic mode cooperativity of more than 110 in practicable systems. See Ref.~\cite {Kroeze2023hcu} for the definition of synthetic mode cooperativity and more details.}\BibitemShut {Stop}%
\bibitem [{\citenamefont {Gopalakrishnan}\ \emph {et~al.}(2009)\citenamefont {Gopalakrishnan}, \citenamefont {Lev},\ and\ \citenamefont {Goldbart}}]{Gopalakrishnan2009eca}%
  \BibitemOpen
  \bibfield  {author} {\bibinfo {author} {\bibfnamefont {S.}~\bibnamefont {Gopalakrishnan}}, \bibinfo {author} {\bibfnamefont {B.~L.}\ \bibnamefont {Lev}}, \ and\ \bibinfo {author} {\bibfnamefont {P.~M.}\ \bibnamefont {Goldbart}},\ }\enquote {\bibinfo {title} {{Emergent crystallinity and frustration with Bose–Einstein condensates in multimode cavities}},}\ \href {\doibase 10.1038/nphys1403} {\bibfield  {journal} {\bibinfo  {journal} {Nature Phys}\ }\textbf {\bibinfo {volume} {5}},\ \bibinfo {pages} {845} (\bibinfo {year} {2009})}\BibitemShut {NoStop}%
\bibitem [{\citenamefont {Kirton}\ \emph {et~al.}(2018)\citenamefont {Kirton}, \citenamefont {Roses}, \citenamefont {Keeling},\ and\ \citenamefont {Dalla~Torre}}]{Kirton2018itt}%
  \BibitemOpen
  \bibfield  {author} {\bibinfo {author} {\bibfnamefont {P.}~\bibnamefont {Kirton}}, \bibinfo {author} {\bibfnamefont {M.~M.}\ \bibnamefont {Roses}}, \bibinfo {author} {\bibfnamefont {J.}~\bibnamefont {Keeling}}, \ and\ \bibinfo {author} {\bibfnamefont {E.~G.}\ \bibnamefont {Dalla~Torre}},\ }\enquote {\bibinfo {title} {{Introduction to the Dicke Model: From Equilibrium to Nonequilibrium, and Vice Versa}},}\ \href {\doibase 10.1002/qute.201800043} {\bibfield  {journal} {\bibinfo  {journal} {Adv. Quantum Technol.}\ }\textbf {\bibinfo {volume} {2}},\ \bibinfo {pages} {1800043} (\bibinfo {year} {2018})}\BibitemShut {NoStop}%
\bibitem [{\citenamefont {Mivehvar}\ \emph {et~al.}(2021)\citenamefont {Mivehvar}, \citenamefont {Piazza}, \citenamefont {Donner},\ and\ \citenamefont {Ritsch}}]{Mivehvar2021cqw}%
  \BibitemOpen
  \bibfield  {author} {\bibinfo {author} {\bibfnamefont {F.}~\bibnamefont {Mivehvar}}, \bibinfo {author} {\bibfnamefont {F.}~\bibnamefont {Piazza}}, \bibinfo {author} {\bibfnamefont {T.}~\bibnamefont {Donner}}, \ and\ \bibinfo {author} {\bibfnamefont {H.}~\bibnamefont {Ritsch}},\ }\enquote {\bibinfo {title} {{Cavity QED with quantum gases: new paradigms in many-body physics}},}\ \href {\doibase 10.1080/00018732.2021.1969727} {\bibfield  {journal} {\bibinfo  {journal} {Adv. Phys.}\ }\textbf {\bibinfo {volume} {70}},\ \bibinfo {pages} {1} (\bibinfo {year} {2021})}\BibitemShut {NoStop}%
\bibitem [{\citenamefont {Erba}\ \emph {et~al.}(2021)\citenamefont {Erba}, \citenamefont {Pastore},\ and\ \citenamefont {Rotondo}}]{Erba2021sgp}%
  \BibitemOpen
  \bibfield  {author} {\bibinfo {author} {\bibfnamefont {V.}~\bibnamefont {Erba}}, \bibinfo {author} {\bibfnamefont {M.}~\bibnamefont {Pastore}}, \ and\ \bibinfo {author} {\bibfnamefont {P.}~\bibnamefont {Rotondo}},\ }\enquote {\bibinfo {title} {{Self-Induced Glassy Phase in Multimodal Cavity Quantum Electrodynamics}},}\ \href {\doibase 10.1103/physrevlett.126.183601} {\bibfield  {journal} {\bibinfo  {journal} {Phys. Rev. Lett.}\ }\textbf {\bibinfo {volume} {126}},\ \bibinfo {pages} {183601} (\bibinfo {year} {2021})}\BibitemShut {NoStop}%
\bibitem [{\citenamefont {J\"ager}\ \emph {et~al.}(2022)\citenamefont {J\"ager}, \citenamefont {Schmit}, \citenamefont {Morigi}, \citenamefont {Holland},\ and\ \citenamefont {Betzholz}}]{Jager2022lme}%
  \BibitemOpen
  \bibfield  {author} {\bibinfo {author} {\bibfnamefont {S.~B.}\ \bibnamefont {J\"ager}}, \bibinfo {author} {\bibfnamefont {T.}~\bibnamefont {Schmit}}, \bibinfo {author} {\bibfnamefont {G.}~\bibnamefont {Morigi}}, \bibinfo {author} {\bibfnamefont {M.~J.}\ \bibnamefont {Holland}}, \ and\ \bibinfo {author} {\bibfnamefont {R.}~\bibnamefont {Betzholz}},\ }\enquote {\bibinfo {title} {{Lindblad Master Equations for Quantum Systems Coupled to Dissipative Bosonic Modes}},}\ \href {\doibase 10.1103/physrevlett.129.063601} {\bibfield  {journal} {\bibinfo  {journal} {Phys. Rev. Lett.}\ }\textbf {\bibinfo {volume} {129}},\ \bibinfo {pages} {063601} (\bibinfo {year} {2022})}\BibitemShut {NoStop}%
\bibitem [{\citenamefont {Vidal}\ and\ \citenamefont {Werner}(2002)}]{Vidal2002cmo}%
  \BibitemOpen
  \bibfield  {author} {\bibinfo {author} {\bibfnamefont {G.}~\bibnamefont {Vidal}}\ and\ \bibinfo {author} {\bibfnamefont {R.~F.}\ \bibnamefont {Werner}},\ }\enquote {\bibinfo {title} {{Computable measure of entanglement}},}\ \href {\doibase 10.1103/physreva.65.032314} {\bibfield  {journal} {\bibinfo  {journal} {Phys. Rev. A}\ }\textbf {\bibinfo {volume} {65}},\ \bibinfo {pages} {032314} (\bibinfo {year} {2002})}\BibitemShut {NoStop}%
\bibitem [{\citenamefont {Suzuki}(1991)}]{Suzuki1991gto}%
  \BibitemOpen
  \bibfield  {author} {\bibinfo {author} {\bibfnamefont {M.}~\bibnamefont {Suzuki}},\ }\enquote {\bibinfo {title} {{General theory of fractal path integrals with applications to many-body theories and statistical physics}},}\ \href {\doibase 10.1063/1.529425} {\bibfield  {journal} {\bibinfo  {journal} {J. Math. Phys.}\ }\textbf {\bibinfo {volume} {32}},\ \bibinfo {pages} {400} (\bibinfo {year} {1991})}\BibitemShut {NoStop}%
\bibitem [{\citenamefont {Trotter}(1959)}]{Trotter1959otp}%
  \BibitemOpen
  \bibfield  {author} {\bibinfo {author} {\bibfnamefont {H.~F.}\ \bibnamefont {Trotter}},\ }\enquote {\bibinfo {title} {{On the product of semi-groups of operators}},}\ \href {\doibase 10.1090/s0002-9939-1959-0108732-6} {\bibfield  {journal} {\bibinfo  {journal} {Proc. Amer. Math. Soc.}\ }\textbf {\bibinfo {volume} {10}},\ \bibinfo {pages} {545} (\bibinfo {year} {1959})}\BibitemShut {NoStop}%
\bibitem [{\citenamefont {Lai}\ and\ \citenamefont {Goldschmidt}(1990)}]{Lai1990mcs}%
  \BibitemOpen
  \bibfield  {author} {\bibinfo {author} {\bibfnamefont {P.-Y.}\ \bibnamefont {Lai}}\ and\ \bibinfo {author} {\bibfnamefont {Y.~Y.}\ \bibnamefont {Goldschmidt}},\ }\enquote {\bibinfo {title} {{Monte Carlo Studies of the Ising Spin-Glass in a Transverse Field}},}\ \href {\doibase 10.1209/0295-5075/13/4/001} {\bibfield  {journal} {\bibinfo  {journal} {Europhys. Lett.}\ }\textbf {\bibinfo {volume} {13}},\ \bibinfo {pages} {289} (\bibinfo {year} {1990})}\BibitemShut {NoStop}%
\bibitem [{\citenamefont {Young}(1983)}]{Young1983ddo}%
  \BibitemOpen
  \bibfield  {author} {\bibinfo {author} {\bibfnamefont {A.~P.}\ \bibnamefont {Young}},\ }\enquote {\bibinfo {title} {{Direct Determination of the Probability Distribution for the Spin-Glass Order Parameter}},}\ \href {\doibase 10.1103/physrevlett.51.1206} {\bibfield  {journal} {\bibinfo  {journal} {Phys. Rev. Lett.}\ }\textbf {\bibinfo {volume} {51}},\ \bibinfo {pages} {1206} (\bibinfo {year} {1983})}\BibitemShut {NoStop}%
\bibitem [{\citenamefont {Binder}(1981)}]{Binder1981cpf}%
  \BibitemOpen
  \bibfield  {author} {\bibinfo {author} {\bibfnamefont {K.}~\bibnamefont {Binder}},\ }\enquote {\bibinfo {title} {{Critical Properties from Monte Carlo Coarse Graining and Renormalization}},}\ \href {\doibase 10.1103/physrevlett.47.693} {\bibfield  {journal} {\bibinfo  {journal} {Phys. Rev. Lett.}\ }\textbf {\bibinfo {volume} {47}},\ \bibinfo {pages} {693} (\bibinfo {year} {1981})}\BibitemShut {NoStop}%
\bibitem [{\citenamefont {M\'ezard}\ \emph {et~al.}(1984)\citenamefont {M\'ezard}, \citenamefont {Parisi}, \citenamefont {Sourlas}, \citenamefont {Toulouse},\ and\ \citenamefont {Virasoro}}]{Mezard1984rsb}%
  \BibitemOpen
  \bibfield  {author} {\bibinfo {author} {\bibfnamefont {M.}~\bibnamefont {M\'ezard}}, \bibinfo {author} {\bibfnamefont {G.}~\bibnamefont {Parisi}}, \bibinfo {author} {\bibfnamefont {N.}~\bibnamefont {Sourlas}}, \bibinfo {author} {\bibfnamefont {G.}~\bibnamefont {Toulouse}}, \ and\ \bibinfo {author} {\bibfnamefont {M.}~\bibnamefont {Virasoro}},\ }\enquote {\bibinfo {title} {{Replica symmetry breaking and the nature of the spin glass phase}},}\ \href {\doibase 10.1051/jphys:01984004505084300} {\bibfield  {journal} {\bibinfo  {journal} {J. Phys. France}\ }\textbf {\bibinfo {volume} {45}},\ \bibinfo {pages} {843} (\bibinfo {year} {1984})}\BibitemShut {NoStop}%
\bibitem [{\citenamefont {Bhatt}\ and\ \citenamefont {Young}(1989)}]{Bhatt1989uit}%
  \BibitemOpen
  \bibfield  {author} {\bibinfo {author} {\bibfnamefont {R.~N.}\ \bibnamefont {Bhatt}}\ and\ \bibinfo {author} {\bibfnamefont {A.~P.}\ \bibnamefont {Young}},\ }\enquote {\bibinfo {title} {{Ultrametricity in the infinite-range Ising spin glass}},}\ \href {\doibase 10.1088/0953-8984/1/18/005} {\bibfield  {journal} {\bibinfo  {journal} {J. Phys.: Condens. Matter}\ }\textbf {\bibinfo {volume} {1}},\ \bibinfo {pages} {2997} (\bibinfo {year} {1989})}\BibitemShut {NoStop}%
\bibitem [{\citenamefont {Katzgraber}\ and\ \citenamefont {Hartmann}(2009)}]{Katzgraber2009uac}%
  \BibitemOpen
  \bibfield  {author} {\bibinfo {author} {\bibfnamefont {H.~G.}\ \bibnamefont {Katzgraber}}\ and\ \bibinfo {author} {\bibfnamefont {A.~K.}\ \bibnamefont {Hartmann}},\ }\enquote {\bibinfo {title} {{Ultrametricity and Clustering of States in Spin Glasses: A One-Dimensional View}},}\ \href {\doibase 10.1103/physrevlett.102.037207} {\bibfield  {journal} {\bibinfo  {journal} {Phys. Rev. Lett.}\ }\textbf {\bibinfo {volume} {102}},\ \bibinfo {pages} {037207} (\bibinfo {year} {2009})}\BibitemShut {NoStop}%
\bibitem [{\citenamefont {Katzgraber}\ \emph {et~al.}(2012)\citenamefont {Katzgraber}, \citenamefont {J\"org}, \citenamefont {Krz\c{a}kała},\ and\ \citenamefont {Hartmann}}]{Katzgraber2012upo}%
  \BibitemOpen
  \bibfield  {author} {\bibinfo {author} {\bibfnamefont {H.~G.}\ \bibnamefont {Katzgraber}}, \bibinfo {author} {\bibfnamefont {T.}~\bibnamefont {J\"org}}, \bibinfo {author} {\bibfnamefont {F.}~\bibnamefont {Krz\c{a}kała}}, \ and\ \bibinfo {author} {\bibfnamefont {A.~K.}\ \bibnamefont {Hartmann}},\ }\enquote {\bibinfo {title} {{Ultrametric probe of the spin-glass state in a field}},}\ \href {\doibase 10.1103/physrevb.86.184405} {\bibfield  {journal} {\bibinfo  {journal} {Phys. Rev. B}\ }\textbf {\bibinfo {volume} {86}},\ \bibinfo {pages} {184405} (\bibinfo {year} {2012})}\BibitemShut {NoStop}%
\bibitem [{\citenamefont {Dalla~Torre}\ \emph {et~al.}(2013)\citenamefont {Dalla~Torre}, \citenamefont {Diehl}, \citenamefont {Lukin}, \citenamefont {Sachdev},\ and\ \citenamefont {Strack}}]{Torre2013kaf}%
  \BibitemOpen
  \bibfield  {author} {\bibinfo {author} {\bibfnamefont {E.~G.}\ \bibnamefont {Dalla~Torre}}, \bibinfo {author} {\bibfnamefont {S.}~\bibnamefont {Diehl}}, \bibinfo {author} {\bibfnamefont {M.~D.}\ \bibnamefont {Lukin}}, \bibinfo {author} {\bibfnamefont {S.}~\bibnamefont {Sachdev}}, \ and\ \bibinfo {author} {\bibfnamefont {P.}~\bibnamefont {Strack}},\ }\enquote {\bibinfo {title} {{Keldysh approach for nonequilibrium phase transitions in quantum optics: Beyond the Dicke model in optical cavities}},}\ \href {\doibase 10.1103/physreva.87.023831} {\bibfield  {journal} {\bibinfo  {journal} {Phys. Rev. A}\ }\textbf {\bibinfo {volume} {87}},\ \bibinfo {pages} {023831} (\bibinfo {year} {2013})}\BibitemShut {NoStop}%
\bibitem [{\citenamefont {Lukin}\ \emph {et~al.}(2001)\citenamefont {Lukin}, \citenamefont {Fleischhauer}, \citenamefont {Cote}, \citenamefont {Duan}, \citenamefont {Jaksch}, \citenamefont {Cirac},\ and\ \citenamefont {Zoller}}]{Lukin2001dba}%
  \BibitemOpen
  \bibfield  {author} {\bibinfo {author} {\bibfnamefont {M.~D.}\ \bibnamefont {Lukin}}, \bibinfo {author} {\bibfnamefont {M.}~\bibnamefont {Fleischhauer}}, \bibinfo {author} {\bibfnamefont {R.}~\bibnamefont {Cote}}, \bibinfo {author} {\bibfnamefont {L.~M.}\ \bibnamefont {Duan}}, \bibinfo {author} {\bibfnamefont {D.}~\bibnamefont {Jaksch}}, \bibinfo {author} {\bibfnamefont {J.~I.}\ \bibnamefont {Cirac}}, \ and\ \bibinfo {author} {\bibfnamefont {P.}~\bibnamefont {Zoller}},\ }\enquote {\bibinfo {title} {{Dipole Blockade and Quantum Information Processing in Mesoscopic Atomic Ensembles}},}\ \href {\doibase 10.1103/physrevlett.87.037901} {\bibfield  {journal} {\bibinfo  {journal} {Phys. Rev. Lett.}\ }\textbf {\bibinfo {volume} {87}},\ \bibinfo {pages} {037901} (\bibinfo {year} {2001})}\BibitemShut {NoStop}%
\bibitem [{\citenamefont {Ningyuan}\ \emph {et~al.}(2016)\citenamefont {Ningyuan}, \citenamefont {Georgakopoulos}, \citenamefont {Ryou}, \citenamefont {Schine}, \citenamefont {Sommer},\ and\ \citenamefont {Simon}}]{Ningyuan2016oac}%
  \BibitemOpen
  \bibfield  {author} {\bibinfo {author} {\bibfnamefont {J.}~\bibnamefont {Ningyuan}}, \bibinfo {author} {\bibfnamefont {A.}~\bibnamefont {Georgakopoulos}}, \bibinfo {author} {\bibfnamefont {A.}~\bibnamefont {Ryou}}, \bibinfo {author} {\bibfnamefont {N.}~\bibnamefont {Schine}}, \bibinfo {author} {\bibfnamefont {A.}~\bibnamefont {Sommer}}, \ and\ \bibinfo {author} {\bibfnamefont {J.}~\bibnamefont {Simon}},\ }\enquote {\bibinfo {title} {{Observation and characterization of cavity Rydberg polaritons}},}\ \href {\doibase 10.1103/physreva.93.041802} {\bibfield  {journal} {\bibinfo  {journal} {Phys. Rev. A}\ }\textbf {\bibinfo {volume} {93}} (\bibinfo {year} {2016}),}\BibitemShut {NoStop}%
\bibitem [{\citenamefont {Torggler}\ \emph {et~al.}(2017)\citenamefont {Torggler}, \citenamefont {Kr\"amer},\ and\ \citenamefont {Ritsch}}]{Torggler2017qaw}%
  \BibitemOpen
  \bibfield  {author} {\bibinfo {author} {\bibfnamefont {V.}~\bibnamefont {Torggler}}, \bibinfo {author} {\bibfnamefont {S.}~\bibnamefont {Kr\"amer}}, \ and\ \bibinfo {author} {\bibfnamefont {H.}~\bibnamefont {Ritsch}},\ }\enquote {\bibinfo {title} {{Quantum annealing with ultracold atoms in a multimode optical resonator}},}\ \href {\doibase 10.1103/physreva.95.032310} {\bibfield  {journal} {\bibinfo  {journal} {Phys. Rev. A}\ }\textbf {\bibinfo {volume} {95}},\ \bibinfo {pages} {032310} (\bibinfo {year} {2017})}\BibitemShut {NoStop}%
\bibitem [{\citenamefont {Rotondo}\ \emph {et~al.}(2018)\citenamefont {Rotondo}, \citenamefont {Marcuzzi}, \citenamefont {Garrahan}, \citenamefont {Lesanovsky},\ and\ \citenamefont {M\"uller}}]{Rotondo2018oqg}%
  \BibitemOpen
  \bibfield  {author} {\bibinfo {author} {\bibfnamefont {P.}~\bibnamefont {Rotondo}}, \bibinfo {author} {\bibfnamefont {M.}~\bibnamefont {Marcuzzi}}, \bibinfo {author} {\bibfnamefont {J.~P.}\ \bibnamefont {Garrahan}}, \bibinfo {author} {\bibfnamefont {I.}~\bibnamefont {Lesanovsky}}, \ and\ \bibinfo {author} {\bibfnamefont {M.}~\bibnamefont {M\"uller}},\ }\enquote {\bibinfo {title} {{Open quantum generalisation of Hopfield neural networks}},}\ \href {\doibase 10.1088/1751-8121/aaabcb} {\bibfield  {journal} {\bibinfo  {journal} {J. Phys. A: Math. Theor.}\ }\textbf {\bibinfo {volume} {51}},\ \bibinfo {pages} {115301} (\bibinfo {year} {2018})}\BibitemShut {NoStop}%
\bibitem [{\citenamefont {Fiorelli}\ \emph {et~al.}(2019)\citenamefont {Fiorelli}, \citenamefont {Rotondo}, \citenamefont {Marcuzzi}, \citenamefont {Garrahan},\ and\ \citenamefont {Lesanovsky}}]{Fiorelli2019qaa}%
  \BibitemOpen
  \bibfield  {author} {\bibinfo {author} {\bibfnamefont {E.}~\bibnamefont {Fiorelli}}, \bibinfo {author} {\bibfnamefont {P.}~\bibnamefont {Rotondo}}, \bibinfo {author} {\bibfnamefont {M.}~\bibnamefont {Marcuzzi}}, \bibinfo {author} {\bibfnamefont {J.~P.}\ \bibnamefont {Garrahan}}, \ and\ \bibinfo {author} {\bibfnamefont {I.}~\bibnamefont {Lesanovsky}},\ }\enquote {\bibinfo {title} {{Quantum accelerated approach to the thermal state of classical all-to-all connected spin systems with applications to pattern retrieval in the Hopfield neural network}},}\ \href {\doibase 10.1103/physreva.99.032126} {\bibfield  {journal} {\bibinfo  {journal} {Phys. Rev. A}\ }\textbf {\bibinfo {volume} {99}},\ \bibinfo {pages} {032126} (\bibinfo {year} {2019})}\BibitemShut {NoStop}%
\bibitem [{\citenamefont {Torggler}\ \emph {et~al.}(2019)\citenamefont {Torggler}, \citenamefont {Aumann}, \citenamefont {Ritsch},\ and\ \citenamefont {Lechner}}]{Torggler2019aqn}%
  \BibitemOpen
  \bibfield  {author} {\bibinfo {author} {\bibfnamefont {V.}~\bibnamefont {Torggler}}, \bibinfo {author} {\bibfnamefont {P.}~\bibnamefont {Aumann}}, \bibinfo {author} {\bibfnamefont {H.}~\bibnamefont {Ritsch}}, \ and\ \bibinfo {author} {\bibfnamefont {W.}~\bibnamefont {Lechner}},\ }\enquote {\bibinfo {title} {{A Quantum N-Queens Solver}},}\ \href {\doibase 10.22331/q-2019-06-03-149} {\bibfield  {journal} {\bibinfo  {journal} {Quantum}\ }\textbf {\bibinfo {volume} {3}},\ \bibinfo {pages} {149} (\bibinfo {year} {2019})}\BibitemShut {NoStop}%
\bibitem [{\citenamefont {Fiorelli}\ \emph {et~al.}(2020)\citenamefont {Fiorelli}, \citenamefont {Marcuzzi}, \citenamefont {Rotondo}, \citenamefont {Carollo},\ and\ \citenamefont {Lesanovsky}}]{Fiorelli2020soa}%
  \BibitemOpen
  \bibfield  {author} {\bibinfo {author} {\bibfnamefont {E.}~\bibnamefont {Fiorelli}}, \bibinfo {author} {\bibfnamefont {M.}~\bibnamefont {Marcuzzi}}, \bibinfo {author} {\bibfnamefont {P.}~\bibnamefont {Rotondo}}, \bibinfo {author} {\bibfnamefont {F.}~\bibnamefont {Carollo}}, \ and\ \bibinfo {author} {\bibfnamefont {I.}~\bibnamefont {Lesanovsky}},\ }\enquote {\bibinfo {title} {{Signatures of Associative Memory Behavior in a Multimode Dicke Model}},}\ \href {\doibase 10.1103/physrevlett.125.070604} {\bibfield  {journal} {\bibinfo  {journal} {Phys. Rev. Lett.}\ }\textbf {\bibinfo {volume} {125}},\ \bibinfo {pages} {070604} (\bibinfo {year} {2020})}\BibitemShut {NoStop}%
\bibitem [{\citenamefont {Marsh}\ \emph {et~al.}(2023{\natexlab{b}})\citenamefont {Marsh}, \citenamefont {Kroeze}, \citenamefont {Ganguli}, \citenamefont {Gopalakrishnan}, \citenamefont {Keeling},\ and\ \citenamefont {Lev}}]{Marsh2023data}%
  \BibitemOpen
  \bibfield  {author} {\bibinfo {author} {\bibfnamefont {B.}~\bibnamefont {Marsh}}, \bibinfo {author} {\bibfnamefont {R.}~\bibnamefont {Kroeze}}, \bibinfo {author} {\bibfnamefont {S.}~\bibnamefont {Ganguli}}, \bibinfo {author} {\bibfnamefont {S.}~\bibnamefont {Gopalakrishnan}}, \bibinfo {author} {\bibfnamefont {J.}~\bibnamefont {Keeling}}, \ and\ \bibinfo {author} {\bibfnamefont {B.}~\bibnamefont {Lev}},\ }\enquote {\bibinfo {title} {Entanglement and replica symmetry breaking in a driven-dissipative quantum spin glass (dataset)},}\ \href {\doibase https://doi.org/10.7910/DVN/IBQDXA} {\  (\bibinfo {year} {2023}{\natexlab{b}}),\ https://doi.org/10.7910/DVN/IBQDXA}\BibitemShut {NoStop}%
\bibitem [{\citenamefont {Zhang}(2005)}]{Zhang2005sca}%
  \BibitemOpen
  \bibfield  {author} {\bibinfo {author} {\bibfnamefont {F.}~\bibnamefont {Zhang}},\ }\href {\doibase 10.1007/b105056} {\emph {\bibinfo {title} {{The Schur complement and its applications}}}}\ (\bibinfo  {publisher} {Springer Science \& Business Media},\ \bibinfo {year} {2005})\BibitemShut {NoStop}%
\bibitem [{\citenamefont {Goldschmidt}\ and\ \citenamefont {Lai}(1990)}]{Goldschmidt1990isg}%
  \BibitemOpen
  \bibfield  {author} {\bibinfo {author} {\bibfnamefont {Y.~Y.}\ \bibnamefont {Goldschmidt}}\ and\ \bibinfo {author} {\bibfnamefont {P.-Y.}\ \bibnamefont {Lai}},\ }\enquote {\bibinfo {title} {{Ising spin glass in a transverse field: Replica-symmetry-breaking solution}},}\ \href {\doibase 10.1103/physrevlett.64.2467} {\bibfield  {journal} {\bibinfo  {journal} {Phys. Rev. Lett.}\ }\textbf {\bibinfo {volume} {64}},\ \bibinfo {pages} {2467} (\bibinfo {year} {1990})}\BibitemShut {NoStop}%
\bibitem [{\citenamefont {Charbonneau}\ \emph {et~al.}(2023)\citenamefont {Charbonneau}, \citenamefont {Marinari}, \citenamefont {M\'{e}zard}, \citenamefont {Parisi}, \citenamefont {Ricci-Tersenghi}, \citenamefont {Sicuro},\ and\ \citenamefont {Zamponi}}]{Charbonneau2023sgt}%
  \BibitemOpen
  \bibfield  {author} {\bibinfo {author} {\bibfnamefont {P.}~\bibnamefont {Charbonneau}}, \bibinfo {author} {\bibfnamefont {E.}~\bibnamefont {Marinari}}, \bibinfo {author} {\bibfnamefont {M.}~\bibnamefont {M\'{e}zard}}, \bibinfo {author} {\bibfnamefont {G.}~\bibnamefont {Parisi}}, \bibinfo {author} {\bibfnamefont {F.}~\bibnamefont {Ricci-Tersenghi}}, \bibinfo {author} {\bibfnamefont {G.}~\bibnamefont {Sicuro}}, \ and\ \bibinfo {author} {\bibfnamefont {F.}~\bibnamefont {Zamponi}},\ }\href {\doibase https://doi.org/10.1142/13341} {\emph {\bibinfo {title} {{Spin Glass Theory and Far Beyond}}}}\ (\bibinfo  {publisher} {World Scientific},\ \bibinfo {year} {2023})\BibitemShut {NoStop}%
\bibitem [{\citenamefont {Yamamoto}\ and\ \citenamefont {Ishii}(1987)}]{Yamamoto1987ape}%
  \BibitemOpen
  \bibfield  {author} {\bibinfo {author} {\bibfnamefont {T.}~\bibnamefont {Yamamoto}}\ and\ \bibinfo {author} {\bibfnamefont {H.}~\bibnamefont {Ishii}},\ }\enquote {\bibinfo {title} {{A perturbation expansion for the Sherrington-Kirkpatrick model with a transverse field}},}\ \href {\doibase 10.1088/0022-3719/20/35/020} {\bibfield  {journal} {\bibinfo  {journal} {J. Phys. C: Solid State Phys.}\ }\textbf {\bibinfo {volume} {20}},\ \bibinfo {pages} {6053} (\bibinfo {year} {1987})}\BibitemShut {NoStop}%
\bibitem [{\citenamefont {Young}(2017)}]{Young2017sot}%
  \BibitemOpen
  \bibfield  {author} {\bibinfo {author} {\bibfnamefont {A.~P.}\ \bibnamefont {Young}},\ }\enquote {\bibinfo {title} {{Stability of the quantum Sherrington-Kirkpatrick spin glass model}},}\ \href {\doibase 10.1103/physreve.96.032112} {\bibfield  {journal} {\bibinfo  {journal} {Phys. Rev. E}\ }\textbf {\bibinfo {volume} {96}},\ \bibinfo {pages} {032112} (\bibinfo {year} {2017})}\BibitemShut {NoStop}%
\bibitem [{\citenamefont {Binder}\ and\ \citenamefont {Young}(1986)}]{Binder1986sge}%
  \BibitemOpen
  \bibfield  {author} {\bibinfo {author} {\bibfnamefont {K.}~\bibnamefont {Binder}}\ and\ \bibinfo {author} {\bibfnamefont {A.~P.}\ \bibnamefont {Young}},\ }\enquote {\bibinfo {title} {{Spin glasses: Experimental facts, theoretical concepts, and open questions}},}\ \href {\doibase 10.1103/revmodphys.58.801} {\bibfield  {journal} {\bibinfo  {journal} {Rev. Mod. Phys.}\ }\textbf {\bibinfo {volume} {58}},\ \bibinfo {pages} {801} (\bibinfo {year} {1986})}\BibitemShut {NoStop}%
\bibitem [{\citenamefont {Davison}\ and\ \citenamefont {Hinkley}(1997)}]{Davison1997bma}%
  \BibitemOpen
  \bibfield  {author} {\bibinfo {author} {\bibfnamefont {A.~C.}\ \bibnamefont {Davison}}\ and\ \bibinfo {author} {\bibfnamefont {D.~V.}\ \bibnamefont {Hinkley}},\ }\href {\doibase 10.1017/cbo9780511802843} {\emph {\bibinfo {title} {{Bootstrap Methods and their Application}}}},\ \bibinfo {number} {1}\ (\bibinfo  {publisher} {Cambridge University Press},\ \bibinfo {year} {1997})\BibitemShut {NoStop}%
\bibitem [{\citenamefont {Siegman}(1986)}]{Siegman1986l}%
  \BibitemOpen
  \bibfield  {author} {\bibinfo {author} {\bibfnamefont {A.~E.}\ \bibnamefont {Siegman}},\ }\href {https://www.osapublishing.org/books/bookshelf/lasers.cfm} {\emph {\bibinfo {title} {{Lasers}}}}\ (\bibinfo  {publisher} {University Science Books},\ \bibinfo {address} {Sausolito, CA},\ \bibinfo {year} {1986})\BibitemShut {NoStop}%
\end{thebibliography}

%

\end{document}